\documentclass[%
 reprint,
superscriptaddress,
nofootinbib,
 amsmath,amssymb,
 aps,
prb,
]{revtex4-2}

\usepackage{physics}
\usepackage{xcolor}
\usepackage{graphicx}
\usepackage{bm}
\usepackage{hyperref}

\newtheorem{theorem}{Theorem}
\newtheorem{lemma}{Lemma}

\hypersetup{
    colorlinks=true,
    linkcolor=blue,
    urlcolor=cyan,
    citecolor=red,
    }
\usepackage[normalem]{ulem}
\usepackage{verbatim}
\usepackage{enumitem}
\usepackage{booktabs}
\usepackage{multirow}

\makeatletter
\def\l@subsubsection#1#2{}
\makeatother

\usepackage[english]{babel}

\usepackage{amsmath,amssymb}

\begin{document}

\title{Deep thermalization under charge-conserving quantum dynamics} %
\author{Rui-An Chang}
\affiliation{Department of Physics, The University of Texas at Austin, Austin, TX 78712, USA}
\author{Harshank Shrotriya}
\affiliation{Centre for Quantum Technologies, National University of Singapore,   Singapore 117543}
\author{Wen Wei Ho}
\affiliation{Centre for Quantum Technologies, National University of Singapore,   Singapore 117543}
\affiliation{Department of Physics, National University of Singapore, Singapore 117551}
\author{Matteo Ippoliti}
\affiliation{Department of Physics, The University of Texas at Austin, Austin, TX 78712, USA}

\begin{abstract}
``Deep thermalization'' describes the emergence of universal wavefunction distributions in quantum many-body dynamics, appearing on a local subsystem upon measurement of its environment.
In this work, we study in detail the effect of continuous internal symmetries and associated conservation laws on deep thermalization. Concretely, we consider quantum spin systems with a $U(1)$ symmetry associated with the conservation of magnetization (or `charge'), and analyze how the choice of initial states (specifically, their degree of charge fluctuations) and the choice of measurement basis (specifically, whether or not it can reveal information about the local charge density) determine the ensuing universal wavefunction distributions. 
We put forth a universal ansatz for the limiting form of the projected ensemble, motivated by maximum-entropy principles rooted in statistical physics and quantum information theory. This limiting form depends on a polynomial amount of data on the initial state and measurement basis, a `coarse-graining' that is an essential feature of {\it bona fide} thermodynamic ensembles.
We support our ansatz with three complementary approaches:
(i) a rigorous proof in the simplest case of no charge fluctuations in either the initial state or the measurement basis; 
(ii) analytical calculations using a `replica limit' approach, applicable when charge fluctuations are allowed in either the input state or the measurement basis but not both;
(iii) extensive numerical simulations of finite-sized systems in the most general case. 
Our findings demonstrate a rich interplay between symmetries and the information extracted by measurements, which allows deep thermalization to exhibit a range of universal behaviors far beyond regular thermalization. 
\end{abstract}

\maketitle

\tableofcontents


\section{Introduction}
The advent of quantum computers and simulators has opened up new ways to address fundamental questions in quantum many-body physics. Among these questions is the emergence of irreversible equilibration from the reversible unitary dynamics of isolated quantum many-body systems.   
This apparent paradox is resolved by recognizing that equilibration within an isolated quantum system should be considered only within a {\it local} subsystem, such that there is a large complementary subsystem (the `bath') with which the former can
exchange information, energy, and any other conserved quantities.
Mathematically, this is captured by  the reduced density matrix $\rho_A = {\rm Tr}_B( \ketbra{\Psi} )$ --- a description of the local subsystem $A$ in which the complement $B$ (`the bath') is ignored. This state quickly becomes mixed due to the build-up of quantum entanglement, and typically converges to the form of a Gibbs state, which is universal and insensitive to details of the initial state, up to information about any conserved quantities. This is what is conventionally known as `quantum thermalization'~\cite{srednicki_chaos_1994,rigol_thermalization_2008,nandkishore_many-body_2015,dalessio_quantum_2016,kaufman_quantum_2016,abanin_colloquium_2019}.  

However, in modern quantum simulator experiments, it is possible for an observer to obtain information about {\it all} of the system's degrees of freedom at once, e.g.~by single-atom resolved fluorescence. This yields `snapshots' of the different classical configurations that the entire system can be found in, typically in the form of bit-strings (each bit denoting, e.g., the occupation of a lattice site).
In this scenario, the separation between a subsystem of interest $A$ and the bath $B$ is arbitrary and artificial, and even if we wish to study local properties of $A$, we need not discard the data from $B$. 
Indeed, one can study the local equilibration of a quantum many-body system {\it conditioned} upon the measurement outcome of the complement, as was done in Refs.~\cite{choi_preparing_2023,cotler_emergent_2023,ho_exact_2022,ippoliti_dynamical_2023,ippoliti_solvable_2022,bhore_deep_2023,lucas_generalized_2023,mark_maximum_2024,liu_deep_2024,zhang2025holographicdeepthermalizationtheory,milekhin2024observableprojectedensembles,mok2024optimalconversionclassicalquantum, du2024embeddedcomplexityquantumcircuit,matchgate2024}. The relevant theoretical construct is the \textit{projected ensemble} --- an ensemble of pure states of $A$, conditioned and weighted by the outcomes of projective measurements performed on the bath $B$, as sketched in Fig.~\ref{fig:system_AB}.
One may view the projected ensemble as a particular, physically motivated unraveling of the reduced density matrix into a set of constituent pure states. 

The novelty and interest in the projected ensemble lies in the fact that it can exhibit new forms of universal equilibration physics, dubbed ``deep thermalization'', which goes beyond the framework of standard quantum thermalization. 
Initial studies concentrated on the conceptually simplest case of dynamics without conservation laws or at infinite temperature~\cite{cotler_emergent_2023}. There it was found (numerically, analytically, and also experimentally~\cite{choi_preparing_2023}) that the projected ensemble tends at late times to a distribution in which quantum states are drawn uniformly randomly from the Hilbert space, i.e., the Haar distribution~\cite{mele_introduction_2024}. In quantum information jargon, the projected ensemble at late but finite times is said to form a `quantum state-design'~\cite{renes_symmetric_2004,ambainis_quantum_2007}. This was later proven in various dynamical settings including `maximally-chaotic' systems in one spatial dimension~\cite{ho_exact_2022,claeys_emergent_2022,ippoliti_dynamical_2023} and random-matrix models~\cite{ippoliti_solvable_2022,wilming_high-temperature_2022}. 
In comparison, studies of quantum dynamics with symmetries and associated conservation laws --- such as  energy conservation generated by a time-independent Hamiltonian --- are at a more preliminary stage. It  has been claimed that the projected ensemble does still converge to a universal limiting form, namely one in which the Haar distribution is `distorted' by the conserved charges~\cite{cotler_emergent_2023,mark_maximum_2024}.  
In either case, the emergence of these universal distributions has  been argued to be   underpinned by {\it maximum-entropy principles} similar to, but going beyond, the second law of thermodynamics~\cite{mark_maximum_2024}, which endow them with special quantum information theoretic properties.
We note also that recent works have generalized the findings of deep thermalization beyond the setting of quantum spin chains (which is the setting of the aforementioned works), to the physically distinct arenas of fermionic systems~\cite{lucas_generalized_2023, matchgate2024} and bosonic continuous-variable quantum systems~\cite{liu_deep_2024}. While the form of the limiting distributions changes, it was found that they nevertheless follow the same maximum entropy principles. 

In this work, we substantially advance the nascent understanding of deep thermalization in systems that obey conservation laws. Conservation laws significantly affect aspects of quantum many-body dynamics such as thermalization, operator spreading, and entanglement~\cite{khemani_operator_2018,rakovszky_diffusive_2018,hunter-jones_operator_2018,agrawal_entanglement_2022,rath_entanglement_2023,ares_entanglement_2023,jonay_slow_2024,langlett_entanglement_2024,turkeshi_quantum_2024,poyhonen_scalable_2024}, and we find their impact on deep thermalization to be just as striking. 
Focusing on the paradigmatic case of $U(1)$ symmetry, we formulate a universal ansatz for the limiting form of the projected ensemble under fully general choices of input states and local measurement bases. 
An essential feature of our ansatz is that it depends on a small (polynomial in system size) amount of data about the system, namely the distribution of charge in the input state and measurement basis states; all other microscopic details of the system and its dynamics are discarded, as befits a genuine thermodynamic ensemble. In this sense, our conjectured limiting ensemble can be understood as a generalization of the familiar Gibbs state that describes quantum thermalization with conserved quantities. While the Gibbs state depends only on the average value of the conserved quantity, via the chemical potential or fugacity, our limiting ensemble depends nontrivially on the entire input charge distribution. Additionally, it depends on the amount of information revealed by the measurements about the value of the charge. It thus reveals a much richer variety of behavior, while still being governed by universal principles of maximum entropy.

Our conjecture is backed by three complementary approaches: a rigorous theorem, proven in the simplest setting of no charge fluctuations in either the input state or the measurement basis; 
analytical calculations within a `replica limit' approach, tractable (under certain approximations) when charge fluctuations are present in either the input state or the measurement basis, but not both; 
and finally, in the most general case, extensive numerical simulations of $U(1)$-symmetric random quantum circuits~\cite{khemani_operator_2018,rakovszky_diffusive_2018,hunter-jones_operator_2018,fisher_random_2023}, a well-explored testbed for symmetry in chaotic many-body dynamics. All approaches support the validity of our universal ansatz across a wide variety of physical situations. 

We note that while our present work focuses on the effect of a $U(1)$ symmetry as a key case study of the effect of conservation laws in quantum dynamics and deep thermalization, the framework we develop, being based on universal entropic principles, is formulated in a very general way that makes it readily applicable to understanding deep thermalization in the presence of  more complex symmetries (e.g.~non-Abelian) and conservation laws. Our results therefore shed some light along these other interesting fronts too. 
Lastly, we also note that while higher-moment statistics and formation of quantum state designs in systems with $U(1)$ symmetry have recently been considered in Refs.~\cite{liu2024symmetry,li2024designs,li2023sudsymmetric,hearth_unitary_2023,li2024efficientquantumpseudorandomnessconservation,yu2025symmetrybreakingdynamicsquantum},
our work is the first to study these concepts in the context of the projected ensemble and of deep thermalization, where randomness emerges from quantum measurements rather than from the choice of random unitary interactions.

The rest of our article is structure as follows. In Sec.~\ref{sec:review}, we   briefly review the concepts of the projected ensemble and the physical principles underpinning deep thermalization and previous results on deep thermalization with and without conservation laws. 
In Sec.~\ref{sec:rmt} we present our first main result: a rigorous proof of deep thermalization for random symmetric states. Building on this result, we present a completely general theory for  the limiting form of the projected ensemble under arbitrary initial states and measurement bases in Sec.~\ref{sec:limiting_ensemble}. Our general ansatz is supported by analytical calculations in a `replica' approach in Sec.~\ref{sec:replica} and by extensive numerical evidence in Sec.~\ref{sec:numerics}. We conclude by discussing our results and possible avenues for future work in Sec.~\ref{sec:discussion}.


\section{Review of deep thermalization \label{sec:review}} 

We begin by reviewing in Sec.~\ref{sec:PE} the theoretical foundations of the projected ensemble framework first introduced in Refs.~\cite{choi_preparing_2023,cotler_emergent_2023}, as well as the basic idea  of deep thermalization in Sec.~\ref{sec:deep_therm} --- namely, the emergence of universal limiting distributions on a local subsystem ---  that can arise within the framework. 
Then, in Sec.~\ref{sec:limiting_review}, we discuss at a high level the form of the limiting ensembles that arise, how they can be affected by symmetries, as well as the physical principles underpinning their emergence.

\subsection{Projected ensemble}
\label{sec:PE}

\begin{figure}
    \includegraphics[width=0.45\textwidth]{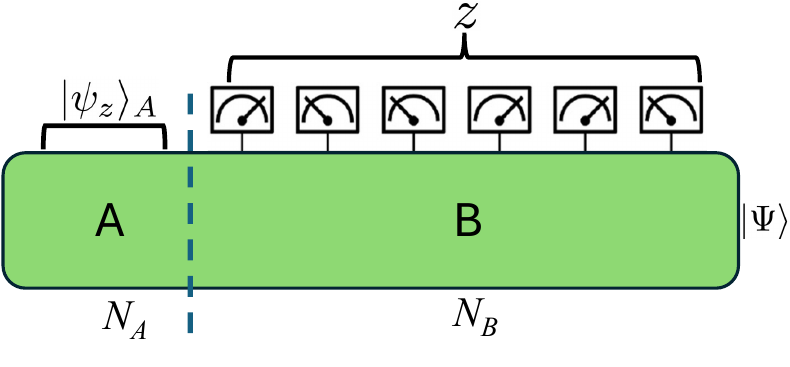}
    \caption{Construction of the projected ensemble in a  one-dimensional spin chain consisting of   subsystem $A$ and bath $B$. $|\Psi\rangle$ is a many-body wavefunction on the entire system $AB$. $|\psi_{\text{z}}\rangle_A$ is the projected state of the subsystem $A$ appearing in the projected ensemble. It is the conditional state on $A$ after a measurement with outcome $z$ is performed on the bath $B$, and occurs with Born probability $p(z)$.}
    \label{fig:system_AB}               
\end{figure}

Consider a pure quantum many-body state $|\Psi\rangle$ on $N$ qubits which is partitioned into two subsystems $A$ and $B$ with $N_A$ and $N_B$ qubits respectively, as shown in Fig.~\ref{fig:system_AB}. 
We assume $N_A \ll N_B$, and are interested in the state of the region $A$, which we take to be geometrically local. The complementary region $B$ thus plays the role of a ``bath''---though crucially we assume that some information about its particular state can be obtained through a quantum measurement, for example through a projective measurement in the computational basis $z$ (though we will relax this assumption in our work).
This allows us to generate an ensemble $\mathcal{E}$ of pure states $|\psi_{\text{z}}\rangle_A$ on the subsystem $A$, each of which is correlated with a measurement outcome $z\in\{0,1\}^{N_B}$ on the bath that occurs with probability $p(z)$, and physically represents the {\it conditional} post-measurement state on $A$.
This ensemble,
\begin{equation}
   \mathcal{E} \equiv \{(p(z),|\psi_z\rangle_A):\;z = 1,...,d_B=2^{N_B}\},
   \label{eq:p_ens_def}
\end{equation}
is called the projected ensemble~\cite{choi_preparing_2023,cotler_emergent_2023}.
Precisely, we have
\begin{equation}
    p(z) = \|_{B}\langle z|\Psi\rangle_{AB} \|^{2}, \;\;\;|\psi_z\rangle_A =\frac{_{B}\langle z|\Psi\rangle_{AB} }{\sqrt{p(z)}}.
\end{equation}

One may view a state ensemble as a probability distribution over the Hilbert space of $A$, and as such can be characterized by its statistical moments. The $k^{\text{th}}$ `moment operator' of the projected ensemble $\mathcal{E}$, capturing all $k$-point correlations within the ensemble, is given by 
\begin{equation}
   \rho^{(k)}_{\mathcal E} = \sum_{z} p(z) (|\psi_z\rangle\langle\psi_z|)^{\otimes k},
\end{equation}
where the subscript $A$ is omitted for brevity.  $\rho^{(k)}$ can be understood as a density matrix in a $k$-fold replicated Hilbert space $\mathcal{H}_{A}^{\otimes k}$, describing $k$ copies of a state randomly drawn from the ensemble. 
It can be readily checked that the mean, $k=1$, is just the reduced density matrix of subsystem $A$: 
\begin{equation}
   \rho^{(1)}_{\mathcal E} = \rho_A = \text{Tr}_{B}(|\Psi\rangle \langle\Psi|).
\end{equation}

A quantity that can be represented by the projected ensemble but not by the reduced density matrix is, for example, the variance of $\langle O\rangle_z = \bra{\psi_z} O \ket{\psi_z}$, the {\it conditional} expectation value of an operator $O$ on $A$ given the outcome $z$ of measurements performed on $B$:
\begin{align} 
    \text{var}_{z\sim \mathcal E} [\langle O \rangle_z] 
    & = \mathbb{E}_{\mathcal{E}}[\langle O\rangle^2_z] - \mathbb{E}_{\mathcal E}[\langle O \rangle_z]^2 \nonumber \\ 
    & = \sum_z p(z) \langle\psi_z|O|\psi_z\rangle^2 - {\rm Tr}(O \rho)^2 
    \nonumber \\
   & = {\rm Tr}[(\rho^{(2)}_{\mathcal E} - \rho^{\otimes 2}) O^{\otimes 2}]. \label{eq:high-order}
\end{align}
This quantity can thus be expressed as the expectation value of a ``higher-order'' observable in a two-fold replicated Hilbert space. Similarly, higher moment fluctuations of $\langle O \rangle_z$ would involve an expectation value on the state $\rho_{\mathcal E}^{(k)}$ on the $k$-fold replicated Hilbert space. This example also highlights the difference between $\rho^{(k)}$ ($k$-th moment operator) and $\rho^{\otimes k}$ ($k$ copies of the first-moment operator, i.e.~the reduced density matrix). 

Measuring higher-order observables like the one in Eq.~\eqref{eq:high-order} is nontrivial, but can be done with either a large amount of experimental sampling or the assistance of classical computation~\cite{garratt_probing_2024,mcginley_postselection-free_2024,hoke_measurement-induced_2023}.
The quantity in Eq.~\eqref{eq:high-order} has in fact been measured in a Rydberg-atom based quantum simulator~\cite{choi_preparing_2023}.

\subsection{Deep thermalization}
\label{sec:deep_therm}

The standard theoretical approach to thermal equilibrium in isolated interacting many-body systems is based on expectation values of local observables  tending toward equilibrium values at late times. In a quantum system, such information is captured by the reduced density matrix $\rho_A$ for a local subsystem, obtained by tracing out the degrees of freedom in $B$ which now plays the role of a bath:
\begin{equation}
\langle O_A(t) \rangle = {\rm Tr}(O_A \rho_A(t)), 
\quad 
   \rho_A(t) = \text{Tr}_{B}(\ketbra{\Psi(t)}). 
\end{equation}
`Standard' quantum thermalization is thus equivalent to the approach of the reduced density matrix to an equilibrium ensemble, $\rho_A(t) \rightarrow {\rm Tr}_B(\rho_{\text{eq}})$. 
Typically, the limiting density matrix $\rho_{\text{eq}}$ has the form of a thermal Gibbs state with a Lagrange multiplier corresponding to each globally conserved quantity---e.g., energy, charge, etc.
If energy is the only conserved quantity, we would thus have 
\begin{equation}
\label{eq:gibbs}
   \rho_{\text{eq}} \propto 
   e^{-\beta H}
\end{equation}
where $\beta$ is the inverse temperature. The reason  the Gibbs state can be expected to appear is because of the {\it principle of maximum entropy} --- it is the state that maximizes a suitable notion of information entropy (i.e., maximizes our ignorance of the system) subject to the known information (in this example the value of the energy, controlled by the inverse temperature $\beta$). 
This can be expected since complex, interacting dynamics should hide all microscopic information and only preserve macroscopic data. 
Precisely, Eq.~\eqref{eq:gibbs} is the state $\rho_A$ on $A$ which maximizes the von Neumann entropy $S(\rho_A) = -\text{Tr}(\rho_A \log \rho_A)$ subject to the constraint that all conserved quantities have the same local expectation value as the initial state~\cite{alhambra_quantum_2022}. This also has the interpretation of minimization of the free energy in quantum thermodynamics. 

Since $\rho_A$ equals the first moment of the projected ensemble $\rho^{(1)}$, as we saw, the statement of standard quantum thermalization can be understood as  a statement about the universal limiting form of the {\it first moment} of the projected ensemble.
It is natural to ask whether this approach toward a universal equilibrium form extends beyond the first moment, to the ensemble as a whole.

The idea of {\it deep thermalization} is precisely that the full projected ensemble $\mathcal{E}$ does converge, at late time in dynamics, to a universal form that we will denote as the `target' ensemble $\mathcal{E}_{\rm target}$. As we will see below, the target ensemble can depend on some coarse features of the dynamics and of the initial state, but should otherwise be insensitive to microscopic details and thus universal. 
Quantitatively, convergence to a given target ensemble $\mathcal{E}_{\rm target}$ over the course of quantum dynamics can be captured at the level of each statistical moment $k$ by the distance measure
\begin{equation} \label{eq:distance_measure}
   \Delta^{(k)} = \frac{1}{2} \|\rho^{(k)}_{\mathcal E} -\rho_{\rm target}^{(k)}\|_{\rm tr}.
\end{equation}
Here $\|\cdot\|_{\rm tr}$ is the trace norm, which has an operational meaning in terms of optimal quantum state discrimination. 
The vanishing of  $\Delta^{(k)}$ as a function of time in dynamics, for a suitable choice of target ensemble and moment $k\geq 2$, is what defines deep thermalization. Standard quantum thermalization is recovered for $k=1$ and using the appropriate Gibbs state as the target. 

\subsection{Limiting ensembles in deep thermalization}
\label{sec:limiting_review}

What should one expect as the limiting target ensembles $\mathcal{E}_\text{target}$ in deep thermalization, and what are the physical principles underlying their emergence? 

\subsubsection{Absence of conservation laws}
\label{sec:deep_thermalization_absence}

The easiest scenario to start from is that of a system dynamically evolving in the absence of symmetries or other constraints.
At the level of the first moment, standard quantum thermalization and the principle of maximum entropy dictate equilibration of a local subsystem to the maximally-mixed state, $\rho_A \xrightarrow{t\rightarrow\infty} \mathbb{I}_A / 2^{N_A}$. This corresponds to Eq.~\eqref{eq:gibbs} with infinite temperature ($\beta = 0$). 
However, this fact does not determine the target ensemble as there are many pure state ensembles whose first moment is maximally mixed. Separate and generalized maximum entropy principle(s) are needed in principle to pin down the form of the full distribution.

Now, because there are no constraints from symmetries, it is reasonable to expect that each measurement outcome on the bath $B$ generates a random state on $A$ {\it without bias}, i.e., that the target ensemble should feature all states in the Hilbert space of $A$ on an {\it equal footing}. This property is characteristic of the Haar ensemble~\cite{mele_introduction_2024}. 
Intuitively, this is the pure state ensemble with `maximal ignorance' of the local subsystem (since no particular state on $A$ stands out). This intuitive property can be precisely and quantitatively captured by a quantum information theoretic quantity called {\it accessible information}~\cite{jozsa_lower_1994} --- the maximal amount of classical information extractable from  measurements on a collection of quantum states. The Haar ensemble is indeed the unique distribution across all pure state distributions which has minimal accessible information, may thus may be viewed as a `maximally entropic' ensemble in this sense.

The Haar ensemble correctly reproduces the first moment $\rho_A \xrightarrow{t\rightarrow\infty} \mathbb{I}_A / 2^{N_A}$, but also makes nontrivial predictions about the form of all higher moments:
$\rho^{(k)}_{\mathcal E} (t) \xrightarrow{t\rightarrow\infty} \rho_{\rm Haar}^{(k)} $, where $\rho_{\rm Haar}^{(k)}$ are the moments of the Haar ensemble, whose forms are known analytically. 
In this specific case, there is a dedicated nomenclature for the condition $\Delta^{(k)} \leq \epsilon$: the ensemble is said to form  an ``$\epsilon$-approximate quantum state $k$-design''~\cite{renes_symmetric_2004,ambainis_quantum_2007,roberts_chaos_2017,mele_introduction_2024}. This concept is of interest in quantum information science, specifically in the context of randomized algorithms or protocols, where it serves as an efficient alternative to the Haar distribution~\cite{knill_randomized_2008,dankert_exact_2009,huang_predicting_2020,lancien_weak_2020}.
The emergence of the Haar ensemble in deep thermalization without conservation laws is supported by exact results on typical random states~\cite{cotler_emergent_2023} and on certain random circuit models in one dimension~\cite{ho_exact_2022,ippoliti_dynamical_2023}, as well as analytical and numerical calculations on other random circuit models~\cite{claeys_emergent_2022,ippoliti_dynamical_2023,ippoliti_solvable_2022}.

\subsubsection{Presence of conservation laws}
\label{sec:deep_therm_conservation}

In the presence of symmetries and associated conservation laws, the picture becomes substantially richer and more complex. 
This complication arises from two fronts:
(i) the distribution of the conserved quantity in the initial state, and 
(ii) the nature of the measurements on the `bath' $B$, namely whether or not they reveal information about the conserved quantity.

The first aspect is, to an extent, already present in the picture of conventional thermalization: the system retains memory of the value of any conserved quantities in the initial state, which at late time is stored in the various Lagrange multipliers (inverse temperature, chemical potential, etc) of the Gibbs state, Eq.~\eqref{eq:gibbs}, that arises from standard thermalization. Since the Gibbs state sets the first moment of the projected ensemble, this immediately raises the question of what the appropriate target ensemble should be away from infinite temperature (where $\rho^{(1)} \neq \mathbb{I}_A / d_A$); clearly,  the Haar ensemble is not a consistent candidate.

Now, an important point to note is that the equilibrium density matrix $\rho_{\rm eq}$ depends only on the {\it expectation value} of the conserved quantities in the initial state, not on their {\it quantum fluctuations}. 
This is because we conserved charges are 
scrambled non-locally across the whole system $AB$ and thus rendered inaccessible to observations on the local subsystem $A$. 
However, this is not necessarily true in deep thermalization, since the projected ensemble contains information that goes beyond local expectation values---it captures higher statistical moments and is thus sensitive to global data via the measurement outcomes on $B$. 
It is plausible, then, that the universal limiting ensembles of deep thermalization might depend on the whole distribution of conserved quantities in the initial state, and not just their expectation values. 
This already opens up a much wider variety of possibilities for universality compared to the scenario of standard thermalization.

The second aspect, i.e., the dependence on the choice of measurement basis of $B$, does not have an analog in conventional thermalization, where the bath is simply discarded (traced out). 
Distinct physical consequeces arise when there is a conserved quantity $Q$ with a local density $q_i$ (i.e., $Q = \sum_i q_i$ with each $q_i$ operator supported on a finite neighborhood of qubit $i$), then there is a question of whether the measurement basis is  compatible with the basis of the conserved quantity.

To elaborate, let us focus for concreteness on local product-state measurement bases, where each qubit is measured along some direction $\hat{n}$ in the Bloch sphere. Measuring the operator $\hat{n}\cdot \vec{\sigma}_i$ might reveal information about the value of charge densities $q_j$ at neighboring locations $j$. Specifically, we may decompose each measured operator $\hat{n}\cdot\vec{\sigma}_i$ into a basis of operators that contains the conserved densities $\{q_j\}$: $\hat{n}\cdot\vec{\sigma}_i = \sum_j \alpha_{ij} q_j + (\text{orthogonal operators})$; then, depending on the overlap coefficients $\{\alpha_{ij}\}$, the measurement outcome $\hat{n}\cdot\vec{\sigma}_i = \pm 1$ may correlate nontrivially with the total value of the conserved quantity  $Q_B = \sum_{j\in B} q_j$ on $B$. 
More generally, a complete measurement on $B$ provides information on $2^{N_B}$ distinct operators, $\ketbra{b}$, where $b \in\{0,1\}^{N_B}$ indexes the measurement basis; each of these may in principle correlate with $Q_B$. 
This has an important effect on deep thermalization, since this knowledge about the conserved quantity in the bath, $Q_B$, can translate into knowledge of the charge on the local subsystem, $Q_A$. For instance, if the total charge $Q$ is known and our measurement of $B$ indicates an unusually small value of $Q_B$, then the post-measurement state of $A$ must have an unusually large value of $Q_A$ to compensate. 

Different local measurement bases will generally provide different amounts of information about the conserved quantity. In special cases that will be relevant to this work, it is possible to span the entire range from {\it `non-revealing'} to {\it `maximally revealing'} bases. 
A `non-revealing' basis is one where the local measurements are perfectly uncorrelated with the conserved quantity, $\bra{b}q_j\ket{b} = 0$ for all basis vectors $b$ and local densities $q_j$. In this case the measurement outcome on $B$ does not systematically bias the post-measurement state on $A$. 
On the opposite end, a `maximally revealing' basis is one where $q_i = \hat{n}\cdot \vec{\sigma}_i$---i.e., we measure the conserved density directly (possible only when the symmetry is on-site and Abelian). In this case we expect different measurement outcomes to have strong, systematic impacts on the post-measurement states and hence on the limiting forms of the projected ensemble. 
Of course, we may have intermediate measurement bases where some information is revealed but not all; accounting for the gamut of cases constitutes a key problem that we address in this work.  

The effect  of different measurement bases has been investigated to a limited degree in recent works for the case of energy-conserving, Hamiltonian systems  (consistent results were found also for symmetries other than time translation, such as fermionic parity~\cite{bhore_deep_2023} or spatial symmetries~\cite{varikuti_unraveling_2024}). 
Now,  with energy conservation, we expect from standard quantum thermalization that the first moment of the projected ensemble converges to the appropriate Gibbs state $\propto {\rm Tr}_B(e^{-\beta H})$, with an average energy dictated by the inverse temperature $\beta$. The conserved quantity (energy) has a local density given by the individual Hamiltonian terms: $H = \sum_i h_i$.
Ref.~\cite{cotler_emergent_2023} focused on the special case of a Hamiltonian $H$ that admits an `energy non-revealing basis'. This is the case when $H$ (in what follows we take the case of a Hamiltonian on qubits as an example) features only two types of Pauli matrices, for example $X$ and $Y$ like in the mixed-field Ising model $H = J \sum_i X_{i}X_{i+1} + h \sum_i X_i + g \sum_i Y_i$, so that $\bra{z}h_i\ket{z} = 0$ for all sites $i$ and all computational basis states $z\in \{0,1\}^N$. 
Under this condition, it was found that the projected ensemble is well described by the so-called {\it Scrooge ensemble}~\cite{jozsa_lower_1994}.

The Scrooge ensemble is a particular unraveling of a density matrix $\rho$ into pure states. It can be characterized via a two-step procedure termed `$\rho$-distortion' of the Haar ensemble~\cite{goldstein_distribution_2006,goldstein_universal_2016}: 
\begin{equation}
    \mathcal{E}_{\text{Scrooge}}[\rho] = \left\{ \frac{\rho^{1/2}\ket{\psi}}{\| \rho^{1/2}\ket{\psi}\|}:\ {\rm d}\psi = d_A\bra{\psi}\rho\ket{\psi}\, {\rm d}\psi_{\rm Haar}  \right\}. 
    \label{eq:scrooge}
\end{equation}
In words, first a random state is drawn from the measure\footnote{A physical way of implementing this sampling is to apply the POVM $\{d_A \ketbra{\psi}:\, \psi\sim{\rm Haar}\}$ to the state $\rho$; this POVM corresponds to a projective measurement in a Haar-random basis. The measurement outcome $\ketbra{\psi}$ has the desired distribution. } $d_A \bra{\psi} \rho \ket{\psi} {\rm d}\psi_{\rm Haar}$, then it is distorted by application of $\rho^{1/2}$ (and normalized).
This prescription returns the Haar measure itself when $\rho = \mathbb{I}_A / d_A$ i.e.~when $\beta = 0$, but when $\beta > 0$ it biases the distribution toward lower-energy states, so that the mean state (first moment) is $\rho$. 

The Scrooge ensemble with $\rho \propto {\rm Tr}_B(e^{-\beta H})$ (at the appropriate temperature was found (at least consistently within numerics) to describe the limiting projected ensembles formed from Hamiltonian eigenstates as well as Hamiltonian quench dynamics at late times, again in the case of energy non-revealing measurements~\cite{cotler_emergent_2023}.
Importantly, out of all state ensembles with the {\it same} first moment $\rho$ (in this case the canonical density matrix), the Scrooge ensemble is the one that minimizes the  ``accessible information'' --- exactly like the Haar ensemble modulo the constraint on $\rho$.
This is again in line with the intuition that a complex, scrambling bath  hides information of a local subsystem subject to any global conservation laws, and makes the information maximally difficult to extract~\cite{mark_maximum_2024,liu_deep_2024}. 
The fact that the Scrooge ensemble emerges in deep thermalization with energy conservation and measurements in an energy non-revealing basis may thus be rationalized as follows:
the only information that remains accessible under the dynamics is the conserved quantity (energy); 
we know its initial value, which fixes the average reduced density matrix $\rho \propto \text{Tr}_B(e^{-\beta H})$; 
since the measurements on the bath $B$ are energy-non-revealing, they cannot provide any more information about the local state on $A$, so we simply `unravel' the reduced density matrix into pure states in a ``maximally-entropic'' fashion. This last step is governed by the idea that the ensemble carries the least  information which is accessible while respecting the constraint, and may be regarded as a generalized maximum entropy principle (see too the discussion in Ref.~\cite{liu_deep_2024}). 

Moving to the generic case of energy-revealing measurements (e.g.~in the $|x\rangle$ basis in the mixed-field Ising example) gives an even richer phenomenology. 
In this case a measurement outcome on $B$ may reveal an above- or below-average amount of energy in the bath, which will have to be compensated by an opposite energy fluctuation in the system, systematically biasing the post-measurement state toward low- or high-energy parts of the Hilbert space respectively.
Thus, instead of a unique Scrooge ensemble specified only by the global temperature, Ref.~\cite{mark_maximum_2024} identified as the limiting form of the projected ensemble a composite ensemble dubbed the `generalized Scrooge ensemble' (GSE), where each possible measurement outcome on the bath yields a state on the system drawn from a {\it different} Scrooge ensemble. Each of these Scrooge ensembles harbors an average energy consistent with the difference between the conserved total energy and the inferred energy from the measurement outcome.
Precisely, the GSE can be rationalized as follows: 
a measurement $B$ updates our knowledge about the average energy on $A$, and thus the resulting projected state should be understood as drawn from some underlying distribution consistent with that mean inferred energy; 
following again the ``maximum entropy'' principle, it is reasonable to assume this underlying distribution is a Scrooge ensemble at the mean inferred energy. Finally, one   averages over all possible measurement outcomes with respective probabilities, leading to a {\it stochastic} mixture of Scrooge ensembles, i.e., the GSE. 

Notably, the limiting GSE in the energy-conserving scenario studied by Ref.~\cite{mark_maximum_2024}  in principle contains an {\it exponential} amount of data: each of the $2^{N_B}$ possible measurement outcomes on $B$ yields different information about  the subsystem $A$'s energy and defines a different Scrooge ensemble, which then enters into the construction of the GSE.
This huge amount of data, comparable to a full microscopic description of the dynamics, calls into question the validity of the GSE as a universal thermodynamic ensemble. In the spirit of thermodynamics, it would be desirable to find a universal prediction for the limiting ensemble that depends only on a drastically reduced amount of data (e.g. polynomial instead of exponential in system size). 
In what follows, we will present our general theory achieving this goal for the case of deep thermalization under a $U(1)$ symmetry, which further paves the way to generalizations to more complex symmetries.


\section{ Projected ensemble of a random $U(1)$-symmetric state\label{sec:rmt}}
 
Before introducing our universal ansatz for the target ensemble in Sec.~\ref{sec:limiting_ensemble}, we focus on a special case that will be helpful to build some intuition on the physics that arises. 
This is the conceptually cleanest scenario of a  global Haar random state with definite charge $Q_0$ measured with maximally charge-revealing ($z$ basis)  on $B$. 
Note that since we are considering a single instance of a random state, there is no explicit quantum dynamics involved --- the connection is that, inspired by random matrix theory, we may imagine the statistical properties of an initial state with fixed charge $Q_0$ evolved by a deep $U(1)$ conserving circuit at large times to be well captured by a Haar random state\footnote{Refs.~\cite{marvian_restrictions_2022,marvian_theory_2023} show that locality poses nontrivial restrictions on the possibility of realizing arbitrary unitaries with a given symmetry. However, discrepancies with random matrix theory arise only at high statistical moments~\cite{hearth_unitary_2023} which are not expected to impact the physics to a first approximation.}.
We adopt a convention where without loss of generality the charge $Q_0$ assumes values from $0$ to $N$ (this can be achieved with a suitable global shift in the measurement values; for example, the local magnetization $M=\sum_{i=1}^N Z_i$ is related to $Q_0$ via $Q_0=M+N/2$). 

Despite its relative simplicity, this scenario is of great practical interest. In fact, it is the only allowed case when the symmetry is a symmetry of the universe that obeys a superselection rule (e.g., the $U(1)$ of electromagnetism) and not just of the system under consideration: then, coherent charge fluctuations in the (pure) initial state as well as in the measurement basis are forbidden.
In this setting, complications due to fluctuations of charge in the initial state do not arise. Further, the nature of the measurements is such that upon observing a bit-string $z$ which carries charge $Q_B$, the projected state is guaranteed to have charge $Q_A =Q_0 - Q_B$. Repeating this process many times, it should be expected that states from any {\it fixed}  charge sector $Q_A$ should be sampled uniformly; then, how often a state with charge $Q_A$ is sampled relative to a state with charge $Q_A'$ should be set by the ratio of the sizes of the charge sectors. 
Indeed, we have:

\begin{theorem} \label{thm:1}
Let $|\Psi\rangle$ be a Haar random state on $N$ qubits with definite charge $Q_0$, so that the charge density is $\sigma \equiv Q_0/N \in [0,1]$ and the state lives in a Hilbert space with dimension $d_{Q_0} = \binom{N}{Q_0}$.
Construct  the projected ensemble $\mathcal{E}_A$ on $N_A$ qubits, using  measurements of $B$ in the computational $(z)$ basis. 
Then as long as 
\begin{align}
    N_B H(\sigma) = \Omega\left(k N_A + \log\left({\epsilon}^{-1} \right) + \log \log\left({\delta}^{-1} \right) \right)
\end{align}
where $H(\sigma) = -\sigma \log \sigma - (1-\sigma)\log(1-\sigma)$ is the binary entropy, 
with probability at least  $1-\delta$ the $k$-th moment of the projected ensemble  obeys 
\begin{align}
    \frac{1}{2} \| \rho^{(k)} - \rho^{(k)}_{\text{DS},Q_0} \|_\text{tr} \leq \epsilon,
\end{align}
where $\rho^{(k)}_{\text{DS},Q_0}$ is the $k$-th moment of the `direct sum (DS)' ensemble, 
\begin{align}
    \rho^{(k)}_{\text{DS},Q_0} = \bigoplus_{Q_A=0}^{N_A} \pi(Q_A|Q_0) \rho^{(k)}_{\text{Haar},A,Q_A}.
\end{align}
Above, $\rho^{(k)}_{\text{Haar},A,Q_A}$ is the $k$-th moment of the Haar ensemble on states on $A$ with fixed charge $Q_A$,   
$d_{Q_A} = \binom{N_A}{Q_A}$ is the dimension of charge sector $Q_A$ on $N_A$ qubits, and  
 the probability $\pi(Q_A|Q_0) = d_{Q_A} d_{Q_B}/d_{Q_0}$, where $d_{Q_B} = \binom{N-N_A}{Q_0-Q_A}$. 
\end{theorem}

Theorem~\ref{thm:1} in fact yields more than just the aforementioned expected outcome:
it states that, with high probability, a {\it single instance} of a charge-symmetric Haar random state of a large enough system will generate a projected ensemble very close to the `direct sum' ensemble. The latter consists of a direct sum of Haar random distributions within each charge sector $Q_A$, weighted by  the frequency $\pi(Q_A|Q_0)$ with which $Q_A$ appears --- a purely combinatorial factor. Our Theorem~\ref{thm:1} can be understood as the analog of Theorem 1 of Ref.~\cite{cotler_emergent_2023}.

The proof of this theorem is technical but straightforward, and left to Appendix~\ref{app:thm1}. We quickly sketch the logic here: 
first, we show that on average over input states $\ket{\Psi}$, the projected ensemble is the direct sum ensemble. Second, we show that the projected ensemble is Lifshitz continuous, and thus (by Levy's lemma~\cite{ledoux_concentration_2001}) it concentrates around this expected value with high probability in the limit of large bath size. 

Thus, we see  from this simple example how symmetries can endow the limiting form of the projected ensemble with more structure than just a featureless Haar distribution over the full Hilbert space. However, the emergence of the `direct sum' ensemble in this case is to be expected; a more interesting question arises when considering initial states with non-definite charge, as well as general measurement bases. We address this more general question in the following.

\section{Theory of universal limiting ensemble for general initial states and measurement bases} \label{sec:limiting_ensemble}

\begin{figure*}
    \centering
    \includegraphics[width=\textwidth]{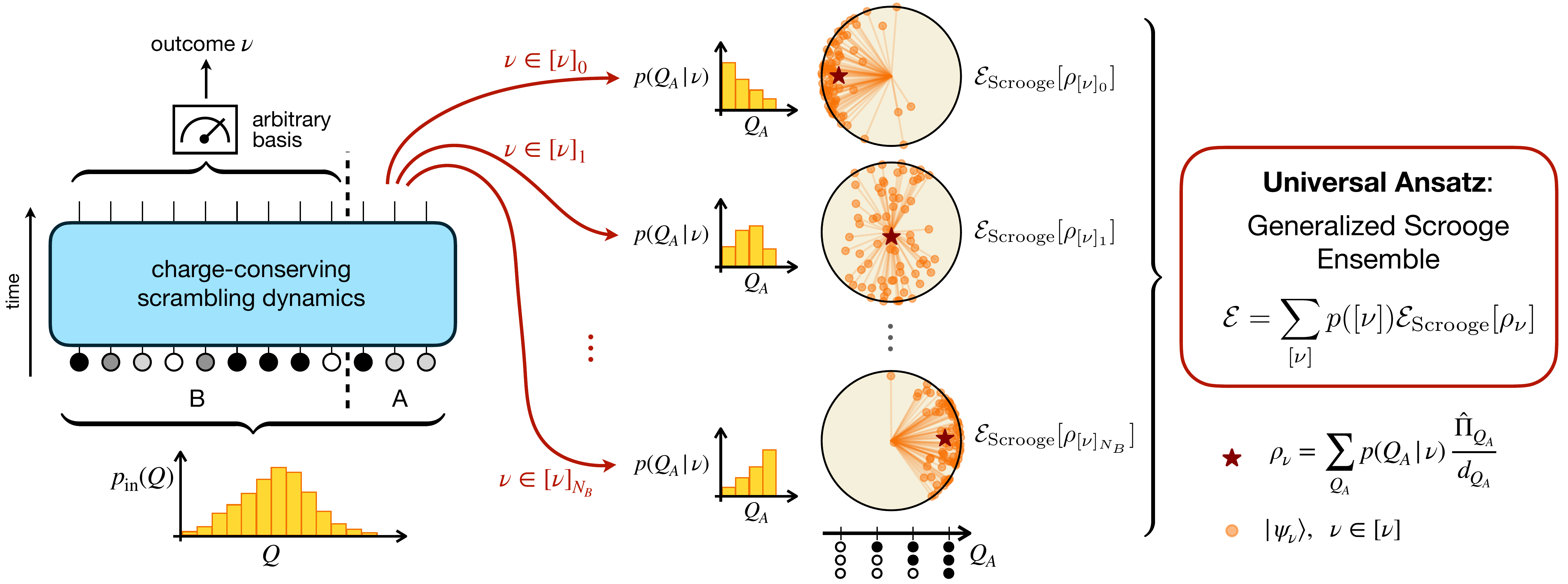}
    \caption{ Schematic representation of our universal ansatz for the limiting form of the projected ensemble. An input state with charge distribution $p_{\rm in}(Q)$ on a bipartite system $AB$ undergoes $U(1)$-symmetric scrambling dynamics. Subsystem $B$ is then measured in an arbitrary basis, yielding an outcome $\nu$ (left). Our theory predicts that measurement outcomes can be grouped into equivalence classes $[\nu]$ based on how they update our knowledge of the charge on subsystem $A$, i.e., based on the conditional distribution $p(Q_A|\nu)$. There are at most $N_B+1$ such equivalence classes. For each equivalence class, we conjecture that the post-measurement states $\{ \ket{\psi_{\nu}}: \nu\in [\nu]\}$ form a Scrooge ensemble (middle):
    we represent the Hilbert space of subsystem $A$ by a Bloch sphere for simplicity; dots represent individual projected states $\ket{\psi_\nu}$ and stars represent the density matrix $\rho_{[\nu]} = \sum_{Q_A} p(Q_A|\nu) \hat{\Pi}_{Q_A} / d_{Q_A}$ that defines each Scrooge ensemble. The overall projected ensemble is the stochastic mixture of all these Scrooge ensembles weighted by the probability $p([\nu])$ of observing each equivalence class. This is the generalized Scrooge ensemble (GSE) (right). 
    \label{fig:concept}}
\end{figure*}

Having addressed the case of a symmetric initial state subject to $z$-basis measurements, we now move on to the general case. From this point forward, we assume the $U(1)$ symmetry does not obey a superselection rule, so coherent superpositions of different charge sectors are allowed.

Since the rigorous proof techniques in Sec.~\ref{sec:rmt} do not straightforwardly extend to this more general case, we will adopt a different strategy. We will present a general {ansatz}, motivated by principles of maximum entropy and minimum accessible information, for the equilibrium form of the projected ensemble in the most general case; 
we will then support this ansatz with different approaches, both analytical and numerical. Our ansatz is illustrated in Fig.~\ref{fig:concept}.

Our aim is to formulate a prediction for the limiting projected ensemble generated by a {\it single} quantum many-body state $|\Psi\rangle = \sum_{Q=0}^N \sqrt{p_{\rm in}(Q)}|\Phi_Q\rangle$ with initial charge distribution $p_{\rm in}(Q)$ (here, each $|\Phi_Q\rangle$ carries definite charge $Q$), time-evolved by $U(1)$-symmetric scrambling dynamics, and measured in a basis $\{|\nu\rangle\}$ on $B$. We will take $\{|\nu\rangle\}$ to be  a tensor product of single-qubit measurement bases that point in a common (but otherwise arbitrary) direction for all qubits in $B$\footnote{This consideration is physically motivated and will prove advantageous, but can in principle be relaxed.}. Thus, $\{ |\nu\rangle \langle \nu| \}$ describes a set of rank-1 measurements.

The first key assumption entering our theory is that there is an underlying {\it class} of states that $|\Psi\rangle$ itself belongs to, which we call the ``generator global state ensemble'' $\mathcal{F}$. 
The precise form of $\mathcal{F}$ depends on the class of dynamics under consideration; for example, $\mathcal{F}$ for the states $|\Psi\rangle$ discussed in Sec.~\ref{sec:rmt} was precisely  the ensemble of Haar-random $U(1)$-symmetric states with fixed charge $Q_0$ (this is the special case $p_\text{in}(Q) = \delta_{Q,Q_0}$). 
In the general case,   $|\Psi\rangle$ has some arbitrary initial charge distribution $p_{\rm in}(Q)$ which stays invariant over time; however we note the charged states $|\Phi_Q\rangle$ fluctuate over time due to dynamics. It is  thus natural to take the generator ensemble that late-time states $|\Psi\rangle$ are drawn from to be
\begin{equation}
    \mathcal{F} = \left\{ \sum_{Q=0}^N \sqrt{p_{\rm in}(Q)} \ket{\Phi_Q}:\ \ket{\Phi_Q} \sim {\rm Haar}(\mathcal{H}_Q) \right\}.
    \label{eq:gen_ensemble_u1}
\end{equation}
This is an ensemble describing fixed charge distribution $p_{\rm in}(Q)$ and  wavefunction components in each charge sector $Q$ that are completely randomized in their respective Hilbert space $\mathcal{H}_Q$. 
In Fig.~\ref{fig:concept}, the generator ensemble is represented by the input charge distribution and the $U(1)$-symmetric scrambling dynamics (left).

Next, consider two microscopic measurement outcomes $\nu_1, \nu_2$. They may produce two different projected states $|\psi_{\nu_1}\rangle, |\psi_{\nu_2}\rangle$ respectively for a given generator state $|\Psi\rangle$, but they may nevertheless behave identically on average (sampled over $\ket{\Psi}$ in $\mathcal{F}$). This allows us to define an equivalence relation $(\sim)$. More precisely, consider the $\mathcal{F}$-averaged conditional post-measurement state upon obtaining measurement outcome $\nu$:   
\begin{equation}
        \rho_\nu = {\rm Tr}_B (\bar{\rho} \ \mathbb{I}_A \otimes \ketbra{\nu}_B) / p(\nu),
    \end{equation}
where $\bar{\rho} = \mathbb{E}_{\Psi \sim \mathcal{F}} \ketbra{\Psi}$. 
This measurement outcome occurs with probability
    $
        p(\nu) = {\rm Tr}(\bar{\rho} \ \mathbb{I}_A \otimes \ketbra{\nu}_B)
    $ again on average over $\mathcal F$.
Note that this conditional state $\rho_\nu$ is in general mixed, unlike  $|\psi_\nu\rangle$ which is pure (by virtue of $\ketbra{\nu}$ being a rank-1 measurement and $|\Psi\rangle$ being pure). 
We say that $\nu_1 \sim \nu_2$ if $\rho_{\nu_1} = \rho_{\nu_2}$, and  thus say they belong to the same equivalence class $[\nu]$. The class itself occurs with probability $p([\nu])=\sum_{\nu \in [\nu]}p(\nu)$.

Our second key assumption is that the ensemble of projected states associated to outcomes $\nu$ in the same equivalence class $[\nu]$,
\begin{equation} 
\mathcal{E}_{[\nu]} \equiv \{ p(\nu) / p([\nu]), |\psi_{\nu}\rangle  \}_{\nu \in [\nu]}
\end{equation}
generated from a {\it fixed} $|\Psi\rangle$, 
is statistically similar to the {\it maximally-entropic} pure state distribution with mean $\rho_\nu$ for some $\nu \in [\nu]$. 
As discussed in Sec.~\ref{sec:review}, we take `maximally entropic' to concretely mean the property of having the least accessible information; this corresponds to the Scrooge ensemble $\mathcal{E}_{\rm Scrooge}[\rho_\nu]$ constructed from an unraveling of $\rho_\nu$ into underlying pure states as described in Eq.~\eqref{eq:scrooge}.
Further, we expect the two ensembles to become statistically more similar as the bath size $|B|$ increases (fixing local system size $|A|$).
The separation of measurement outcomes into different equivalence classes and the emergence of Scrooge ensembles in each equivalence class is sketched in the middle section of Fig.~\ref{fig:concept}. 

Combining all ingredients, this finally leads us to our ansatz for the limiting form of the projected ensemble:  
\begin{equation}
    \mathcal{E} = \sum_{[\nu]} p([\nu]) \mathcal{E}_{\rm Scrooge}[\rho_\nu], 
    \label{eq:gen_scrooge_compressed}
\end{equation}
a stochastic mixture of Scrooge ensembles over different equivalence classes  of measurement outcomes, i.e., a generalized Scrooge ensemble. Crucially, if the {\it number} of equivalence classes $[\nu]$ is only polynomial in system size, this amounts to a massive compression of information from the exponential amount we would have needed had we kept track of all microscopic measurement outcomes. Such compression would allow us to interpret Eq.~\eqref{eq:gen_scrooge_compressed} as a meaningful thermodynamic  equilibrium state. As we shall see in a number of examples below, such compression of data does indeed happen with many parent ensembles $\mathcal{F}$s, including the case of $U(1)$-symmetric dynamics, the focus of this work. 

{\bf Example 1. Haar-random generator ensemble.} Let $\mathcal{F}$ be the Haar ensemble on the whole Hilbert space and $|\Psi\rangle$ be a single instance of a state drawn from $\mathcal{F}$. This ensemble captures dynamics of physical systems evolving without any conservation laws or at infinite temperature.  
Our construction yields $\bar{\rho} = \mathbb{E}_{\Psi \sim \mathcal{F}} \ketbra{\Psi} = \mathbb{I}_{AB} / d_A d_B$ (maximally mixed state) and thus $\rho_\nu = \mathbb{I}_A / d_A$ for all $\nu$. Thus there is a single equivalence class $[\nu]$ comprising all measurement outcomes, with $p([\nu]) = 1$, and unraveling $\rho_\nu$ yields the prediction
\begin{equation}
    \mathcal{E} 
    = \mathcal{E}_{\rm Scrooge}[\mathbb{I}_A/d_A]  = \mathcal{E}_{\rm Haar}
\end{equation}
(note the Scrooge ensemble associated to the maximally mixed state is nothing more than the Haar ensemble).
This recovers the expected outcome of deep thermalization without conservation laws. 

{\bf Example 2. Generator ensemble with arbitrary on-site symmetry.} Consider a global state $|\Psi\rangle$ drawn from the ensemble $\mathcal{F}$ obtained from a reference state $\ket{\Psi_0}$ by acting with unitaries $U$ chosen Haar-randomly\footnote{Unitaries that respect the symmetry form a compact subgroup of the overall unitary group $U(d)$ and so admit a Haar measure.} among unitaries that commute with an arbitrary {\it on-site} symmetry: $[U,\bigotimes_i g_i] = 0$ where $g_i$ implements the symmetry transformation on qubit $i$. 
Physically relevant examples include parity conservation ($\mathbb{Z}_2$ symmetry, $g_i = X_i$), charge conservation ($U(1)$ symmetry, $g_i = e^{-i\phi Z_i}$), and spin conservation ($SU(2)$ symmetry, $g_i = e^{-i \mathbf{a} \cdot \boldsymbol{\sigma}_i /2}$). 
In all these cases, the generator ensemble $\mathcal{F}$ is permutation-invariant: the SWAP unitary clearly commutes with the on-site symmetry, and thus is part of the unitary subgroup that defines $\mathcal{F}$. It follows that {\it any two bitstrings $\nu_1$, $\nu_2$ with the same number of `0' outcomes (in the arbitrary local basis) are in the same equivalence class $\nu_1 \sim \nu_2$,} as they can be turned into each other by a permutation, which leaves $\bar{\rho} = \mathbb{E}_{\Psi \sim \mathcal{F}} \ketbra{\Psi}$ unchanged. Thus there are at most $N_B+1$ equivalence classes $[\nu]$, yielding a {\it bona fide} universal thermodynamic ensemble as discussed.

\subsection{Prediction for $U(1)$ symmetric dynamics}\label{sec:limiting_ensemble_U1}

We now present our construction in full detail for the case of dynamics with $U(1)$ symmetry, the main focus of this paper, though we note  that our framework lends itself immediately to generalizations in multiple directions (discrete or continuous, Abelian or non-Abelian symmetries, Hamiltonian and Floquet dynamics, etc) and thus opens up interesting avenues for future work. 

The generator ensemble we consider is as in Eq.~\eqref{eq:gen_ensemble_u1}, which is specified by an arbitrary input charge distribution $p_{\rm in}(Q)$. The average state over $\mathcal{F}$ is thus 
\begin{equation}
    \bar{\rho} = \sum_{Q=0}^N p_{\rm in}(Q) \frac{\hat\Pi_Q}{\binom{N}{Q}},
\end{equation}
with $\hat{\Pi}_Q$ the projector on charge sector $Q$ of dimension $d_Q = \binom{N}{Q}$. 
The probability of a measurement outcome $\nu$ is
\begin{equation}
    p(\nu) = \sum_{Q=0}^N p_{\rm in}(Q) \sum_{Q_B = 0}^{N_B} \frac{\binom{N_A}{Q-Q_B}}{\binom{N}{Q}} \bra{\nu} \hat{\Pi}_{Q_B} \ket{\nu},
    \label{eq:pnu}
\end{equation}
where we used the fact that $\hat\Pi_Q = \sum_{Q_B} \hat{\Pi}_{Q-Q_B}^{(A)} \otimes \hat{\Pi}^{(B)}_{Q_B}$ in terms of charge sector projectors for subsystems $A$ and $B$ (we generally omit the superscripts ${}^{(A/B)}$ in the following to lighten the notation). 
The quantity 
\begin{equation} 
\bra{\nu} \hat{\Pi}_{Q_B} \ket{\nu} \equiv p_{\rm out}(Q_B|\nu) \label{eq:p_out}
\end{equation}
is the charge distribution of the measurement basis state $\ket{\nu}$. 
Collecting together all bitstrings that give rise to the same distribution $p_{\rm out}(Q_B|\nu)$ into an equivalence class $[\nu]$, we now obtain
\begin{equation}
p([\nu]) = \sum_{Q,Q_A} p_{\rm in}(Q) \frac{\binom{N_A}{Q_A} |[\nu]|}{\binom{N}{Q}} p_{\rm out}(Q-Q_A|\nu),
\label{eqn:pnu_equivalence}
\end{equation}
where $|[\nu]|$ denotes the number of bitstrings $\nu$ in the equivalence class.
Then, the associated density matrix $\rho_\nu$ on $A$ is given by
\begin{align}
    \rho_\nu 
    & = \frac{ {\rm Tr}_B (\bar\rho \ \mathbb{I}_A \otimes \ketbra{\nu}_B) }{ p(\nu) }
    = \sum_{Q_A} p(Q_A|\nu) \frac{\hat \Pi_{Q_A}}{\binom{N_A}{Q_A}}, 
    \label{eqn:rho_nu}
\end{align}
in terms of a charge probability distribution
\begin{equation}
    p(Q_A|\nu) =  \frac{\sum_{Q} p_{\rm in}(Q) p_{\rm out}(Q-Q_A|\nu) \binom{N_A}{Q_A} \binom{N}{Q}^{-1} }{p(\nu)}.\label{eq:pqa_nu}
\end{equation}

Our theory predicts that the limiting form of the projected ensemble in this case is given by the generalized Scrooge ensemble Eq.~\eqref{eq:gen_scrooge_compressed}, with $p([\nu])$ given by Eq.~\eqref{eqn:pnu_equivalence} and $\rho_\nu$ given by Eq.~\eqref{eqn:rho_nu}.
This has an intuitive physical interpretation in terms of Bayesian probability: each  measurement outcome $\nu$ updates our belief about the charge present on subsystem $A$, $Q_A \sim p(Q_A|\nu)$; 
a pure  state on $A$ is then drawn from the maximally entropic (in the quantum information theoretic sense) distribution, i.e., the Scrooge ensemble $\mathcal{E}_{\rm Scrooge}[\rho_\nu]$, where $\rho_\nu$ is the maximally-mixed density matrix with this updated belief, Eq.~\eqref{eqn:rho_nu}. The overall ensemble is then a stochastic mixture of these Scrooge ensembles, each drawn with probability $p(\nu)$; this is further compressed down to a stochastic mixture over equivalence classes of measurement outcomes $[\nu]$.

An important feature of our prediction is that there are only at most $N_B+1$ different equivalence classes $[\nu]$ entering the definition of the GSE: in the most general case, absent any additional degeneracies, the equivalence classes are defined by $\sum_i \nu_i$ (number of `1' outcomes in the arbitrary local basis). This is a consequence of permutation invariance of the generator ensemble $\mathcal{F}$, which holds for any on-site symmetry, as discussed earlier. 
Thus, overall our GSE ansatz is built out of a polynomial amount of data: the charge distribution of the input state $p_{\rm in}(Q)$ (specified by $N$ independent numbers), and the charge distributions in the measurement basis $p_{{\rm out}}(Q_B|\nu)$ (specified by at most $N_B(N_B+1)$ independent numbers). This dependence on a limited amount of data is an essential feature of a genuine thermodynamic ensemble, based on discarding microscopic details in favor of coarse-grained information on the system.

\begin{figure}
    \centering
   \includegraphics[width=0.99\columnwidth]{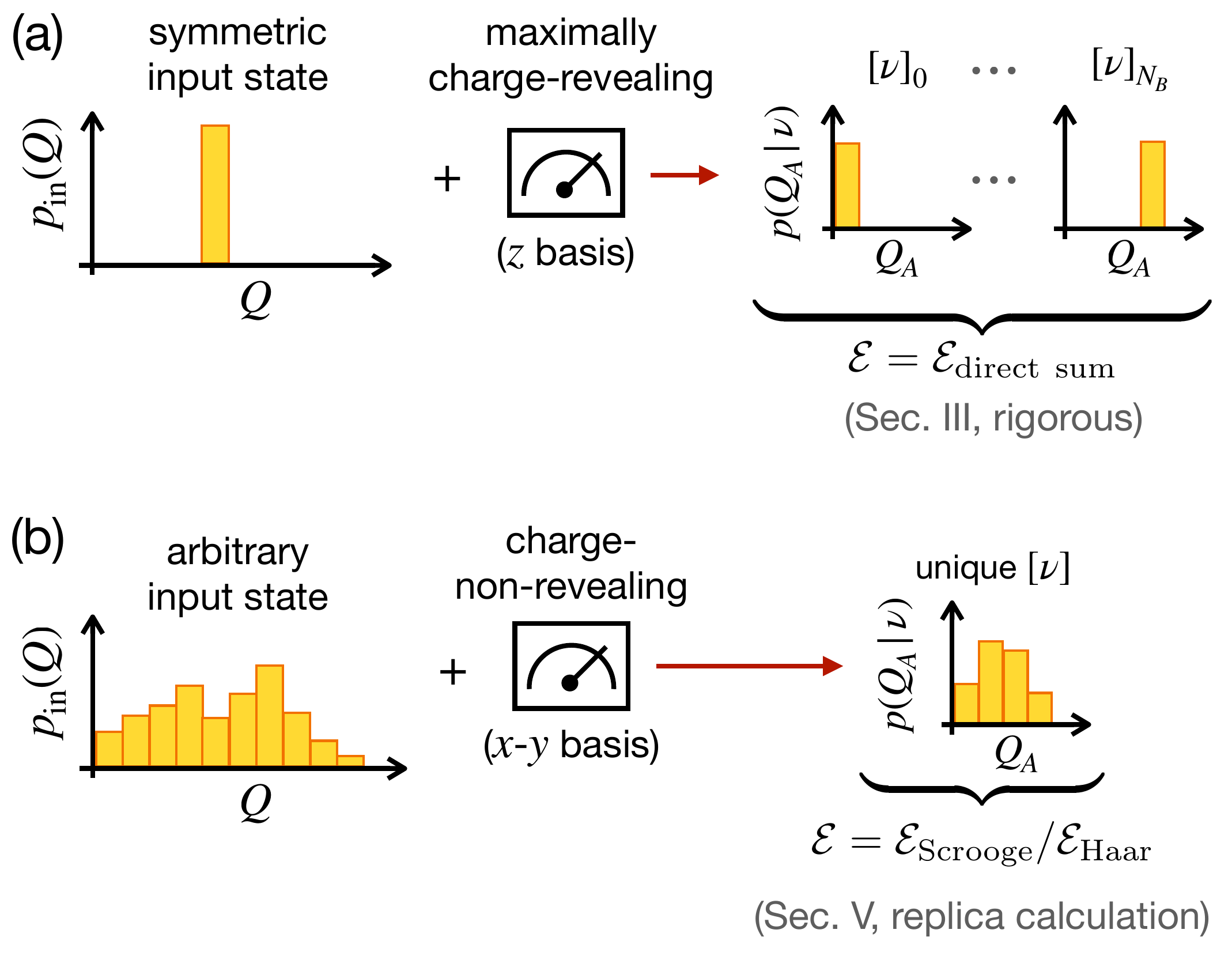}
    \caption{Special limits of the universal GSE ansatz. 
    (a) For symmetric input states under maximally charge-revealing ($z$-basis) measurements the GSE reduces to the `direct sum' ensemble of Sec.~\ref{sec:rmt}.
    (b) For charge-non-revealing measurements ($x$-$y$ basis) on arbitrary input states, all measurement outcomes fall into the same equivalence class and the GSE reduces to a single Scrooge ensemble. In turn this can reduce to the Haar ensemble if the input state is charge-neutral on average. }
    \label{fig:special_cases}
\end{figure}

We close this general discussion by noting that our universal ansatz reduces to even simpler forms in several cases of physical interest [Fig.~\ref{fig:special_cases}].
One such case is the one studied in Sec.~\ref{sec:rmt} of symmetric input states ($p_{\rm in}(Q) = \delta_{Q,Q_0}$) measured in the ``maximally charge-revealing'' ($z$) basis. In that basis, measurement outcomes uniquely identify a value of the charge: $p_{\rm out}(Q_B|\nu) = \delta_{Q_B,\nu}$. Our ansatz then reduces to a mixture of ``Scrooge ensembles'' specified by the charge distribution $p(Q_A|\nu) = \delta_{Q_0-\nu}$; but since these are simply Haar ensembles in the individual charge sector, thus recovering the `diagonal ensemble' of Sec.~\ref{sec:rmt}, see Fig.~\ref{fig:special_cases}(a). 
Another limiting case is when the measurement basis $\{|\nu\rangle\}$ is that of a   maximally ``charge non-revealing'' basis --- then, the number of  equivalence class further drastically simplifies from polynomial to unity, as all bit-strings behave statistically the same (no one bit-string reveals any more information than another!), see Fig.~\ref{fig:special_cases}(b). Consequently, the GSE simply reduces to the Haar ensemble or single Scrooge ensemble, as we will discuss in detail in Sec.~\ref{sec:replica}.

\section{Analytical results \label{sec:replica}}

In the previous section, we presented a physically motivated ansatz for the limiting form of the projected ensemble in scrambling $U(1)$-conserving quantum dynamics: the GSE, Eq.~\eqref{eq:gen_scrooge_compressed}. The direct sum (DS) ensemble of Sec.~\ref{sec:rmt} constituted one particular instance of the GSE, where the global state had definite charge and the measurements were maximally charge-revealing. That setting allowed for a rigorous proof of deep thermalization, Theorem~\ref{thm:1}.
We now turn to the question of whether our ansatz can be analytically derived from first principles in other cases of input states and measurement bases, thus supporting our general prediction.

To begin, we remind the reader we start from $\ket{\Psi} = \sum_Q p_{\rm in} (Q) \ket{\Phi_Q}$,  a global state with fixed but arbitrary initial charge distribution $p_{\rm in}(Q)$ where component states $\ket{\Phi_Q}$ have definite charge $Q$. We then evolve it with an interacting $U(1)$-conserving unitary. 
As explained in Sec.~\ref{sec:limiting_ensemble}, we will again make the first logical leap that the statistics of global states at late-times is captured well by the generator ensemble $\mathcal{F}$ given by Eq.~\eqref{eq:gen_ensemble_u1}, thus the $k$-th moment of the projected ensemble is given by
\begin{equation}
    \rho_{\mathcal E}^{(k)} \xrightarrow{t\to\infty} \mathop{\mathbb{E}}_{\Psi \sim \mathcal{F} } \sum_\nu \frac{(\braket{\nu }{\Psi} \braket{\Psi}{\nu})^{\otimes k}}{(\braket{\Psi}{\nu} \braket{\nu}{\Psi})^{k-1}}.
    \label{eq:pe_limit_ansatz}
\end{equation}
We will now carry out the averaging on the right hand side of Eq.~\eqref{eq:pe_limit_ansatz} within an approximation to be specified below. As a result we will find an explicit, analytic form of the $k$-th moment operator $\rho^{(k)}_{\mathcal{E}}$. From knowledge of all moments one can in principle reverse-engineer the ensemble $\mathcal{E}$ itself, and thus test our general conjecture, Sec.~\ref{sec:limiting_ensemble}. 

For arbitrary $p_{\rm in}(Q)$ and general measurement bases, a direct evaluation of Eq.~\eqref{eq:pe_limit_ansatz} is challenging. However, significant progress can be made 
in the scenarios where measurements are made in a basis which fully reveals charge information ($\nu = z$-basis) or in a basis that never reveals charge information ($\nu = x,y$-basis).
Further, to deal with the technical difficulty of averaging a rational function that contains $\ket{\Psi}$ in the denominator, we resort to a `replica trick', where one treats the power $1-k\equiv n$ of the denominator as a positive integer, carries out the Haar averages by well established Weingarten calculus methods~\cite{mele_introduction_2024}, and finally takes the `replica limit' $n \to 1-k$. This method is not mathematically rigorous but is widely applied in statistical physics and often yields correct answers in practice. Nonetheless, the answers should be checked by independent methods. 
In addition to the replica limit, a further subtlety is the fact that our derivations will require a limit of large subsystem size $N_A\to\infty$, in addition to the standard thermodynamic limit $N\to\infty$. 
We will benchmark the results against exact numerical calculations (in Sec.~\ref{sec:numerics}) and we will show that the rigorous result of Sec.~\ref{sec:rmt} is recovered in the appropriate limit. Our replica approach is described in detail in Appendix~\ref{app:ansatz}.

\subsection{Maximally charge-revealing measurements \label{sec:ensemble_z}}

We first focus on the case of measurements in the computational ($z$) basis. These measurements fully reveal the local charge density (0 or 1) at each measured site, and   thus are   `maximally charge-revealing' [see Sec.~\ref{sec:deep_therm_conservation}].

We carry out the replica calculation for this case in Appendix~\ref{app:charge_revealing}.
Even within the `replica trick', we find that the problem is not fully analytically solvable. Analytical progress requires also a limit of large Hilbert space dimension for the charge sectors of the whole system, $\binom{N}{Q}$, {\it and} of subsystem $A$, $\binom{N_A}{Q_A}$. Further, the thermodynamic limit $N,N_A\to\infty$ must be taken {\it before} the replica limit. This is a reversal of the correct order that we expect will introduce an error in our results.
To estimate this error, in Appendix~\ref{app:approximation_error} we retain the leading correction in the inverse Hilbert space dimension $\binom{N_A}{Q_A}^{-1}$ and show that, after taking the replica limit, the correction is proportional to $k^2 / \binom{N_A}{Q_A}$, $k$ being the statistical moment under consideration. Based on this result, we expect the result to be accurate only in the limit of a large subsystem, $N_A\gg \ln(k)$, a stronger requirement than just the thermodynamic limit $N\to\infty$. 
Within this approximation, the result of our calculation is found to be in the form of a generalized Scrooge ensemble (GSE)~\cite{mark_maximum_2024}, in agreement with our general conjecture Sec.~\ref{sec:limiting_ensemble}. 

Specifically, we show  in Appendix~\ref{app:charge_revealing} that (within the approximations mentioned above) one computes 
\begin{equation}
    \rho^{(k)}_{\mathcal{E}} \simeq \sum_\nu p(\nu) \rho^{(k)}_{\rm Scrooge}[\rho_\nu],
\end{equation}
with
\begin{align}
    \rho_\nu 
    & = \sum_{Q_A} p(Q_A|\nu) \frac{\hat\Pi_{Q_A}}{ \binom{N_A}{Q_A}},  \label{eq:maxrev_rhonu} \\
    p(\nu) 
    & = \sum_{Q_A'} p_{\rm in}(Q_A'+ Q_B(\nu))  \frac{ \binom{N_A}{Q_A'}  }{ \binom{N}{Q_A'+ Q_B(\nu) } }, \label{eq:maxrev_pnu} \\
    p(Q_A|\nu) 
    & = \frac{p_{\rm in}(Q_A+Q_B(\nu) )}{p(\nu) } \frac{ \binom{N_A}{Q_A}  }{ \binom{N}{Q_A+Q_B(\nu)} }, \label{eq:maxrev_pqanu}
\end{align}
where we used the notation $Q_B(\nu) = \sum_i \nu_i$ (the number of `1' outcomes in the computational basis corresponds to total charge).
This result agrees with our general prediction, Eq.~\eqref{eq:pqa_nu}, in the case of maximally charge-revealing measurements, where $p_{\rm out}(Q_B|\nu) = \delta_{Q_B,Q_B(\nu)}$: the measurement outcome uniquely identifies the charge on subsystem $B$.

Physically, the GSE prescription we obtained can be interpreted as follows:
\begin{itemize}
    \item[(i)] We first draw a value of the charge $Q_B$ on the bath according to the distribution
    \begin{equation}
        Q_B \sim \sum_{Q'} p_{\rm in}(Q') \binom{N_A}{Q'-Q_B} \binom{N_B}{Q_B}\binom{N}{Q'}^{-1} ,
    \end{equation}
    i.e., Eq.~\eqref{eq:maxrev_pnu} summing over the $\binom{N_B}{Q_B}$ bitstrings of the same charge $Q_B$.
    This is the distribution of charge on $B$ assuming that the overall charge of the system $Q \sim p_{\rm in} (Q)$ is completely scrambled.
    \item[(ii)] Knowledge about $Q_B$ from (i) updates our belief about the charge on $A$ to the posterior distribution $p(Q_A|\nu)$ [Eq.~\eqref{eq:maxrev_pqanu}].
    \item[(iii)] Finally the projected state is sampled from the Scrooge ensemble $\mathcal{E}_{\rm Scrooge}[\rho_\nu]$ on $A$, with the density matrix $\rho_\nu$ [Eq.~\eqref{eq:maxrev_rhonu}] reflecting our updated belief. 
\end{itemize}

Two notable limiting cases for this ensemble are discussed next.

\subsubsection{No charge fluctuations\label{sec:ensemble_z_no}}

First we verify that our prediction recovers the case considered in Sec.~\ref{sec:rmt} when  the initial state has a well defined charge, $p_{\rm in} (Q) = \delta_{Q,Q_0}$. 
In that case it is straightforward to see that
\begin{equation}
    p(Q_A|\nu) = \delta_{Q_A, Q_0 - Q_B(\nu) },
\end{equation}
i.e., since we are already certain about $Q = Q_0$, measuring $Q_B$ on $B$ tells us with certainty that $Q_A = Q_0 - Q_B$. 
Thus the average state on $A$ upon measuring bitstring $\nu$ on $B$ is
\begin{equation}
    \rho_\nu = \binom{N_A}{Q_0 - Q_B(\nu) }^{-1} \hat{\Pi}_{Q_0 - Q_B(\nu) },
\end{equation}
i.e., a maximally mixed state of fixed charge. We see that each value of the charge $Q_B = \sum_i \nu_i$ defines a distinct equivalence class $[\nu]$.
The Scrooge ensemble $\mathcal{E}_{\rm Scrooge}[\rho_\nu]$ is hence  simply the Haar measure on the charge sector $Q_A = Q_0 - Q_B(\nu) $. 
The GSE then returns the direct sum of Haar measures across charge sectors weighted according to 
\begin{equation} 
p([\nu]) 
= d_{Q_A}d_{Q_B} / d_Q \equiv \pi(Q_A|Q_0), 
\end{equation}
with $Q_B = Q_B(\nu)$ and $Q_A = Q_0 - Q_B$. 
This recovers the rigorous results of Theorem~\ref{thm:1}. 

\subsubsection{Equilibrium charge fluctuations \label{sec:ensemble_z_eq}}

Another case where the GSE reduces to a simple but physically relevant form is when the initial state has the same charge distribution as a thermal equilibrium state, 
\begin{equation} 
\rho_{\rm eq.} = \frac{e^{\beta \mu \hat{Q}}}{(1+e^{\beta\mu})^N}.
\end{equation}
Here, $\beta$ is the inverse temperature and $\mu$ the chemical potential. Together, they make up the fugacity $z = e^{\beta \mu}$. 
A pure initial state with this charge distribution is, for instance, $(\ket{0} + e^{\beta\mu}\ket{1})^{\otimes N}$, up to normalization.
One has in this case
\begin{equation}
    p_{\rm in} (Q) = {\rm Tr}\left( \hat{\Pi}_Q \frac{z^{\hat Q}}{(1+z)^N} \right) = \frac{z^Q}{(1+z)^N} \binom{N}{Q}.
    \label{eq:equilibrium_fluctuations}
\end{equation}

Plugging this form of $p_{\rm in}(Q)$ into Eq.~\eqref{eq:maxrev_pnu} and summing over all bitstrings $\nu$ of the same charge $Q_B(\nu)$ yields
\begin{align}
    p(Q_B) 
    & = \sum_{Q_A'} \frac{z^{Q_A'+Q_B } }{ (1+z)^N}  \binom{N_A}{Q_A'} \binom{N_B}{Q_B} \nonumber \\
    & =  \frac{z^{Q_B}}{(1+z)^{N_B}}  \binom{N_B}{Q_B}
\end{align}
and the post-measurement conditional distribution of charge on $A$
\begin{equation}
	p(Q_A|\nu) = \frac{z^{Q_A}}{(1+z)^{N_A}} \binom{N_A}{Q_A}.
\end{equation}
Both of these have the same form of a thermal equilibrium charge distribution with the same value of the fugacity $z = e^{\beta\mu}$. 
Qualitatively, this means that $Q_A$ and $Q_B$ separately follow (uncorrelated) thermal equilibrium distributions, and measuring $Q_B$ does not give us any information about $Q_A$. 

It follows that the density matrices used in the GSE construction, Eq.~\eqref{eq:maxrev_rhonu}, are given by
\begin{align}
\rho_\nu    
	& = \sum_{Q_A=0}^{N_A} \hat{\Pi}_{Q_A} \frac{z^{Q_A}}{(1+z)^{N_A}} 
    \propto e^{\beta\mu\hat{Q}_A}. 
\end{align} 
Since all the density matrices $\{ \rho_\nu  \}$ are equal, independent of $\nu$, we have a {\it single} equivalence class $[\nu]$ in this case, comprising all measurement outcomes. The GSE then reduces to the simple Scrooge ensemble induced by the thermal equilibrium state $e^{\beta\mu\hat{Q}_A}$. 
In the special case of infinite temperature  (or zero chemical potential, $\beta\mu=0$), the Scrooge ensemble further reduces to the Haar ensemble over the {entire} Hilbert space of $A$. 
We comment further on the significance of this result in relation to `charge-non-revealing' measurements below.

\subsection{Charge non-revealing measurements \label{sec:ensemble_x}}

We next consider measurements along a direction in the $x$-$y$ plane, which do not reveal any information about the local charge density ($z$ component of the spin).
In this case, our general ansatz predicts the following. For any possible outcome state $\ket{\nu}$ on $B$, the charge distribution $p_{\rm out}$ [Eq.~\eqref{eq:p_out}] reads
\begin{equation}
    p_{\rm out}(Q_B|\nu) = 2^{-N_B} \binom{N_B}{Q_B},
\end{equation}
which is the equilibrium charge distribution at infinite temperature ($\beta \mu  = 0$). 
It is easy to check that $p(\nu) = 2^{-N_B}$ for all $\nu$---this can be seen from Eq.~\eqref{eq:pnu} or by noting that all bitstrings are related via local phase flips, which leave the generator ensemble $\mathcal{F}$ invariant). Then from Eq.~\eqref{eq:pqa_nu} we obtain
\begin{equation}
    p(Q_A|\nu) = \sum_{Q} p_{\rm in}(Q) \frac{\binom{N_Q}{Q_A} \binom{N_B}{Q-Q_A}}{ \binom{N}{Q} }
\end{equation}
and thus the unique, $\nu$-independent density matrix 
\begin{equation}
    \rho_\nu = \sum_{Q,Q_A} p_{\rm in}(Q) \frac{\binom{N_B}{Q-Q_A}}{ \binom{N}{Q} } \hat{\Pi}_{Q_A}.
\end{equation}
We have therefore a single equivalence class $[\nu]$ for all measurement outcomes, and the GSE ansatz reduces to a simple Scrooge ensemble for the $\rho_\nu$ above. 

In Appendix~\ref{app:charge_non_revealing} we show that this result indeed holds, within the same replica approach as in Sec.~\ref{sec:ensemble_z}, when starting from states of a well-defined but arbitrary charge $Q_0$, $p_{\rm in}(Q) = \delta_{Q,Q_0}$. 
Notably, for a charge-neutral initial state ($Q_0 = N/2$), the density matrix $\rho_\nu$ becomes close to the identity, up to corrections of order $N_A/N$ which vanish in the thermodynamic limit, so the Scrooge ensemble reduces to the Haar ensemble.

The fact that we obtain a single Scrooge ensemble rather than a GSE may be understood as a consequence of the fact that $x$-basis measurement outcomes on $B$ do not provide any information on the way that the charge $Q_0$ is partitioned between $A$ and $B$. 
This should be contrasted to the case of charge-revealing measurements, where the measured value of $Q_B$ generically updates our belief about the charge in $A$, giving rise to a GSE---a different Scrooge ensemble for each possible outcome $Q_B$.
An exception is the case studied in Sec.~\ref{sec:ensemble_z_eq}, of states with an equilibrium charge distribution. In that case, since the charge distributions on $A$ and $B$ are fully uncorrelated, {\it all measurement bases are effectively `charge-non-revealing'}---even the $z$ basis.
It is notable that the result of Sec.~\ref{sec:ensemble_z_eq} (a single Scrooge ensemble) is in agreement with the present result for $x$-basis measurements. This shows a universality of deep thermalization across different instances of `charge non-revealing' measurements.

We conclude by comparing our results to the case of Hamiltonian dynamics, also reviewed in Sec.~\ref{sec:deep_therm_conservation}. There Ref.~\cite{cotler_emergent_2023} found that the projected ensemble under `energy-non-revealing' measurements is well described by a single (i.e., not generalized) Scrooge ensemble, while general `partially energy-revealing' measurements give rise to a GSE~\cite{mark_maximum_2024}. This is analogous to the phenomenology we have uncovered in the case of $U(1)$ charge, though we stress again the important difference in complexity of our results: the GSE of Ref.~\cite{mark_maximum_2024} requires an exponential amount of data to construct, while ours only a polynomial amount. This makes our ensembles   physical, as they represent genuine {\it thermodynamic} ensembles at equilibrium: such compression of information is a defining feature of statistical mechanics.

\section{Numerical results \label{sec:numerics}}

\begin{figure}
    \includegraphics[width=\columnwidth]{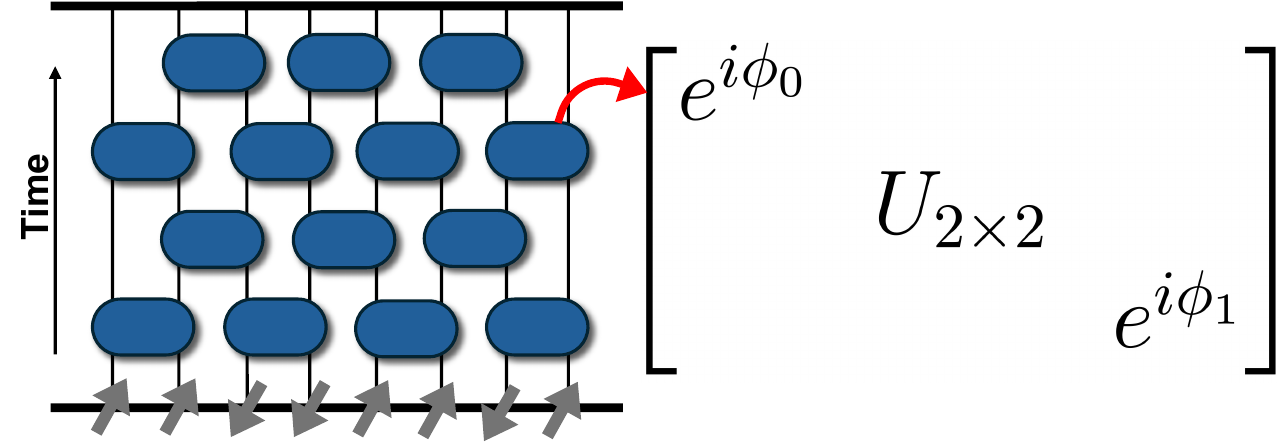}
    \caption{Left: a random $U(1)$-symmetric quantum circuit acting on a $1D$ qubit chain, the setting of our numerical simulations. Right: implementation of the $U(1)$ symmetry (charge conservation) in the circuit. Each two-qubit quantum gate has a block-diagonal form such that  the weights on the one-dimensional charge sectors $\{\ket{00}\}$, $\{\ket{11}\}$ are preserved but are imparted random phases $e^{i\phi_0}, e^{i\phi_1}$, while the two-dimensional charge sector $\{\ket{01},\ket{10}\}$ is Haar randomized by a two-by-two Haar random unitary $U_{2 \times 2}$.}
    \label{fig:RQC}          
\end{figure}

In this section we present data from numerical simulations to corroborate our analytical predictions of Sec.~\ref{sec:limiting_ensemble}.
We focus on $1D$ qubit chains that evolve under random unitary circuits of two-qubit gates applied to the system in a staggered (`brickwork') pattern, as shown in Fig.~\ref{fig:RQC}.
We impose a local $U(1)$ symmetry tied to the conservation of   the $z$ component of the total spin, $\hat{Q} = \sum_i (\mathbb{I}+Z_i)/2$,  
by restricting the gates in the random unitary circuits to have a block-diagonal form, also shown in Fig.~\ref{fig:RQC}.
Specifically the states $\ket{00}$, $\ket{11}$ span one-dimensional charge sectors and thus a local gate simply imparts uniformly random phases $e^{i\phi_0}$, $e^{i\phi_1}$ onto them ($\phi_0$ can be set to zero by fixing the overall phase), while the states $\ket{01}$, $\ket{10}$ span a two-dimensional charge sector and transform within themselves via the action of a Haar random unitary  in $U(2)$. 
This model has been used to study the dynamics of generic chaotic systems with a local conserved density in an analytically tractable way~\cite{khemani_operator_2018,rakovszky_diffusive_2018,hunter-jones_operator_2018}.

To test the theory we presented in Sec.~\ref{sec:limiting_ensemble}, we aim to numerically characterize the projected ensemble of a time-evolved state $\ket{\Psi(t)} = U_t U_{t-1}\cdots U_1\ket{\Psi(0)}$ in the limit $t\to\infty$, where $\{U_t\}$ is a sequence of brickwork layers of unitary gates sampled randomly as described above, and $\ket{\Psi(0)}$ is an arbitrary initial state of the qubit chain. 
For concreteness we consider the following family of initial states, parametrized by an angle $\theta \in [0, \pi/2]$:
\begin{align}\label{eq:init_states}
   \ket{\Psi(0)} & = \ket{+\theta,-\theta}^{\otimes N/2}, \\
   \ket{+\theta} & = \cos \frac{\theta}{2}|0\rangle + \sin\frac{\theta}{2}|1\rangle \label{eq:plus_theta} \\
   \ket{-\theta} & = -\sin\frac{\theta}{2}|0\rangle + \cos\frac{\theta}{2}|1\rangle. \label{eq:minus_theta}
\end{align}
(The total number of qubits $N$ is taken to be even). These are product states with average charge $\langle \hat{Q} \rangle = \bra{\Psi(0)} \hat{Q} \ket{\Psi(0)} = N/2$ for all $\theta$, i.e.~these states are charge-neutral on average.
However, as a function of $\theta$, the states have different fluctuations of the charge: for example, the variance of $Q$ is given by 
\begin{equation}
    \bra{\Psi(0)}\hat{Q}^2 \ket{\Psi(0)} - \bra{\Psi(0)}\hat{Q}\ket{\Psi(0)}^2
    = \frac{N}{4} \sin^2(\theta).
    \label{eq:charge_var}
\end{equation}
For $\theta = 0$ we have $\ket{\Psi(0)} = |01\rangle^{\otimes N/2}$, known as the N\'eel state in the context of magnetism. This state has no charge fluctuations at all, $p(Q) = \delta_{Q,N/2}$.
For $\theta = \pi / 2$ we have $\ket{\Psi(0)} = \ket{+-}^{\otimes N/2}$, which has $p(Q) = 2^{-N}\binom{N}{Q}$, the same distribution as a maximally mixed state.

We benchmark the analytical derivations in Sec.~\ref{sec:replica} by numerically simulating `maximally charge-revealing' ($z$ basis) measurements in Sec.~\ref{sec:numerics_z} and `charge-non-revealing' ($x$ basis) measurements in Sec.~\ref{sec:numerics_x}. Finally we systematically test our general ansatz, Sec.~\ref{sec:limiting_ensemble}, by sweeping over general values of $\theta$ for both the input state and the measurement basis in Sec.~\ref{sec:numerics_general}. 

\subsection{Maximally charge-revealing measurements \label{sec:numerics_z}}

\subsubsection{No charge fluctuations: N\'eel state} \label{sec:neel}

For $\theta = 0$ we recover the N\'eel state, $\ket{01}^{\otimes N/2}$, which is a state of definite total charge $Q_0 = N/2$. Since the measurements are carried out in the $z$ basis, each projected state has a definite charge: $Q_A = Q_0 - Q_B$, where $Q_B$ is the charge on $B$ revealed by the measurement. As discussed in Sec.~\ref{sec:rmt}, the form of the projected ensemble in this case is predicted to be a direct sum of Haar ensembles in each charge sector:
\begin{equation}
    \rho^{(k)}_{{\rm target}} = \rho^{(k)}_{\text{DS},Q_0} = \bigoplus_{Q_A=0}^{N_A} \pi(Q_A|Q_0) \rho^{(k)}_{{\rm Haar},A,Q_A},
    \label{eq:direct_sum_target}
\end{equation}
with $\pi(Q_A|Q_0) = \binom{N_A}{Q_A} \binom{N_B}{Q_0-Q_A} \binom{N}{Q_0}^{-1}$ and, for the N\'eel state, $Q_0 = N/2$. Here $\rho^{(k)}_{\text{Haar},A,Q_A}$ is the $k^{\text{th}}$ moment of the Haar ensemble in the Hilbert space of charge sector $Q_A$. 

We test the predition of Eq.~\eqref{eq:direct_sum_target} by calculating the trace distance $\Delta^{(k)}$ between the $k$-th moment operators, defined in Eq.~\eqref{eq:distance_measure}. 
For simplicity, we fix the size of the subsystem $A$ to $N_A = 2$ (the minimal subsystem with a nontrivial charge sector, $Q_A = 1$) and the statistical moment to $k = 2$.
We vary the total size of the system $N$ between 8 and 24 qubits to investigate the scaling toward the thermodynamic limit $N\to\infty$ as the subsystem $A$ is kept fixed.
We average the distance measure $\Delta^{(2)}$ over 128 realizations of the random circuit. Note that the average takes place {\it after} the calculation of the (non-linear) distance measure $\Delta^{(2)}$, so the emergence of universal random wavefunction distributions (in this case, the `direct sum' ensemble) is due to the randomness inherent in the projected ensemble, not to the randomness of the circuit realization.

\begin{figure}
    \includegraphics[width=0.5\textwidth]{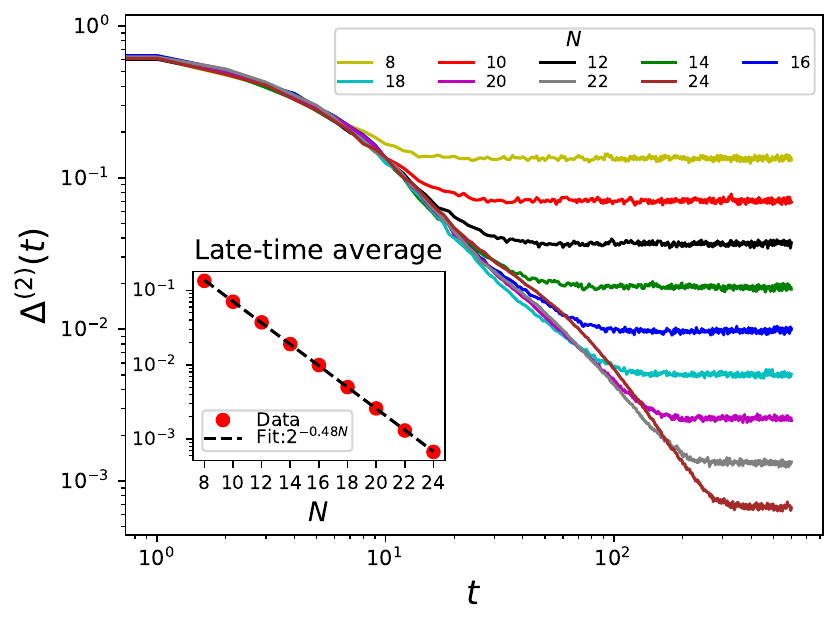}
    \caption{
    Trace distance $\Delta^{(2)}$ between the second moment of the projected ensemble and that of the `direct sum' target ensemble Eq.~\eqref{eq:direct_sum_target} as a function of time, for the initial state $\ket{01}^{\otimes N/2}$ and measurements in the $z$ basis. 
    The total system size $N$ is varied while the subsystem size $N_A$ is fixed to 2. Results are averaged over 128 realizations of the random circuit. 
    Inset: late-time averaged value of $\Delta^{(2)}$ as a function of $N$. Exponential fit (dashed line) yields $2^{-0.48N}$, consistent with the expected $2^{-N/2}$.
    }
    \label{fig:nor_di_Neel}            
\end{figure}

As shown in Fig.~\ref{fig:nor_di_Neel}, the trace distance $\Delta^{(2)}$ decays with time until it reaches a plateau arising from the finite-sized nature of the simulation. As the total system size $N$ increases, this plateau decays exponentially with the total system size $N$, as shown in the inset of Fig.~\ref{fig:nor_di_Neel}. An exponential fit yields $2^{-0.48N}$, consistent with the inverse square root of the total Hilbert space dimension $d = 2^N$. This scaling agrees with the expected local fluctuations in a Haar-random state of $N$ qubits, as the inverse square root of the Hilbert space dimension~\cite{cotler_emergent_2023}.
This suggests $\Delta^{(2)}\to 0$ in the limit of large system size $N$ and long time $t$, which signifies deep thermalization to the `direct sum' form of the projected ensemble, Eq.~\eqref{eq:direct_sum_target}, at the level of the second moment $k = 2$. 
These observations are in agreement with our theoretical predictions, including Theorem~\ref{thm:1}. 

\subsubsection{Equilibrium charge fluctuations: $X$-basis state} \label{sec:all_plus}

\begin{figure}[t]
    \includegraphics[width=0.5\textwidth]{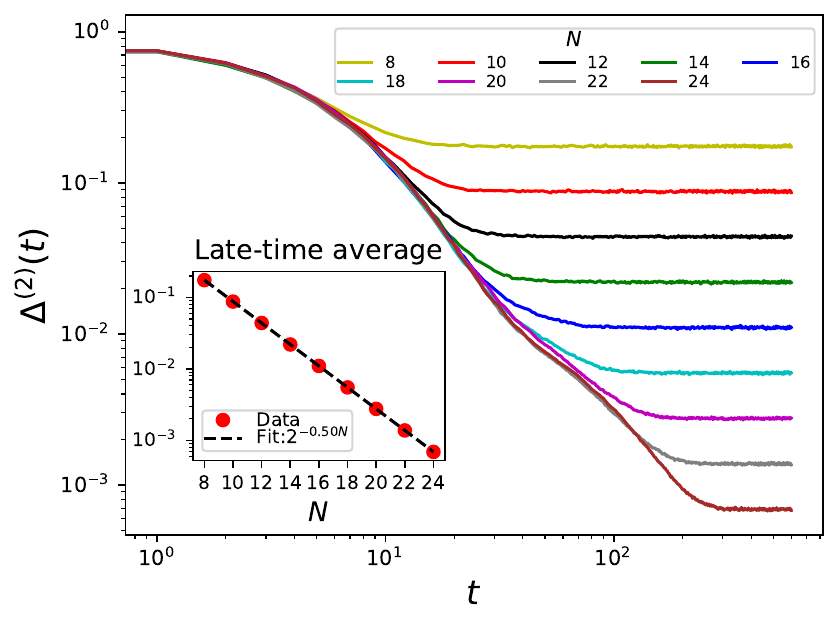}
    \caption{Trace distance $\Delta^{(2)}$ between the second moment of the projected ensemble and that of the Haar ensemble over the entire Hilbert space as a function of time, for the initial state $\ket{+-}^{\otimes N/2}$ and measurements in the $z$ basis. The total system size $N$ is varied while the subsystem size $N_A$ is fixed to 2. Results are averaged over 128 realizations of the random circuit.
    Inset: late-time averaged value of $\Delta^{(2)}$ as a function of $N$. The behavior is consistent with $2^{-N/2}$ (dashed line), indicating convergence to $\Delta^{(2)}=0$ in the thermodynamic limit.
    }
    \label{fig:nor_dis_allplus}       
\end{figure}

For $\theta = \pi/2$ we have the initial state $\ket{+-}^{\otimes N/2}$, whose charge distribution is the same as that of the maximally mixed state. We have seen in Sec.~\ref{sec:ensemble_z_eq} that in this case the limiting form of the projected ensemble should be Haar-random in the full Hilbert space. 
To test if and how the projected ensemble approaches the Haar ensemble over time, we repeat the calculation of Eq.~\eqref{eq:distance_measure} as we did for the N\'eel state above, but using the Haar ensemble as target.  

As shown in Fig.~\ref{fig:nor_dis_allplus}, the distance $\Delta^{(2)}$ between second moment operators decays with time until it eventually saturates to a non-zero plateau arising due to finite-size effects. 
The value of this late-time plateau decays exponentially in $N$, as shown in the inset of Fig.~\ref{fig:nor_dis_allplus}. An exponential fit of the late-time average values vs $N$ yields the scaling $\sim 2^{-0.50 N}$, again consistent with $1/\sqrt{d_A}$. 
We can infer that the distance $\Delta^{(2)}$ approaches zero in the thermodynamic limit $N\rightarrow \infty$ (with fixed $N_A$), which is consistent  with our predictions of deep thermalization to the Haar ensemble. 

\subsubsection{General initial states} \label{sec:arbitrary}

\begin{figure*}
    \centering
    \includegraphics[width=0.9\textwidth]{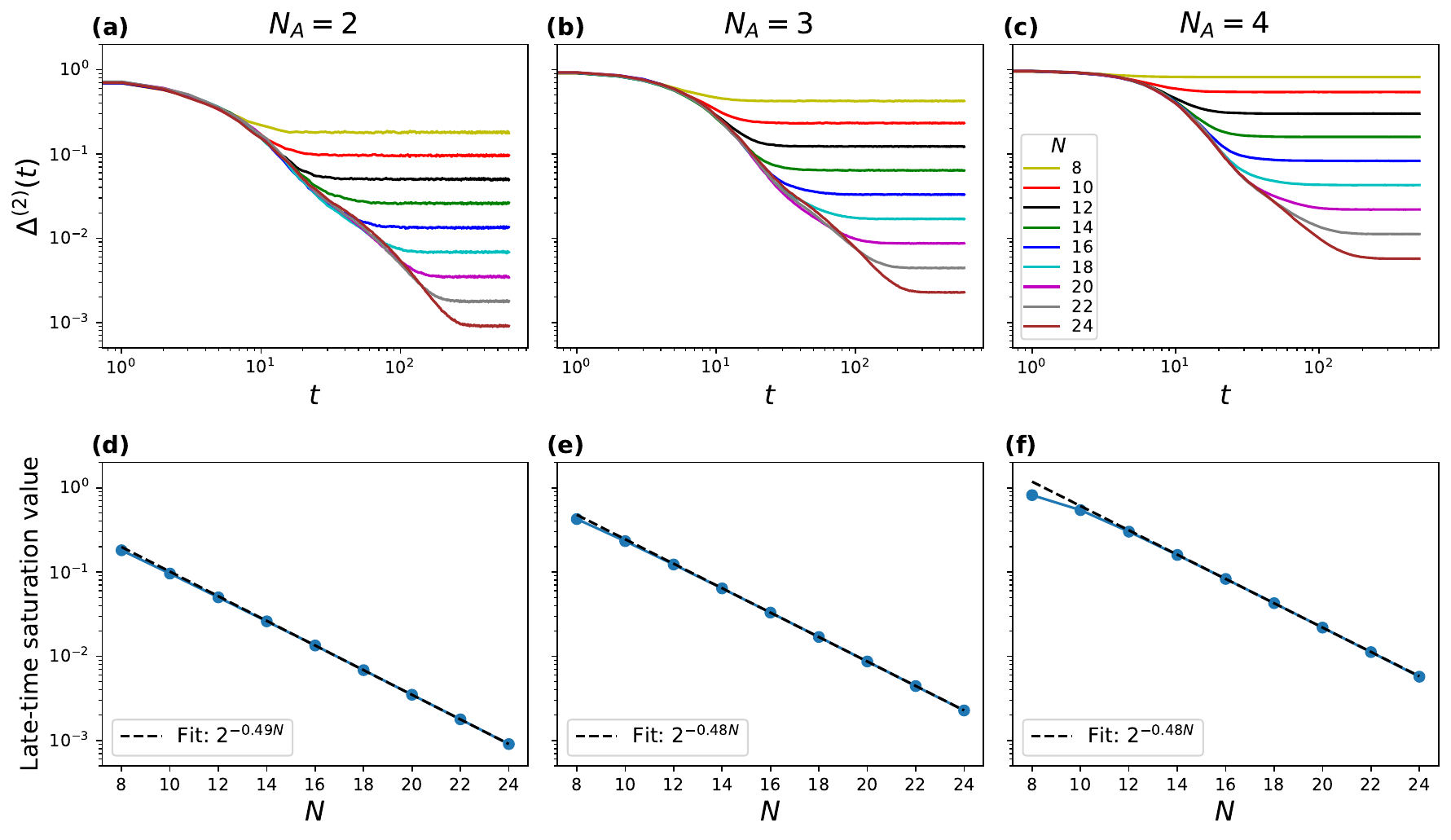}
    \caption{(a)-(c): Trace distance $\Delta^{(2)}$ between the second moment of the projected ensemble and that of the GSE, Eq.~\eqref{eq:gen_scrooge}, as a function of time, for the initial state in Eq.~\eqref{eq:init_states} with $\theta=\pi/10$ and measurements in the $z$ basis. The subsystem size $N_{A}$ is held fixed at $2$, $3$, and $4$ in (a), (b), (c), respectively, while the total system size $N$ is varied. Results are averaged over 128 realizations of the random circuit.
    (d)-(f): Late-time average values of $\Delta^{(2)}$, extracted from (a)-(c), plotted against $N$. As $N_{A}$ is increased, a proportionally larger value of $N_B$ is required in order to converge to a universal ensemble, as expected. Nonetheless we see a decay of the late-time value of $\Delta^{(2)}$ consistent with $2^{-N/2}$ in all cases (dashed lines).
    }
\label{fig:generalized_scrooge_combined}
\end{figure*}

So far we have discussed the two extreme cases of maximal and minimal information about the charge in the initial state: $\theta = 0$ (N\'eel state) and $\theta = \pi/2$ ($x$-basis state). The target ensembles for these two cases reduce to simple forms---respectively the Haar ensemble and the direct sum of charge-sector Haar ensembles. 
For intermediate values of $\theta$, the predicted limiting form of the projected ensemble becomes more complex and is given by the GSE, Eq.~\eqref{eq:gen_scrooge_compressed}. 
To test this prediction, we select an intermediate value of $\theta = \pi/10$ (the reason for this parameter choice is that the GSE's second-moment operator is roughly equidistant from those at $\theta=0$ and $\theta=\pi/2$, as discussed in Appendix~\ref{app:param_choice}). 
We numerically construct the GSE's second moment operator by Monte Carlo sampling $M$ pairs $(Q_B,\ket{\psi})$, where $Q_B = Q_B(\nu)$, $\nu \sim p(\nu)$, and $\ket{\psi} \sim \mathcal{E}_{\rm Scrooge}[\rho_\nu]$, and computing
\begin{equation}
\rho^{(2)}_{GSE} \simeq \frac{1}{M} \sum_{i=1}^M \ketbra{\psi_i}^{\otimes 2}.
\label{eq:montecarlo_moment}
\end{equation}
We find this approximation is numerically converged to the required precision at the largest system size studied ($N_A=4$, $N=24$) with $M = 10^8$ samples, see Appendix~\ref{app:param_choice}.
Finally, we numerically compute the trace distance $\Delta^{(2)}$ between the second moment of the projected ensemble and that of the GSE, Eq.~\eqref{eq:montecarlo_moment}.

Our derivation of the GSE, Appendix~\ref{app:charge_revealing}, relies on an assumption of large Hilbert space dimension not only for the whole system ($N \to\infty$), but also for the subsystem of interest ($N_A\to\infty$, and specifically $N_A\gg \log k$ based on the analysis of Appendix~\ref{app:approximation_error}). To study the possible dependence on subsystem size, we set $N_A = 2, 3, 4$ and compute the trace distance $\Delta^{(2)}$ as a function of time in each case. 
The numerical results are shown in Fig.~\ref{fig:generalized_scrooge_combined}.
We see the same qualitative behavior of the previous cases: $\Delta^{(2)}$ decays with time to a finite-size plateau, which in turn decays exponentially in $N$, consistent with $2^{-N/2}$. 
As $N_A$ increases, the size of the bath needs to increase proportionally---e.g., in the absence of conservation laws, it is known that $N_B\geq kN_A$ is needed for equilibration of the $k$-th moment~\cite{cotler_emergent_2023}; a similar constraint appears in Theorem~\ref{thm:1}. Thus for $N_A> 2$ we see larger numerical values of the distance $\Delta^{(2)}$, but the trends as $N\to\infty$ are unchanged. 

Fig.~\ref{fig:generalized_scrooge_combined} shows that the GSE is a surprisingly accurate description of the limiting ensemble already at the minimum nontrivial subsystem size of $N_A = 2$. This raises the possibility that the $N_A \to\infty$ limit used in our replica limit calculation, App.~\ref{app:ansatz}, may be an artifact of the approach and not in fact necessary for the validity of the result.

\subsection{Charge non-revealing measurements \label{sec:numerics_x}}

Next we verify our predictions for $x$-basis measurements, which are charge-non-revealing. 
We again focus on the second moment ($k=2$) and consider the N\'eel state $\ket{01}^{\otimes N/2}$ and $X$-basis product state $\ket{+-}^{\otimes N/2}$. In the former case, the results of Sec.~\ref{sec:ensemble_x} suggest convergence to the Haar ensemble. The latter case falls outside the ones we could address by replica methods, but our general ansatz predicts the same outcome. 
For this reason we set the target ensemble to be the Haar ensemble and compute the distance $\Delta^{(2)} = \frac{1}{2} \| \rho^{(2)}_{\mathcal E} - \rho_{\rm Haar}^{(2)}\|_{\rm tr}$. 

\begin{figure}
  \includegraphics[width=0.5\textwidth]{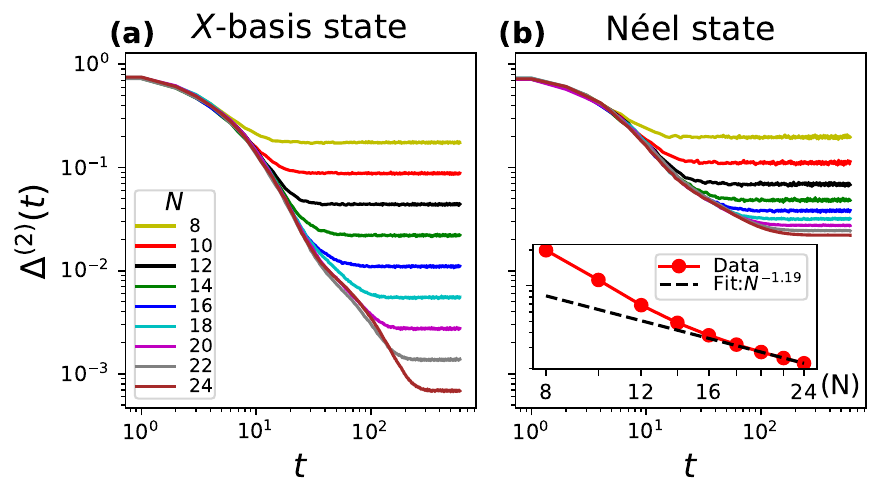}
  \caption{Trace distance $\Delta^{(2)}$ between the second moment of the projected ensemble and the Haar ensemble, for the initial state in Eq.~\eqref{eq:init_states} with measurements in the $x$ basis. (a) $\theta=\pi/2$ ($X$-basis initial state). $\Delta^{(2)}$ decays to a late-time plateau that is exponentially small in the size of the total system $N$. This result is identical to that of $Z$-basis measurement shown in Fig.~\ref{fig:nor_dis_allplus}.
  (b) $\theta = 0$ (N\'eel initial state). The late-time plateau decays much more slowly in $N$. (Inset) A fit indicates a power-law decay consistent with $\sim 1/N$ (the fit exponent $N^{-1.19}$ is likely affected by strong finite-size effects due to a crossover from an exponential behavior at smaller $N$, to the asymptotic power-law behavior at large $N$).}
\label{fig:xbasis_Haar}
\end{figure}

For the $X$-basis state,  Fig.~\ref{fig:xbasis_Haar} (a), we can clearly see that $\Delta^{(2)}$ decays to a late-time plateau that is exponentially small in the size of the total system $N$, consistent with deep thermalization to the Haar ensemble.
For the N\'eel state, on the other hand, the convergence as a function of $N$ appears considerably slower, as shown in Fig. \ref{fig:xbasis_Haar} (b). 
The reason for these strong finite-size effects is due to the first moment: $\rho_A$ (at late time, on average) is diagonal, with entries that depend only on the charge sector $Q_A$ and are given by 
\begin{equation}
    \bra{z} \rho_A \ket{z} = \binom{N_B}{Q_0 - Q_A} \binom{N}{Q_0}^{-1} \label{eq:finiteN_rho}
\end{equation}
where $z$ is any basis state of charge $Q_A$ and $Q_0 = N/2$. These entries become $Q_A$-independent in the limit $N_B\to\infty$, so that $\rho_A\to \mathbb{I} / 2^{N_A}$ and it becomes possible to obtain deep thermalization to the Haar ensemble; but the covergence is only polynomial in $1/N_B$. Indeed we have 
\begin{align}
    \binom{N_B}{Q_0 - Q_A} \binom{N}{Q_0}^{-1}
    & = \frac{N_B!}{N!} \frac{(N/2)!}{(N/2-Q_A)!} \nonumber \\
    & \qquad \times  \frac{(N/2)!}{(N/2-N_A+Q_A)!} 
\end{align}
and, by making repeated use of the `birthday paradox' asymptotics $x! / (x-k)! = x^k \exp [-\binom{k}{2} \frac{1}{x} + O(1/x^2)]$ (valid for $x\gg k$), we can conclude
\begin{align}
    \binom{N_B}{Q_0 - Q_A} \binom{N}{Q_0}^{-1}
    & \simeq 2^{-N_A} e^{O(1/N)},
\end{align}
where the coefficient of the $1/N$ correction depends on $Q_A$ and generically does not vanish.
Therefore the trace distance between $\rho_A$ and $\rho_{\rm Haar}^{(1)} = \mathbb{I}/2^{N_A}$ (first moment of the Haar ensemble) decays only as $\sim 1/N$. This error on the first moment also affects the second moment, and thus slows down the decay of $\Delta^{(2)}$. The inset to Fig.~\ref{fig:xbasis_Haar}(b) indeed confirms a power-law decay in $N$, with an exponent close to $-1$. 

\begin{figure}
  \includegraphics[width=0.5\textwidth]{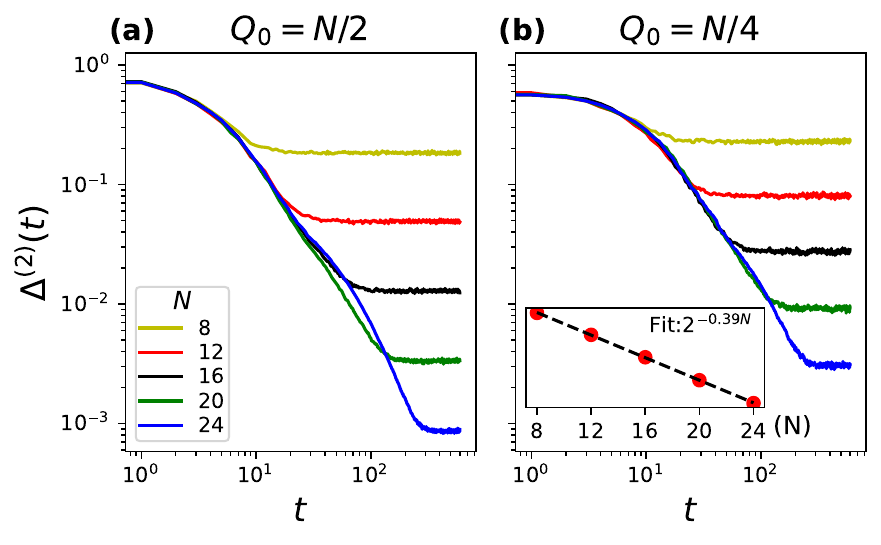}
  \caption{Trace distance $\Delta^{(2)}$ between the second moment of the projected ensemble and the Scrooge ensemble formed from the {finite}-$N$ density matrix $\rho_A$. Measurements are performed in the $x$ basis. 
  (a) N\'eel initial state ($\ket{01}^{\otimes N/2}$), with total charge $Q_0 = N/2$. The late-time value of $\Delta^{(2)}$ decays as $\sim 2^{-N/2}$ (not shown). (b) Initial state $\ket{0001}^{\otimes N/4}$, with total charge $Q_0 = N/4$. Inset: The the late-time value of $\Delta^{(2)}$ decays exponentially in $N$; a fit yields $\sim 2^{-0.39 N}$, consistent with the expectation $\sim 2^{-H_2(\sigma)N/2}$ where $H_2(\sigma)$ is the binary entropy of the charge density $\sigma$, and $H_2(1/4) \simeq 0.81$.}
\label{fig:xbasis_partialfill}
\end{figure}

In light of this, we repeat the calculation for a target ensemble that explicitly depends on $N$: we consider the Scrooge ensemble formed from the {\it finite}-$N$ average density matrix $\rho_A(N)$, with entries given in Eq.~\eqref{eq:finiteN_rho}. 
Results for $\Delta^{(2)}$ with this size-dependent target ensemble are shown in Fig. \ref{fig:xbasis_partialfill}(a), and clearly display an exponential-in-$N$ convergence. Therefore we have exponential-in-$N$ convergence to the Scrooge ensemble $\mathcal{E}_{\rm Scrooge}[\rho_A(N)]$, alongside $\sim 1/N$ convergence of the latter toward the Haar ensemble. 

Finally, we test the predictions of Sec.~\ref{sec:ensemble_x} on states of well-defined charge away from neutrality, $Q_0 \neq N/2$. The conjectured limiting ensemble in that case is the Scrooge ensemble for the appropriate density matrix $\rho_A$. 
We thus repeat our calculation for the initial state $\ket{0001}^{\otimes N/4}$, and again construct an $N$-dependent target ensemble based on the finite-$N$ reduced density matrix $\rho_A(N)$, Eq.~\eqref{eq:finiteN_rho} with $Q_0\mapsto N/4$. Results, shown in Fig.~\ref{fig:xbasis_partialfill}(b), are again consistent with exponential-in-$N$ convergence. Specifically, we see a finite-size decay as $2^{-H_2(\sigma)N/2}$, with $H_2(\sigma) = -\sigma\log_2(\sigma)-(1-\sigma)\log_2(1-\sigma)$ the binary entropy (in base 2) of the charge density $\sigma = Q_0 / N = 1/4$. This is again consistent with the expectation of finite-size effects scaling as the inverse square root of Hilbert space dimension, since $d_{Q_0} = \binom{N}{Q_0} \sim 2^{H_2(\sigma)N+o(N)}$.
These results support the conclusion of deep thermalization to the Scrooge ensemble under charge-non-revealing measurements on symmetric states.

\subsection{General initial states and measurement bases \label{sec:numerics_general}}

\begin{figure}
    \centering
    \includegraphics[width=0.99\linewidth]{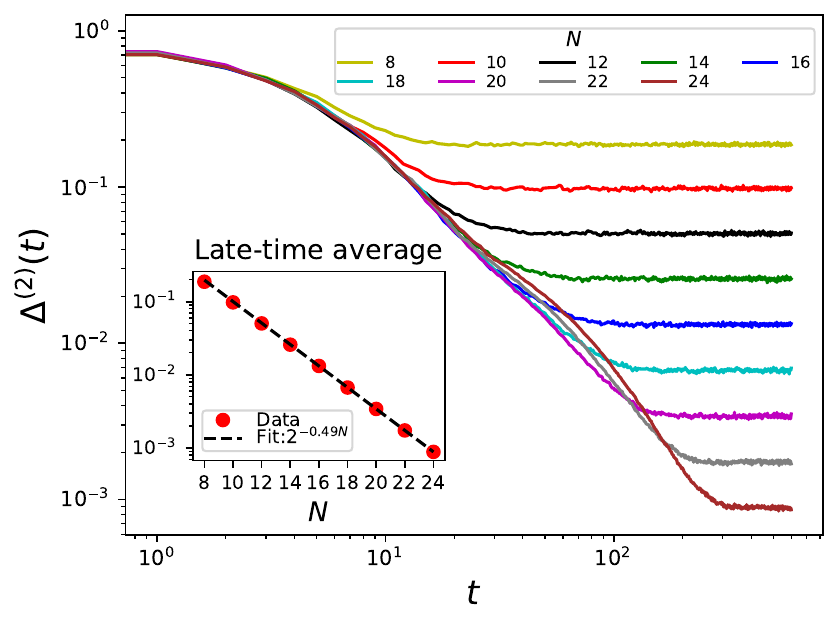}
\caption{Trace distance $\Delta^{(2)}$ as a function of time between the second moment of the projected ensemble and that of the generalized Scrooge ensemble for the initial state $\ket{01}^{\otimes N/2}$. The measurements are performed in a partially charge-revealing basis $\ket{\pm \pi/4}$, halfway between the $z$ and $x$ basis. The total system size $N$ is varied while the subsystem size $N_A$ is fixed to 2. Results are averaged over 128 realizations of the random circuit. Inset: late-time averaged value of $\Delta^{(2)}$ as a function of $N$. Exponential fit (dashed line) yields $2^{-0.49N}$, consistent with the expected $2^{-N/2}$ and indicating convergence to $\Delta^{(2)}=0$ in the thermodynamic limit.}
    \label{fig:partial_revealing}
\end{figure}

Having verified the predictions for the specific cases of maximally charge-revealing measurements and charge non-revealing measurements in Sec.~\ref{sec:numerics_z} and Sec.~\ref{sec:numerics_x}, now we proceed to test the general theory in Sec.~\ref{sec:limiting_ensemble} by focusing on {\it arbitrary} initial states and {\it arbitrary} measurement bases.

First we take the N\'eel state, $\ket{01}^{\otimes N/2}$, of definite total charge $Q_0 = N/2$, as our initial state and perform measurements in a partially charge-revealing basis
\begin{align}
    \ket{+\pi/4}
    & =  \cos \frac{\pi}{8} \ket{0} + \sin \frac{\pi}{8}\ket{1}\\
    \ket{-\pi/4}
    & = -\sin \frac{\pi}{8}\ket{0} + \cos \frac{\pi}{8}\ket{1}.
\end{align}
This basis is halfway between the $z$ and $x$ basis.
To test our general ansatz, we calculate the trace distance between the projected ensemble and the target ensemble as a function of time (circuit depth) with fixed subsystem size $N_A$ and varying total system size $N$. 
Our universal ansatz depends on the charge distributions $p_{\rm in}(Q)$ and $p_{\rm out}(Q_B|\nu)$. 
The former is simply $p_{\rm in}(Q) = \delta_{Q,N/2}$ for the N\'eel state. The latter is calculated in Appendix~\ref{app:P_out} for the whole family of $\ket{\pm \theta}$ product states [Eqs.~(\ref{eq:plus_theta},\ref{eq:minus_theta})] which includes the $\ket{\pm\pi/4}$ basis above. 
In terms of these quantities, the $k$-th moment of the target ensemble is given by 
\begin{equation}
    \sum_{[\nu]} p([\nu]) \rho^{(k)}_{{\rm Scrooge}[\rho_\nu]}, \label{eq:gse_moments}
\end{equation}
with $p([\nu])$ given by Eq.~\eqref{eqn:pnu_equivalence} and $\rho_\nu$ given by Eq.~\eqref{eqn:rho_nu}. Equivalence classes are based on the total number of $\ket{-\pi/4}$ outcomes, $\sum_i \nu_i$.

To construct the Scrooge ensemble moments in Eq.~\eqref{eq:gse_moments}, we again use the Monte Carlo sampling method described in Sec.\ref{sec:arbitrary} with the number of samples $M=10^8$. The numerical results are shown in Fig.~\ref{fig:partial_revealing}.  One can see that, for each total system size $N$,  $\Delta^{(2)}$ decays with time to a finite-size plateau. Moreover, the late-time average of $\Delta^{(2)}$ decays exponentially in $N$, numerically consistent with $2^{-N/2}$, thus showing deep thermalization in the thermodynamic limit and verifying our general ansatz of the limiting ensemble even under partially charge-revealing measurements.

\begin{figure}
    \centering
    \includegraphics[width=0.99\linewidth]{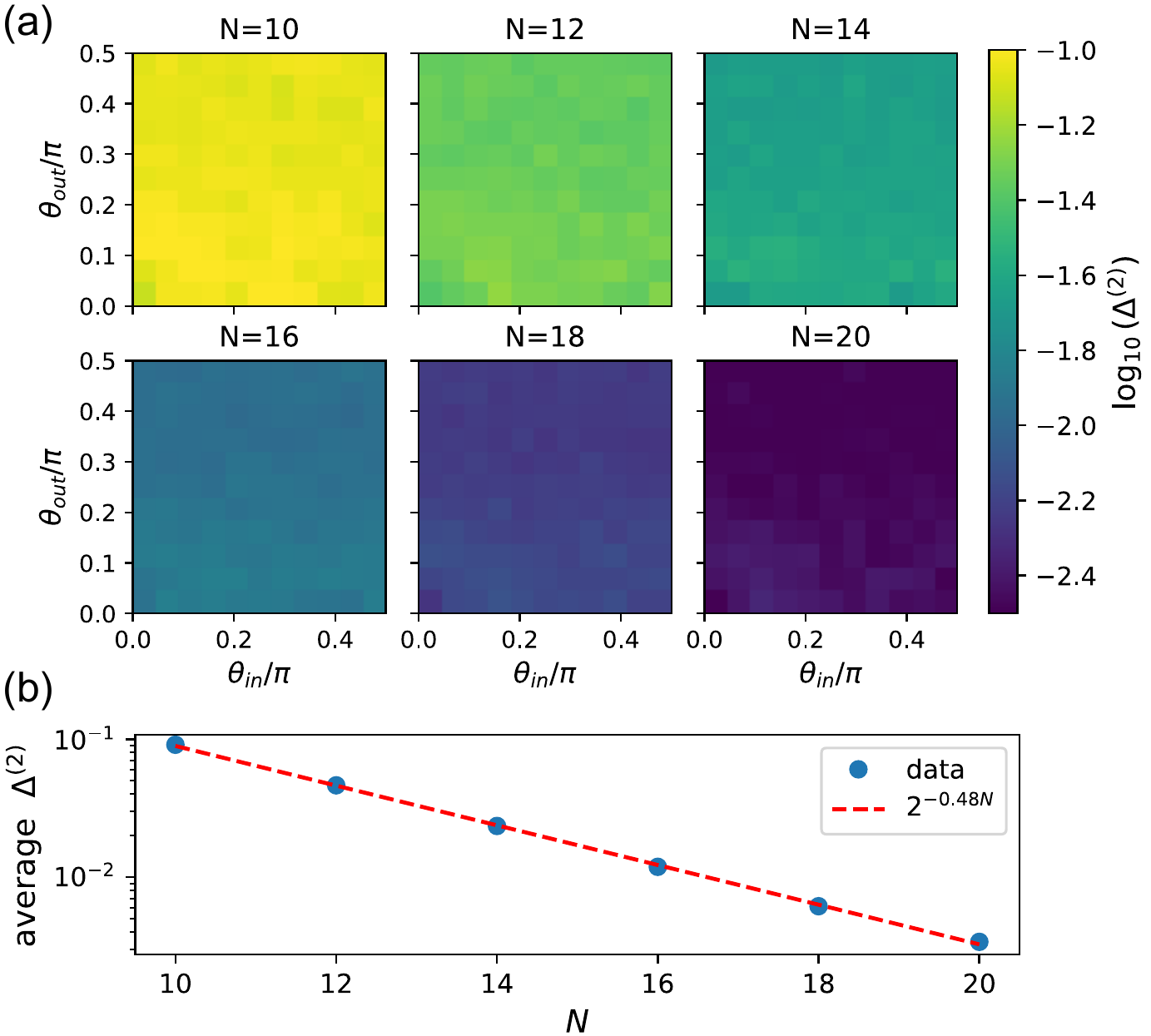}
    \caption{Numerical test of the universal limiting ensemble across different input states and measurement bases. Here, to simulate the late-time limit, we directly apply a global $U(1)$-symmetric random unitary rather than a deep circuit.
    (a) Second-moment trace distance $\Delta^{(2)}$ of the projected ensemble from our GSE ansatz, as a function of angles $\theta_{\rm in}$, $\theta_{\rm out}$ specifying the input state and measurement basis (see main text) for system sizes $N = 10, 12, \dots 20$. Each angle is sampled at intervals of $\pi/20$ in $[0,\pi/2]$. The distance $\Delta^{(2)}$ decays with system size uniformly across choices of input state and measurement basis. 
    (b) Distance $\Delta^{(2)}$ averaged over $\theta_{\rm in, out}$ as a function of system size shows exponential decay consistent with $2^{-N/2}$. }
    \label{fig:sweep}
\end{figure}

Finally, we systematically test our universal ansatz across input states and measurement bases. 
We take the family of input states $\ket{+\theta_{\rm in},-\theta_{\rm in}}^{\otimes N/2}$, in terms of an angle $\theta_{\rm in} \in [0,\pi/2]$ and the single-qubit basis $\ket{\pm \theta}$ in Eqs.~(\ref{eq:plus_theta},\ref{eq:minus_theta}). These states are charge-neutral on average\footnote{We also verify that our results hold away from average charge neutrality, e.g. for states $\ket{+\theta,+\theta,+\theta,-\theta}^{\otimes N/4}$.}
but with variable levels of charge fluctuations. 
We set the measurement basis to be $\ket{\pm \theta_{\rm out}}$ on each qubit in $B$, for $\theta_{\rm out} \in [0,\pi/2]$ and test our ansatz by sweeping $\theta_{\rm in,out}$ independently on a grid of 11 values in $[0,\pi/2]$. 
Given the considerable number of simulations, we limit the system size to $N \leq 20$ and, instead of simulating random $U(1)$-symmetric circuits, we directly jump to the infinite-depth limit by applying a global $U(1)$-symmetric Haar-random unitary to the input state\footnote{In reality we draw $\ket{\Phi_Q} \sim {\rm Haar}(\mathcal{H}_Q)$ and make the generator state $\sum_Q \sqrt{p_{\rm in}(Q)} \ket{\Phi_Q}$ for the appropriate $p_{\rm in}(Q)$, see Appendix~\ref{app:P_out}.}, invoking similar logic as the considerations of Sec.~\ref{sec:rmt}. 

Numerical results are shown in Fig.~\ref{fig:sweep}(a). We see that the distance measure $\Delta^{(2)}$ is nearly uniform across parameter space, indicating that our ansatz correctly describes the projected ensemble in very general combinations of input states and measurement bases. Fig.~\ref{fig:sweep}(b) also shows the average of $\Delta^{(2)}$ over the $\theta_{\rm in,out}$ angles as a function of system size, showing exponential decay $\sim 2^{-N/2}$ that quantitatively agrees with our expectation for states of average charge $N/2$.

\subsection{Higher moments}

So far we have only reported data for the second moment, $k = 2$. To confirm that our predictions hold for higher moments as well, we report additional data in Appendix~\ref{app:higher_k} on moments $k = 3$ and $4$. We focus on the case of states with intermediate charge fluctuations measured in the computational basis, Sec.~\ref{sec:arbitrary}. Our results are again consistent with deep thermalization to the appropriate GSE, suggesing that all results observed in this Section would carry over to higher moments $k>2$.  


 \section{Discussion \label{sec:discussion}}

In this work, we have studied the phenomenon of deep thermalization, a novel form of quantum equilibration, in systems harboring a $U(1)$ conserved charge. We have put forth a general theory [Section.~\ref{sec:limiting_ensemble}], based on statistical-mechanical and quantum information-theoretic principles, for the universal limiting form of the projected ensemble in deep thermalization in the presence of this conserved charge. We have presented extensive supporting evidence in the form of a rigorous theorem, analytical calculations in a replica approach, and exact finite-sized numerical simulations.
Our theory predicts a rich variety of limiting behavior for the projected ensemble, based on the charge fluctuations in the initial state and on the measurement basis, specifically how much information it reveals about the local charge density. 

Our general ansatz takes the form of a generalized Scrooge ensemble (GSE), a stochastic mixture of maximally-entropic ensembles. Each entry in this mixture reflects an updated belief about the state of subsystem $A$ upon obtaining a measurement outcome $\nu$ on $B$. Crucially, we have found that many microscopically distinct outcomes $\nu$ yield equivalent belief updates and so may be collected into equivalence classes $[\nu]$. This effectively compresses the amount of information needed to define our universal ansatz from an exponential to a polynomial amount. Such a compression or coarse-graining of information is a key feature of statistical mechanics, and hence qualifies our ensembles as  universal, thermodynamic ensembles. 

The GSE reduces to several other ensembles of interest in particular cases. These include the `direct sum' ensemble (a mixture of Haar ensembles in the individual charge sectors of $A$), the Scrooge ensemble, and the global Haar ensemble. Each of these can be understood physically in terms of the information revealed by measurements about the conserved charge.
A sharp illustration of the universality of our results is the fact that they recover the same physical outcome in two settings that are microscopically very different: on the one hand, a state with equilibrium charge fluctuations subjected to maximally charge-revealing ($z$) measurements, Sec.~\ref{sec:ensemble_z_eq}; on the other, a state with no charge fluctuations subjected to charge-non-revealing ($x$) measurements, Sec.~\ref{sec:ensemble_x}. Somewhat counterintuitively, both cases fall under the `non-revealing' universality: even though in the former case the measurements fully reveal the charge $Q_B$ of the bath, due to the equilibrium nature of the charge fluctuations, the observer gains no information about the charge $Q_A$ of the subsystem of interest, so that the measurements are effectively non-revealing. 
This result is further evidence that deep thermalization is a universal phenomenon dictated by principles of maximum entropy~\cite{mark_maximum_2024} and minimum accessible information~\cite{jozsa_lower_1994,liu_deep_2024}, and otherwise insensitive to microscopic details.

Our results on deep thermalization with a $U(1)$ conserved charge are closely related to recent works on deep thermalization in energy-conserving (Hamiltonian) systems, where one also obtains a GSE for `energy-revealing' measurements and a Scrooge ensemble (which reduces to the Haar ensemble at infinite temperature) for `energy-non-revealing' measurements~\cite{cotler_emergent_2023,mark_maximum_2024}. A distinct conceptual feature of our setup relative to the Hamiltonian one is that the GSE is built out of a limited (polynomial in $N$) amount of data, namely the distribution $p_{\rm in}(Q)$ and the charge distribution on measurement basis states on the bath, $p_{\rm out}(Q_B|\nu)$. The Hamiltonian case instead appears to depend on an exponential amount of data (e.g.~all possible bitstrings measured on the bath)~\cite{mark_maximum_2024}. 
It is an interesting open question whether this exponential amount of data is necessary or could be compressed to a polynomial amount by a suitable coarse-graining process. We leave this question, as well as a full study of the similarities and differences between the case of Abelian on-site symmetry and energy conservation, for future work. 

Another exciting direction for future research is the effect of {\it non-Abelian} symmetries, such as $SU(2)$, or non-on-site symmetries such as translation~\cite{varikuti_unraveling_2024}.  
These generally do not admit a local, fully-revealing basis---i.e., they do not have a complete basis of disentangled eigenstates, which makes them more similar to the case of energy conservation.
More broadly, non-Abelian symmetry has been shown to have remarkably strong effects on thermalization~\cite{murthy_non-abelian_2023,majidy_non-abelian_2023,majidy_noncommuting_2023} and on the entanglement of post-measurement states~\cite{majidy_critical_2023,feng_charge_2024},
and we expect its effect on deep thermalization to be no less rich. The conceptual framework introduced in this work readily lends itself to generalization in these and other directions. 

From a practical standpoint, the formation of quantum state designs, whether in individual charge sectors or across the whole Hilbert space, has potential applications in quantum information processing on systems (like certain ultra-cold atom-based quantum simulators) whose dynamics obey a particle number conservation law. Applications include classical shadow tomography and more generally randomized measurements for quantum state learning on analog simulators~\cite{tran_measuring_2023,mcginley_shadow_2023,hearth_efficient_2024}, randomized benchmarking~\cite{baldwin_subspace_2020} and process tomography~\cite{marvian_restrictions_2022}, initialization of random states for quantum simulation based on typicality~\cite{goldstein_canonical_2006,mi_time-crystalline_2022,richter_simulating_2021,andersen_thermalization_2024}, etc. The case of Haar ensembles in individual charge sectors is especially interesting as it allows to target (with a polynomial postselection overhead) highly random states of fixed charge, enabling charge-resolved variants of all the aforementioned protocols. 

\begin{acknowledgments}
Numerical simulations were performed in part on HPC resources provided by the Texas Advanced Computing Center (TACC) at the University of Texas at Austin. 
W.~W.~H.~is supported by the National Research Foundation (NRF), Singapore, through the NRF Felllowship NRF-NRFF15-2023-0008, and through the National Quantum Office, hosted in A*STAR, under its Centre for Quantum Technologies Funding Initiative (S24Q2d0009).
\end{acknowledgments}

\bibliography{u1deeptherm}

\begin{thebibliography}{71}%
\makeatletter
\providecommand \@ifxundefined [1]{%
 \@ifx{#1\undefined}
}%
\providecommand \@ifnum [1]{%
 \ifnum #1\expandafter \@firstoftwo
 \else \expandafter \@secondoftwo
 \fi
}%
\providecommand \@ifx [1]{%
 \ifx #1\expandafter \@firstoftwo
 \else \expandafter \@secondoftwo
 \fi
}%
\providecommand \natexlab [1]{#1}%
\providecommand \enquote  [1]{``#1''}%
\providecommand \bibnamefont  [1]{#1}%
\providecommand \bibfnamefont [1]{#1}%
\providecommand \citenamefont [1]{#1}%
\providecommand \href@noop [0]{\@secondoftwo}%
\providecommand \href [0]{\begingroup \@sanitize@url \@href}%
\providecommand \@href[1]{\@@startlink{#1}\@@href}%
\providecommand \@@href[1]{\endgroup#1\@@endlink}%
\providecommand \@sanitize@url [0]{\catcode `\\12\catcode `\$12\catcode
  `\&12\catcode `\#12\catcode `\^12\catcode `\_12\catcode `\%12\relax}%
\providecommand \@@startlink[1]{}%
\providecommand \@@endlink[0]{}%
\providecommand \url  [0]{\begingroup\@sanitize@url \@url }%
\providecommand \@url [1]{\endgroup\@href {#1}{\urlprefix }}%
\providecommand \urlprefix  [0]{URL }%
\providecommand \Eprint [0]{\href }%
\providecommand \doibase [0]{https://doi.org/}%
\providecommand \selectlanguage [0]{\@gobble}%
\providecommand \bibinfo  [0]{\@secondoftwo}%
\providecommand \bibfield  [0]{\@secondoftwo}%
\providecommand \translation [1]{[#1]}%
\providecommand \BibitemOpen [0]{}%
\providecommand \bibitemStop [0]{}%
\providecommand \bibitemNoStop [0]{.\EOS\space}%
\providecommand \EOS [0]{\spacefactor3000\relax}%
\providecommand \BibitemShut  [1]{\csname bibitem#1\endcsname}%
\let\auto@bib@innerbib\@empty
\bibitem [{\citenamefont {Srednicki}(1994)}]{srednicki_chaos_1994}%
  \BibitemOpen
  \bibfield  {author} {\bibinfo {author} {\bibfnamefont {M.}~\bibnamefont
  {Srednicki}},\ }\bibfield  {title} {\bibinfo {title} {Chaos and quantum
  thermalization},\ }\href {https://doi.org/10.1103/PhysRevE.50.888} {\bibfield
   {journal} {\bibinfo  {journal} {Physical Review E}\ }\textbf {\bibinfo
  {volume} {50}},\ \bibinfo {pages} {888} (\bibinfo {year} {1994})}\BibitemShut
  {NoStop}%
\bibitem [{\citenamefont {Rigol}\ \emph {et~al.}(2008)\citenamefont {Rigol},
  \citenamefont {Dunjko},\ and\ \citenamefont
  {Olshanii}}]{rigol_thermalization_2008}%
  \BibitemOpen
  \bibfield  {author} {\bibinfo {author} {\bibfnamefont {M.}~\bibnamefont
  {Rigol}}, \bibinfo {author} {\bibfnamefont {V.}~\bibnamefont {Dunjko}},\ and\
  \bibinfo {author} {\bibfnamefont {M.}~\bibnamefont {Olshanii}},\ }\bibfield
  {title} {\bibinfo {title} {Thermalization and its mechanism for generic
  isolated quantum systems},\ }\href {https://doi.org/10.1038/nature06838}
  {\bibfield  {journal} {\bibinfo  {journal} {Nature}\ }\textbf {\bibinfo
  {volume} {452}},\ \bibinfo {pages} {854} (\bibinfo {year}
  {2008})}\BibitemShut {NoStop}%
\bibitem [{\citenamefont {Nandkishore}\ and\ \citenamefont
  {Huse}(2015)}]{nandkishore_many-body_2015}%
  \BibitemOpen
  \bibfield  {author} {\bibinfo {author} {\bibfnamefont {R.}~\bibnamefont
  {Nandkishore}}\ and\ \bibinfo {author} {\bibfnamefont {D.~A.}\ \bibnamefont
  {Huse}},\ }\bibfield  {title} {\bibinfo {title} {Many-{Body} {Localization}
  and {Thermalization} in {Quantum} {Statistical} {Mechanics}},\ }\href
  {https://doi.org/10.1146/annurev-conmatphys-031214-014726} {\bibfield
  {journal} {\bibinfo  {journal} {Annual Review of Condensed Matter Physics}\
  }\textbf {\bibinfo {volume} {6}},\ \bibinfo {pages} {15} (\bibinfo {year}
  {2015})}\BibitemShut {NoStop}%
\bibitem [{\citenamefont {D'Alessio}\ \emph {et~al.}(2016)\citenamefont
  {D'Alessio}, \citenamefont {Kafri}, \citenamefont {Polkovnikov},\ and\
  \citenamefont {Rigol}}]{dalessio_quantum_2016}%
  \BibitemOpen
  \bibfield  {author} {\bibinfo {author} {\bibfnamefont {L.}~\bibnamefont
  {D'Alessio}}, \bibinfo {author} {\bibfnamefont {Y.}~\bibnamefont {Kafri}},
  \bibinfo {author} {\bibfnamefont {A.}~\bibnamefont {Polkovnikov}},\ and\
  \bibinfo {author} {\bibfnamefont {M.}~\bibnamefont {Rigol}},\ }\bibfield
  {title} {\bibinfo {title} {From quantum chaos and eigenstate thermalization
  to statistical mechanics and thermodynamics},\ }\href
  {https://doi.org/10.1080/00018732.2016.1198134} {\bibfield  {journal}
  {\bibinfo  {journal} {Advances in Physics}\ }\textbf {\bibinfo {volume}
  {65}},\ \bibinfo {pages} {239} (\bibinfo {year} {2016})}\BibitemShut
  {NoStop}%
\bibitem [{\citenamefont {Kaufman}\ \emph {et~al.}(2016)\citenamefont
  {Kaufman}, \citenamefont {Tai}, \citenamefont {Lukin}, \citenamefont
  {Rispoli}, \citenamefont {Schittko}, \citenamefont {Preiss},\ and\
  \citenamefont {Greiner}}]{kaufman_quantum_2016}%
  \BibitemOpen
  \bibfield  {author} {\bibinfo {author} {\bibfnamefont {A.~M.}\ \bibnamefont
  {Kaufman}}, \bibinfo {author} {\bibfnamefont {M.~E.}\ \bibnamefont {Tai}},
  \bibinfo {author} {\bibfnamefont {A.}~\bibnamefont {Lukin}}, \bibinfo
  {author} {\bibfnamefont {M.}~\bibnamefont {Rispoli}}, \bibinfo {author}
  {\bibfnamefont {R.}~\bibnamefont {Schittko}}, \bibinfo {author}
  {\bibfnamefont {P.~M.}\ \bibnamefont {Preiss}},\ and\ \bibinfo {author}
  {\bibfnamefont {M.}~\bibnamefont {Greiner}},\ }\bibfield  {title} {\bibinfo
  {title} {Quantum thermalization through entanglement in an isolated many-body
  system},\ }\href {https://doi.org/10.1126/science.aaf6725} {\bibfield
  {journal} {\bibinfo  {journal} {Science}\ }\textbf {\bibinfo {volume}
  {353}},\ \bibinfo {pages} {794} (\bibinfo {year} {2016})}\BibitemShut
  {NoStop}%
\bibitem [{\citenamefont {Abanin}\ \emph {et~al.}(2019)\citenamefont {Abanin},
  \citenamefont {Altman}, \citenamefont {Bloch},\ and\ \citenamefont
  {Serbyn}}]{abanin_colloquium_2019}%
  \BibitemOpen
  \bibfield  {author} {\bibinfo {author} {\bibfnamefont {D.~A.}\ \bibnamefont
  {Abanin}}, \bibinfo {author} {\bibfnamefont {E.}~\bibnamefont {Altman}},
  \bibinfo {author} {\bibfnamefont {I.}~\bibnamefont {Bloch}},\ and\ \bibinfo
  {author} {\bibfnamefont {M.}~\bibnamefont {Serbyn}},\ }\bibfield  {title}
  {\bibinfo {title} {Colloquium: {Many}-body localization, thermalization, and
  entanglement},\ }\href {https://doi.org/10.1103/RevModPhys.91.021001}
  {\bibfield  {journal} {\bibinfo  {journal} {Reviews of Modern Physics}\
  }\textbf {\bibinfo {volume} {91}},\ \bibinfo {pages} {021001} (\bibinfo
  {year} {2019})}\BibitemShut {NoStop}%
\bibitem [{\citenamefont {Choi}\ \emph {et~al.}(2023)\citenamefont {Choi},
  \citenamefont {Shaw}, \citenamefont {Madjarov}, \citenamefont {Xie},
  \citenamefont {Finkelstein}, \citenamefont {Covey}, \citenamefont {Cotler},
  \citenamefont {Mark}, \citenamefont {Huang}, \citenamefont {Kale},
  \citenamefont {Pichler}, \citenamefont {Brandao}, \citenamefont {Choi},\ and\
  \citenamefont {Endres}}]{choi_preparing_2023}%
  \BibitemOpen
  \bibfield  {author} {\bibinfo {author} {\bibfnamefont {J.}~\bibnamefont
  {Choi}}, \bibinfo {author} {\bibfnamefont {A.~L.}\ \bibnamefont {Shaw}},
  \bibinfo {author} {\bibfnamefont {I.~S.}\ \bibnamefont {Madjarov}}, \bibinfo
  {author} {\bibfnamefont {X.}~\bibnamefont {Xie}}, \bibinfo {author}
  {\bibfnamefont {R.}~\bibnamefont {Finkelstein}}, \bibinfo {author}
  {\bibfnamefont {J.~P.}\ \bibnamefont {Covey}}, \bibinfo {author}
  {\bibfnamefont {J.~S.}\ \bibnamefont {Cotler}}, \bibinfo {author}
  {\bibfnamefont {D.~K.}\ \bibnamefont {Mark}}, \bibinfo {author}
  {\bibfnamefont {H.-Y.}\ \bibnamefont {Huang}}, \bibinfo {author}
  {\bibfnamefont {A.}~\bibnamefont {Kale}}, \bibinfo {author} {\bibfnamefont
  {H.}~\bibnamefont {Pichler}}, \bibinfo {author} {\bibfnamefont {F.~G. S.~L.}\
  \bibnamefont {Brandao}}, \bibinfo {author} {\bibfnamefont {S.}~\bibnamefont
  {Choi}},\ and\ \bibinfo {author} {\bibfnamefont {M.}~\bibnamefont {Endres}},\
  }\bibfield  {title} {\bibinfo {title} {Preparing random states and
  benchmarking with many-body quantum chaos},\ }\href
  {https://doi.org/10.1038/s41586-022-05442-1} {\bibfield  {journal} {\bibinfo
  {journal} {Nature}\ }\textbf {\bibinfo {volume} {613}},\ \bibinfo {pages}
  {468} (\bibinfo {year} {2023})}\BibitemShut {NoStop}%
\bibitem [{\citenamefont {Cotler}\ \emph {et~al.}(2023)\citenamefont {Cotler},
  \citenamefont {Mark}, \citenamefont {Huang}, \citenamefont {Hernandez},
  \citenamefont {Choi}, \citenamefont {Shaw}, \citenamefont {Endres},\ and\
  \citenamefont {Choi}}]{cotler_emergent_2023}%
  \BibitemOpen
  \bibfield  {author} {\bibinfo {author} {\bibfnamefont {J.~S.}\ \bibnamefont
  {Cotler}}, \bibinfo {author} {\bibfnamefont {D.~K.}\ \bibnamefont {Mark}},
  \bibinfo {author} {\bibfnamefont {H.-Y.}\ \bibnamefont {Huang}}, \bibinfo
  {author} {\bibfnamefont {F.}~\bibnamefont {Hernandez}}, \bibinfo {author}
  {\bibfnamefont {J.}~\bibnamefont {Choi}}, \bibinfo {author} {\bibfnamefont
  {A.~L.}\ \bibnamefont {Shaw}}, \bibinfo {author} {\bibfnamefont
  {M.}~\bibnamefont {Endres}},\ and\ \bibinfo {author} {\bibfnamefont
  {S.}~\bibnamefont {Choi}},\ }\bibfield  {title} {\bibinfo {title} {Emergent
  {Quantum} {State} {Designs} from {Individual} {Many}-{Body} {Wave}
  {Functions}},\ }\href {https://doi.org/10.1103/PRXQuantum.4.010311}
  {\bibfield  {journal} {\bibinfo  {journal} {PRX Quantum}\ }\textbf {\bibinfo
  {volume} {4}},\ \bibinfo {pages} {010311} (\bibinfo {year}
  {2023})}\BibitemShut {NoStop}%
\bibitem [{\citenamefont {Ho}\ and\ \citenamefont
  {Choi}(2022)}]{ho_exact_2022}%
  \BibitemOpen
  \bibfield  {author} {\bibinfo {author} {\bibfnamefont {W.~W.}\ \bibnamefont
  {Ho}}\ and\ \bibinfo {author} {\bibfnamefont {S.}~\bibnamefont {Choi}},\
  }\bibfield  {title} {\bibinfo {title} {Exact {Emergent} {Quantum} {State}
  {Designs} from {Quantum} {Chaotic} {Dynamics}},\ }\href
  {https://doi.org/10.1103/PhysRevLett.128.060601} {\bibfield  {journal}
  {\bibinfo  {journal} {Physical Review Letters}\ }\textbf {\bibinfo {volume}
  {128}},\ \bibinfo {pages} {060601} (\bibinfo {year} {2022})}\BibitemShut
  {NoStop}%
\bibitem [{\citenamefont {Ippoliti}\ and\ \citenamefont
  {Ho}(2023)}]{ippoliti_dynamical_2023}%
  \BibitemOpen
  \bibfield  {author} {\bibinfo {author} {\bibfnamefont {M.}~\bibnamefont
  {Ippoliti}}\ and\ \bibinfo {author} {\bibfnamefont {W.~W.}\ \bibnamefont
  {Ho}},\ }\bibfield  {title} {\bibinfo {title} {Dynamical {Purification} and
  the {Emergence} of {Quantum} {State} {Designs} from the {Projected}
  {Ensemble}},\ }\href {https://doi.org/10.1103/PRXQuantum.4.030322} {\bibfield
   {journal} {\bibinfo  {journal} {PRX Quantum}\ }\textbf {\bibinfo {volume}
  {4}},\ \bibinfo {pages} {030322} (\bibinfo {year} {2023})}\BibitemShut
  {NoStop}%
\bibitem [{\citenamefont {Ippoliti}\ and\ \citenamefont
  {Ho}(2022)}]{ippoliti_solvable_2022}%
  \BibitemOpen
  \bibfield  {author} {\bibinfo {author} {\bibfnamefont {M.}~\bibnamefont
  {Ippoliti}}\ and\ \bibinfo {author} {\bibfnamefont {W.~W.}\ \bibnamefont
  {Ho}},\ }\bibfield  {title} {\bibinfo {title} {Solvable model of deep
  thermalization with distinct design times},\ }\href
  {https://doi.org/10.22331/q-2022-12-29-886} {\bibfield  {journal} {\bibinfo
  {journal} {Quantum}\ }\textbf {\bibinfo {volume} {6}},\ \bibinfo {pages}
  {886} (\bibinfo {year} {2022})}\BibitemShut {NoStop}%
\bibitem [{\citenamefont {Bhore}\ \emph {et~al.}(2023)\citenamefont {Bhore},
  \citenamefont {Desaules},\ and\ \citenamefont {Papic}}]{bhore_deep_2023}%
  \BibitemOpen
  \bibfield  {author} {\bibinfo {author} {\bibfnamefont {T.}~\bibnamefont
  {Bhore}}, \bibinfo {author} {\bibfnamefont {J.-Y.}\ \bibnamefont
  {Desaules}},\ and\ \bibinfo {author} {\bibfnamefont {Z.}~\bibnamefont
  {Papic}},\ }\bibfield  {title} {\bibinfo {title} {Deep thermalization in
  constrained quantum systems},\ }\href
  {https://doi.org/10.1103/PhysRevB.108.104317} {\bibfield  {journal} {\bibinfo
   {journal} {Physical Review B}\ }\textbf {\bibinfo {volume} {108}},\ \bibinfo
  {pages} {104317} (\bibinfo {year} {2023})}\BibitemShut {NoStop}%
\bibitem [{\citenamefont {Lucas}\ \emph {et~al.}(2023)\citenamefont {Lucas},
  \citenamefont {Piroli}, \citenamefont {De~Nardis},\ and\ \citenamefont
  {De~Luca}}]{lucas_generalized_2023}%
  \BibitemOpen
  \bibfield  {author} {\bibinfo {author} {\bibfnamefont {M.}~\bibnamefont
  {Lucas}}, \bibinfo {author} {\bibfnamefont {L.}~\bibnamefont {Piroli}},
  \bibinfo {author} {\bibfnamefont {J.}~\bibnamefont {De~Nardis}},\ and\
  \bibinfo {author} {\bibfnamefont {A.}~\bibnamefont {De~Luca}},\ }\bibfield
  {title} {\bibinfo {title} {Generalized deep thermalization for free
  fermions},\ }\href {https://doi.org/10.1103/PhysRevA.107.032215} {\bibfield
  {journal} {\bibinfo  {journal} {Physical Review A}\ }\textbf {\bibinfo
  {volume} {107}},\ \bibinfo {pages} {032215} (\bibinfo {year}
  {2023})}\BibitemShut {NoStop}%
\bibitem [{\citenamefont {Mark}\ \emph {et~al.}(2024)\citenamefont {Mark},
  \citenamefont {Surace}, \citenamefont {Elben}, \citenamefont {Shaw},
  \citenamefont {Choi}, \citenamefont {Refael}, \citenamefont {Endres},\ and\
  \citenamefont {Choi}}]{mark_maximum_2024}%
  \BibitemOpen
  \bibfield  {author} {\bibinfo {author} {\bibfnamefont {D.~K.}\ \bibnamefont
  {Mark}}, \bibinfo {author} {\bibfnamefont {F.}~\bibnamefont {Surace}},
  \bibinfo {author} {\bibfnamefont {A.}~\bibnamefont {Elben}}, \bibinfo
  {author} {\bibfnamefont {A.~L.}\ \bibnamefont {Shaw}}, \bibinfo {author}
  {\bibfnamefont {J.}~\bibnamefont {Choi}}, \bibinfo {author} {\bibfnamefont
  {G.}~\bibnamefont {Refael}}, \bibinfo {author} {\bibfnamefont
  {M.}~\bibnamefont {Endres}},\ and\ \bibinfo {author} {\bibfnamefont
  {S.}~\bibnamefont {Choi}},\ }\bibfield  {title} {\bibinfo {title} {A
  {Maximum} {Entropy} {Principle} in {Deep} {Thermalization} and in
  {Hilbert}-{Space} {Ergodicity}},\ }\href {https://arxiv.org/abs/2403.11970v1}
  {\bibfield  {journal} {\bibinfo  {journal} {arXiv:2403.11970v1}\ } (\bibinfo
  {year} {2024})}\BibitemShut {NoStop}%
\bibitem [{\citenamefont {Liu}\ \emph {et~al.}(2024{\natexlab{a}})\citenamefont
  {Liu}, \citenamefont {Huang},\ and\ \citenamefont {Ho}}]{liu_deep_2024}%
  \BibitemOpen
  \bibfield  {author} {\bibinfo {author} {\bibfnamefont {C.}~\bibnamefont
  {Liu}}, \bibinfo {author} {\bibfnamefont {Q.~C.}\ \bibnamefont {Huang}},\
  and\ \bibinfo {author} {\bibfnamefont {W.~W.}\ \bibnamefont {Ho}},\
  }\bibfield  {title} {\bibinfo {title} {Deep thermalization in gaussian
  continuous-variable quantum systems},\ }\href
  {https://doi.org/10.1103/PhysRevLett.133.260401} {\bibfield  {journal}
  {\bibinfo  {journal} {Phys. Rev. Lett.}\ }\textbf {\bibinfo {volume} {133}},\
  \bibinfo {pages} {260401} (\bibinfo {year} {2024}{\natexlab{a}})}\BibitemShut
  {NoStop}%
\bibitem [{\citenamefont {Zhang}\ \emph {et~al.}(2025)\citenamefont {Zhang},
  \citenamefont {Xu}, \citenamefont {Chen},\ and\ \citenamefont
  {Zhuang}}]{zhang2025holographicdeepthermalizationtheory}%
  \BibitemOpen
  \bibfield  {author} {\bibinfo {author} {\bibfnamefont {B.}~\bibnamefont
  {Zhang}}, \bibinfo {author} {\bibfnamefont {P.}~\bibnamefont {Xu}}, \bibinfo
  {author} {\bibfnamefont {X.}~\bibnamefont {Chen}},\ and\ \bibinfo {author}
  {\bibfnamefont {Q.}~\bibnamefont {Zhuang}},\ }\href
  {https://arxiv.org/abs/2411.03587} {\bibinfo {title} {Holographic deep
  thermalization: theory and experimentation}} (\bibinfo {year} {2025}),\
  \Eprint {https://arxiv.org/abs/2411.03587} {arXiv:2411.03587 [quant-ph]}
  \BibitemShut {NoStop}%
\bibitem [{\citenamefont {Milekhin}\ and\ \citenamefont
  {Murciano}(2024)}]{milekhin2024observableprojectedensembles}%
  \BibitemOpen
  \bibfield  {author} {\bibinfo {author} {\bibfnamefont {A.}~\bibnamefont
  {Milekhin}}\ and\ \bibinfo {author} {\bibfnamefont {S.}~\bibnamefont
  {Murciano}},\ }\href {https://arxiv.org/abs/2410.21397} {\bibinfo {title}
  {Observable-projected ensembles}} (\bibinfo {year} {2024}),\ \Eprint
  {https://arxiv.org/abs/2410.21397} {arXiv:2410.21397 [quant-ph]} \BibitemShut
  {NoStop}%
\bibitem [{\citenamefont {Mok}\ \emph {et~al.}(2024)\citenamefont {Mok},
  \citenamefont {Haug}, \citenamefont {Shaw}, \citenamefont {Endres},\ and\
  \citenamefont {Preskill}}]{mok2024optimalconversionclassicalquantum}%
  \BibitemOpen
  \bibfield  {author} {\bibinfo {author} {\bibfnamefont {W.-K.}\ \bibnamefont
  {Mok}}, \bibinfo {author} {\bibfnamefont {T.}~\bibnamefont {Haug}}, \bibinfo
  {author} {\bibfnamefont {A.~L.}\ \bibnamefont {Shaw}}, \bibinfo {author}
  {\bibfnamefont {M.}~\bibnamefont {Endres}},\ and\ \bibinfo {author}
  {\bibfnamefont {J.}~\bibnamefont {Preskill}},\ }\href
  {https://arxiv.org/abs/2410.05181} {\bibinfo {title} {Optimal conversion from
  classical to quantum randomness via quantum chaos}} (\bibinfo {year}
  {2024}),\ \Eprint {https://arxiv.org/abs/2410.05181} {arXiv:2410.05181
  [quant-ph]} \BibitemShut {NoStop}%
\bibitem [{\citenamefont {Du}\ \emph {et~al.}(2024)\citenamefont {Du},
  \citenamefont {Liu},\ and\ \citenamefont
  {Ma}}]{du2024embeddedcomplexityquantumcircuit}%
  \BibitemOpen
  \bibfield  {author} {\bibinfo {author} {\bibfnamefont {Z.}~\bibnamefont
  {Du}}, \bibinfo {author} {\bibfnamefont {Z.-W.}\ \bibnamefont {Liu}},\ and\
  \bibinfo {author} {\bibfnamefont {X.}~\bibnamefont {Ma}},\ }\href
  {https://arxiv.org/abs/2408.16602} {\bibinfo {title} {Embedded complexity and
  quantum circuit volume}} (\bibinfo {year} {2024}),\ \Eprint
  {https://arxiv.org/abs/2408.16602} {arXiv:2408.16602 [quant-ph]} \BibitemShut
  {NoStop}%
\bibitem [{\citenamefont {Bejan}\ \emph {et~al.}(2024)\citenamefont {Bejan},
  \citenamefont {B\'eri},\ and\ \citenamefont {McGinley}}]{matchgate2024}%
  \BibitemOpen
  \bibfield  {author} {\bibinfo {author} {\bibfnamefont {M.}~\bibnamefont
  {Bejan}}, \bibinfo {author} {\bibfnamefont {B.}~\bibnamefont {B\'eri}},\ and\
  \bibinfo {author} {\bibfnamefont {M.}~\bibnamefont {McGinley}},\ }\bibfield
  {title} {\bibinfo {title} {Matchgate circuits deeply thermalize},\ }\href
  {https://arxiv.org/abs/2412.01884} {\bibfield  {journal} {\bibinfo  {journal}
  {arXiv:2412.01884}\ } (\bibinfo {year} {2024})}\BibitemShut {NoStop}%
\bibitem [{\citenamefont {Mele}(2024)}]{mele_introduction_2024}%
  \BibitemOpen
  \bibfield  {author} {\bibinfo {author} {\bibfnamefont {A.~A.}\ \bibnamefont
  {Mele}},\ }\bibfield  {title} {\bibinfo {title} {Introduction to {Haar}
  {Measure} {Tools} in {Quantum} {Information}: {A} {Beginner}'s {Tutorial}},\
  }\href {https://doi.org/10.22331/q-2024-05-08-1340} {\bibfield  {journal}
  {\bibinfo  {journal} {Quantum}\ }\textbf {\bibinfo {volume} {8}},\ \bibinfo
  {pages} {1340} (\bibinfo {year} {2024})}\BibitemShut {NoStop}%
\bibitem [{\citenamefont {Renes}\ \emph {et~al.}(2004)\citenamefont {Renes},
  \citenamefont {Blume-Kohout}, \citenamefont {Scott},\ and\ \citenamefont
  {Caves}}]{renes_symmetric_2004}%
  \BibitemOpen
  \bibfield  {author} {\bibinfo {author} {\bibfnamefont {J.~M.}\ \bibnamefont
  {Renes}}, \bibinfo {author} {\bibfnamefont {R.}~\bibnamefont {Blume-Kohout}},
  \bibinfo {author} {\bibfnamefont {A.~J.}\ \bibnamefont {Scott}},\ and\
  \bibinfo {author} {\bibfnamefont {C.~M.}\ \bibnamefont {Caves}},\ }\bibfield
  {title} {\bibinfo {title} {Symmetric informationally complete quantum
  measurements},\ }\href {https://doi.org/10.1063/1.1737053} {\bibfield
  {journal} {\bibinfo  {journal} {Journal of Mathematical Physics}\ }\textbf
  {\bibinfo {volume} {45}},\ \bibinfo {pages} {2171} (\bibinfo {year}
  {2004})}\BibitemShut {NoStop}%
\bibitem [{\citenamefont {Ambainis}\ and\ \citenamefont
  {Emerson}(2007)}]{ambainis_quantum_2007}%
  \BibitemOpen
  \bibfield  {author} {\bibinfo {author} {\bibfnamefont {A.}~\bibnamefont
  {Ambainis}}\ and\ \bibinfo {author} {\bibfnamefont {J.}~\bibnamefont
  {Emerson}},\ }\bibfield  {title} {\bibinfo {title} {Quantum t-designs: t-wise
  {Independence} in the {Quantum} {World}},\ }in\ \href
  {https://doi.org/10.1109/CCC.2007.26} {\emph {\bibinfo {booktitle}
  {Twenty-{Second} {Annual} {IEEE} {Conference} on {Computational} {Complexity}
  ({CCC}'07)}}}\ (\bibinfo {year} {2007})\ pp.\ \bibinfo {pages}
  {129--140}\BibitemShut {NoStop}%
\bibitem [{\citenamefont {Claeys}\ and\ \citenamefont
  {Lamacraft}(2022)}]{claeys_emergent_2022}%
  \BibitemOpen
  \bibfield  {author} {\bibinfo {author} {\bibfnamefont {P.~W.}\ \bibnamefont
  {Claeys}}\ and\ \bibinfo {author} {\bibfnamefont {A.}~\bibnamefont
  {Lamacraft}},\ }\bibfield  {title} {\bibinfo {title} {Emergent quantum state
  designs and biunitarity in dual-unitary circuit dynamics},\ }\href
  {https://doi.org/10.22331/q-2022-06-15-738} {\bibfield  {journal} {\bibinfo
  {journal} {Quantum}\ }\textbf {\bibinfo {volume} {6}},\ \bibinfo {pages}
  {738} (\bibinfo {year} {2022})}\BibitemShut {NoStop}%
\bibitem [{\citenamefont {Wilming}\ and\ \citenamefont
  {Roth}(2022)}]{wilming_high-temperature_2022}%
  \BibitemOpen
  \bibfield  {author} {\bibinfo {author} {\bibfnamefont {H.}~\bibnamefont
  {Wilming}}\ and\ \bibinfo {author} {\bibfnamefont {I.}~\bibnamefont {Roth}},\
  }\bibfield  {title} {\bibinfo {title} {High-temperature thermalization
  implies the emergence of quantum state designs},\ }\href
  {http://arxiv.org/abs/2202.01669} {\bibfield  {journal} {\bibinfo  {journal}
  {arXiv:2202.01669 [cond-mat, physics:math-ph, physics:quant-ph]}\ } (\bibinfo
  {year} {2022})}\BibitemShut {NoStop}%
\bibitem [{\citenamefont {Khemani}\ \emph {et~al.}(2018)\citenamefont
  {Khemani}, \citenamefont {Vishwanath},\ and\ \citenamefont
  {Huse}}]{khemani_operator_2018}%
  \BibitemOpen
  \bibfield  {author} {\bibinfo {author} {\bibfnamefont {V.}~\bibnamefont
  {Khemani}}, \bibinfo {author} {\bibfnamefont {A.}~\bibnamefont
  {Vishwanath}},\ and\ \bibinfo {author} {\bibfnamefont {D.~A.}\ \bibnamefont
  {Huse}},\ }\bibfield  {title} {\bibinfo {title} {Operator {Spreading} and the
  {Emergence} of {Dissipative} {Hydrodynamics} under {Unitary} {Evolution} with
  {Conservation} {Laws}},\ }\href {https://doi.org/10.1103/PhysRevX.8.031057}
  {\bibfield  {journal} {\bibinfo  {journal} {Phys. Rev. X}\ }\textbf {\bibinfo
  {volume} {8}},\ \bibinfo {pages} {031057} (\bibinfo {year}
  {2018})}\BibitemShut {NoStop}%
\bibitem [{\citenamefont {Rakovszky}\ \emph {et~al.}(2018)\citenamefont
  {Rakovszky}, \citenamefont {Pollmann},\ and\ \citenamefont {von
  Keyserlingk}}]{rakovszky_diffusive_2018}%
  \BibitemOpen
  \bibfield  {author} {\bibinfo {author} {\bibfnamefont {T.}~\bibnamefont
  {Rakovszky}}, \bibinfo {author} {\bibfnamefont {F.}~\bibnamefont
  {Pollmann}},\ and\ \bibinfo {author} {\bibfnamefont {C.~W.}\ \bibnamefont
  {von Keyserlingk}},\ }\bibfield  {title} {\bibinfo {title} {Diffusive
  {Hydrodynamics} of {Out}-of-{Time}-{Ordered} {Correlators} with {Charge}
  {Conservation}},\ }\href {https://doi.org/10.1103/PhysRevX.8.031058}
  {\bibfield  {journal} {\bibinfo  {journal} {Phys. Rev. X}\ }\textbf {\bibinfo
  {volume} {8}},\ \bibinfo {pages} {031058} (\bibinfo {year}
  {2018})}\BibitemShut {NoStop}%
\bibitem [{\citenamefont {Hunter-Jones}(2018)}]{hunter-jones_operator_2018}%
  \BibitemOpen
  \bibfield  {author} {\bibinfo {author} {\bibfnamefont {N.}~\bibnamefont
  {Hunter-Jones}},\ }\bibfield  {title} {\bibinfo {title} {Operator growth in
  random quantum circuits with symmetry},\ }\bibfield  {journal} {\bibinfo
  {journal} {arXiv:1812.08219}\ }\href
  {https://doi.org/10.48550/arXiv.1812.08219} {10.48550/arXiv.1812.08219}
  (\bibinfo {year} {2018})\BibitemShut {NoStop}%
\bibitem [{\citenamefont {Agrawal}\ \emph {et~al.}(2022)\citenamefont
  {Agrawal}, \citenamefont {Zabalo}, \citenamefont {Chen}, \citenamefont
  {Wilson}, \citenamefont {Potter}, \citenamefont {Pixley}, \citenamefont
  {Gopalakrishnan},\ and\ \citenamefont {Vasseur}}]{agrawal_entanglement_2022}%
  \BibitemOpen
  \bibfield  {author} {\bibinfo {author} {\bibfnamefont {U.}~\bibnamefont
  {Agrawal}}, \bibinfo {author} {\bibfnamefont {A.}~\bibnamefont {Zabalo}},
  \bibinfo {author} {\bibfnamefont {K.}~\bibnamefont {Chen}}, \bibinfo {author}
  {\bibfnamefont {J.~H.}\ \bibnamefont {Wilson}}, \bibinfo {author}
  {\bibfnamefont {A.~C.}\ \bibnamefont {Potter}}, \bibinfo {author}
  {\bibfnamefont {J.~H.}\ \bibnamefont {Pixley}}, \bibinfo {author}
  {\bibfnamefont {S.}~\bibnamefont {Gopalakrishnan}},\ and\ \bibinfo {author}
  {\bibfnamefont {R.}~\bibnamefont {Vasseur}},\ }\bibfield  {title} {\bibinfo
  {title} {Entanglement and {Charge}-{Sharpening} {Transitions} in {U}(1)
  {Symmetric} {Monitored} {Quantum} {Circuits}},\ }\href
  {https://doi.org/10.1103/PhysRevX.12.041002} {\bibfield  {journal} {\bibinfo
  {journal} {Physical Review X}\ }\textbf {\bibinfo {volume} {12}},\ \bibinfo
  {pages} {041002} (\bibinfo {year} {2022})}\BibitemShut {NoStop}%
\bibitem [{\citenamefont {Rath}\ \emph {et~al.}(2023)\citenamefont {Rath},
  \citenamefont {Vitale}, \citenamefont {Murciano}, \citenamefont {Votto},
  \citenamefont {Dubail}, \citenamefont {Kueng}, \citenamefont {Branciard},
  \citenamefont {Calabrese},\ and\ \citenamefont
  {Vermersch}}]{rath_entanglement_2023}%
  \BibitemOpen
  \bibfield  {author} {\bibinfo {author} {\bibfnamefont {A.}~\bibnamefont
  {Rath}}, \bibinfo {author} {\bibfnamefont {V.}~\bibnamefont {Vitale}},
  \bibinfo {author} {\bibfnamefont {S.}~\bibnamefont {Murciano}}, \bibinfo
  {author} {\bibfnamefont {M.}~\bibnamefont {Votto}}, \bibinfo {author}
  {\bibfnamefont {J.}~\bibnamefont {Dubail}}, \bibinfo {author} {\bibfnamefont
  {R.}~\bibnamefont {Kueng}}, \bibinfo {author} {\bibfnamefont
  {C.}~\bibnamefont {Branciard}}, \bibinfo {author} {\bibfnamefont
  {P.}~\bibnamefont {Calabrese}},\ and\ \bibinfo {author} {\bibfnamefont
  {B.}~\bibnamefont {Vermersch}},\ }\bibfield  {title} {\bibinfo {title}
  {Entanglement {Barrier} and its {Symmetry} {Resolution}: {Theory} and
  {Experimental} {Observation}},\ }\href
  {https://doi.org/10.1103/PRXQuantum.4.010318} {\bibfield  {journal} {\bibinfo
   {journal} {PRX Quantum}\ }\textbf {\bibinfo {volume} {4}},\ \bibinfo {pages}
  {010318} (\bibinfo {year} {2023})}\BibitemShut {NoStop}%
\bibitem [{\citenamefont {Ares}\ \emph {et~al.}(2023)\citenamefont {Ares},
  \citenamefont {Murciano},\ and\ \citenamefont
  {Calabrese}}]{ares_entanglement_2023}%
  \BibitemOpen
  \bibfield  {author} {\bibinfo {author} {\bibfnamefont {F.}~\bibnamefont
  {Ares}}, \bibinfo {author} {\bibfnamefont {S.}~\bibnamefont {Murciano}},\
  and\ \bibinfo {author} {\bibfnamefont {P.}~\bibnamefont {Calabrese}},\
  }\bibfield  {title} {\bibinfo {title} {Entanglement asymmetry as a probe of
  symmetry breaking},\ }\href {https://doi.org/10.1038/s41467-023-37747-8}
  {\bibfield  {journal} {\bibinfo  {journal} {Nature Communications}\ }\textbf
  {\bibinfo {volume} {14}},\ \bibinfo {pages} {2036} (\bibinfo {year}
  {2023})}\BibitemShut {NoStop}%
\bibitem [{\citenamefont {Jonay}\ \emph {et~al.}(2024)\citenamefont {Jonay},
  \citenamefont {Rodriguez-Nieva},\ and\ \citenamefont
  {Khemani}}]{jonay_slow_2024}%
  \BibitemOpen
  \bibfield  {author} {\bibinfo {author} {\bibfnamefont {C.}~\bibnamefont
  {Jonay}}, \bibinfo {author} {\bibfnamefont {J.~F.}\ \bibnamefont
  {Rodriguez-Nieva}},\ and\ \bibinfo {author} {\bibfnamefont {V.}~\bibnamefont
  {Khemani}},\ }\bibfield  {title} {\bibinfo {title} {Slow thermalization and
  subdiffusion in \${U}(1)\$ conserving {Floquet} random circuits},\ }\href
  {https://doi.org/10.1103/PhysRevB.109.024311} {\bibfield  {journal} {\bibinfo
   {journal} {Physical Review B}\ }\textbf {\bibinfo {volume} {109}},\ \bibinfo
  {pages} {024311} (\bibinfo {year} {2024})}\BibitemShut {NoStop}%
\bibitem [{\citenamefont {Langlett}\ and\ \citenamefont
  {Rodriguez-Nieva}(2024)}]{langlett_entanglement_2024}%
  \BibitemOpen
  \bibfield  {author} {\bibinfo {author} {\bibfnamefont {C.~M.}\ \bibnamefont
  {Langlett}}\ and\ \bibinfo {author} {\bibfnamefont {J.~F.}\ \bibnamefont
  {Rodriguez-Nieva}},\ }\bibfield  {title} {\bibinfo {title} {Entanglement
  patterns of quantum chaotic {Hamiltonians} with a scalar {U}(1) charge},\
  }\href {https://arxiv.org/abs/2403.10600v1} {\bibfield  {journal} {\bibinfo
  {journal} {arXiv:2403.10600v1}\ } (\bibinfo {year} {2024})}\BibitemShut
  {NoStop}%
\bibitem [{\citenamefont {Turkeshi}\ \emph {et~al.}(2024)\citenamefont
  {Turkeshi}, \citenamefont {Calabrese},\ and\ \citenamefont
  {De~Luca}}]{turkeshi_quantum_2024}%
  \BibitemOpen
  \bibfield  {author} {\bibinfo {author} {\bibfnamefont {X.}~\bibnamefont
  {Turkeshi}}, \bibinfo {author} {\bibfnamefont {P.}~\bibnamefont
  {Calabrese}},\ and\ \bibinfo {author} {\bibfnamefont {A.}~\bibnamefont
  {De~Luca}},\ }\bibfield  {title} {\bibinfo {title} {Quantum {Mpemba} {Effect}
  in {Random} {Circuits}},\ }\bibfield  {journal} {\bibinfo  {journal}
  {arXiv:2405.14514}\ }\href {https://doi.org/10.48550/arXiv.2405.14514}
  {10.48550/arXiv.2405.14514} (\bibinfo {year} {2024})\BibitemShut {NoStop}%
\bibitem [{\citenamefont {Poyhonen}\ \emph {et~al.}(2024)\citenamefont
  {Poyhonen}, \citenamefont {Moghaddam}, \citenamefont {Ivaki},\ and\
  \citenamefont {Ojanen}}]{poyhonen_scalable_2024}%
  \BibitemOpen
  \bibfield  {author} {\bibinfo {author} {\bibfnamefont {K.}~\bibnamefont
  {Poyhonen}}, \bibinfo {author} {\bibfnamefont {A.~G.}\ \bibnamefont
  {Moghaddam}}, \bibinfo {author} {\bibfnamefont {M.~N.}\ \bibnamefont
  {Ivaki}},\ and\ \bibinfo {author} {\bibfnamefont {T.}~\bibnamefont
  {Ojanen}},\ }\bibfield  {title} {\bibinfo {title} {Scalable approach to
  monitored quantum dynamics and entanglement phase transitions},\ }\bibfield
  {journal} {\bibinfo  {journal} {arXiv:2406.19052}\ }\href
  {https://doi.org/10.48550/arXiv.2406.19052} {10.48550/arXiv.2406.19052}
  (\bibinfo {year} {2024})\BibitemShut {NoStop}%
\bibitem [{\citenamefont {Fisher}\ \emph {et~al.}(2023)\citenamefont {Fisher},
  \citenamefont {Khemani}, \citenamefont {Nahum},\ and\ \citenamefont
  {Vijay}}]{fisher_random_2023}%
  \BibitemOpen
  \bibfield  {author} {\bibinfo {author} {\bibfnamefont {M.~P.~A.}\
  \bibnamefont {Fisher}}, \bibinfo {author} {\bibfnamefont {V.}~\bibnamefont
  {Khemani}}, \bibinfo {author} {\bibfnamefont {A.}~\bibnamefont {Nahum}},\
  and\ \bibinfo {author} {\bibfnamefont {S.}~\bibnamefont {Vijay}},\ }\bibfield
   {title} {\bibinfo {title} {Random {Quantum} {Circuits}},\ }\href
  {https://doi.org/10.1146/annurev-conmatphys-031720-030658} {\bibfield
  {journal} {\bibinfo  {journal} {Annual Review of Condensed Matter Physics}\
  }\textbf {\bibinfo {volume} {14}},\ \bibinfo {pages} {335} (\bibinfo {year}
  {2023})}\BibitemShut {NoStop}%
\bibitem [{\citenamefont {Liu}\ \emph {et~al.}(2024{\natexlab{b}})\citenamefont
  {Liu}, \citenamefont {Zhang}, \citenamefont {Yin},\ and\ \citenamefont
  {Zhang}}]{liu2024symmetry}%
  \BibitemOpen
  \bibfield  {author} {\bibinfo {author} {\bibfnamefont {S.}~\bibnamefont
  {Liu}}, \bibinfo {author} {\bibfnamefont {H.-K.}\ \bibnamefont {Zhang}},
  \bibinfo {author} {\bibfnamefont {S.}~\bibnamefont {Yin}},\ and\ \bibinfo
  {author} {\bibfnamefont {S.-X.}\ \bibnamefont {Zhang}},\ }\bibfield  {title}
  {\bibinfo {title} {Symmetry restoration and quantum mpemba effect in
  symmetric random circuits},\ }\href
  {https://doi.org/10.1103/PhysRevLett.133.140405} {\bibfield  {journal}
  {\bibinfo  {journal} {Phys. Rev. Lett.}\ }\textbf {\bibinfo {volume} {133}},\
  \bibinfo {pages} {140405} (\bibinfo {year} {2024}{\natexlab{b}})}\BibitemShut
  {NoStop}%
\bibitem [{\citenamefont {Li}\ \emph {et~al.}(2024{\natexlab{a}})\citenamefont
  {Li}, \citenamefont {Zheng}, \citenamefont {Liu}, \citenamefont {Jiang},\
  and\ \citenamefont {Liu}}]{li2024designs}%
  \BibitemOpen
  \bibfield  {author} {\bibinfo {author} {\bibfnamefont {Z.}~\bibnamefont
  {Li}}, \bibinfo {author} {\bibfnamefont {H.}~\bibnamefont {Zheng}}, \bibinfo
  {author} {\bibfnamefont {J.}~\bibnamefont {Liu}}, \bibinfo {author}
  {\bibfnamefont {L.}~\bibnamefont {Jiang}},\ and\ \bibinfo {author}
  {\bibfnamefont {Z.-W.}\ \bibnamefont {Liu}},\ }\bibfield  {title} {\bibinfo
  {title} {Designs from local random quantum circuits with $\mathrm{SU}(d)$
  symmetry},\ }\href {https://doi.org/10.1103/PRXQuantum.5.040349} {\bibfield
  {journal} {\bibinfo  {journal} {PRX Quantum}\ }\textbf {\bibinfo {volume}
  {5}},\ \bibinfo {pages} {040349} (\bibinfo {year}
  {2024}{\natexlab{a}})}\BibitemShut {NoStop}%
\bibitem [{\citenamefont {Li}\ \emph {et~al.}(2023)\citenamefont {Li},
  \citenamefont {Zheng}, \citenamefont {Wang}, \citenamefont {Jiang},
  \citenamefont {Liu},\ and\ \citenamefont {Liu}}]{li2023sudsymmetric}%
  \BibitemOpen
  \bibfield  {author} {\bibinfo {author} {\bibfnamefont {Z.}~\bibnamefont
  {Li}}, \bibinfo {author} {\bibfnamefont {H.}~\bibnamefont {Zheng}}, \bibinfo
  {author} {\bibfnamefont {Y.}~\bibnamefont {Wang}}, \bibinfo {author}
  {\bibfnamefont {L.}~\bibnamefont {Jiang}}, \bibinfo {author} {\bibfnamefont
  {Z.-W.}\ \bibnamefont {Liu}},\ and\ \bibinfo {author} {\bibfnamefont
  {J.}~\bibnamefont {Liu}},\ }\href@noop {} {\bibinfo {title} {Su(d)-symmetric
  random unitaries: Quantum scrambling, error correction, and machine
  learning}} (\bibinfo {year} {2023}),\ \Eprint
  {https://arxiv.org/abs/2309.16556} {arXiv:2309.16556 [quant-ph]} \BibitemShut
  {NoStop}%
\bibitem [{\citenamefont {Hearth}\ \emph {et~al.}(2023)\citenamefont {Hearth},
  \citenamefont {Flynn}, \citenamefont {Chandran},\ and\ \citenamefont
  {Laumann}}]{hearth_unitary_2023}%
  \BibitemOpen
  \bibfield  {author} {\bibinfo {author} {\bibfnamefont {S.~N.}\ \bibnamefont
  {Hearth}}, \bibinfo {author} {\bibfnamefont {M.~O.}\ \bibnamefont {Flynn}},
  \bibinfo {author} {\bibfnamefont {A.}~\bibnamefont {Chandran}},\ and\
  \bibinfo {author} {\bibfnamefont {C.~R.}\ \bibnamefont {Laumann}},\
  }\bibfield  {title} {\bibinfo {title} {Unitary k-designs from random
  number-conserving quantum circuits},\ }\href
  {https://arxiv.org/abs/2306.01035v1} {\bibfield  {journal} {\bibinfo
  {journal} {arXiv:2306.01035v1}\ } (\bibinfo {year} {2023})}\BibitemShut
  {NoStop}%
\bibitem [{\citenamefont {Li}\ \emph {et~al.}(2024{\natexlab{b}})\citenamefont
  {Li}, \citenamefont {Zheng},\ and\ \citenamefont
  {Liu}}]{li2024efficientquantumpseudorandomnessconservation}%
  \BibitemOpen
  \bibfield  {author} {\bibinfo {author} {\bibfnamefont {Z.}~\bibnamefont
  {Li}}, \bibinfo {author} {\bibfnamefont {H.}~\bibnamefont {Zheng}},\ and\
  \bibinfo {author} {\bibfnamefont {Z.-W.}\ \bibnamefont {Liu}},\ }\href
  {https://arxiv.org/abs/2411.04893} {\bibinfo {title} {Efficient quantum
  pseudorandomness under conservation laws}} (\bibinfo {year}
  {2024}{\natexlab{b}}),\ \Eprint {https://arxiv.org/abs/2411.04893}
  {arXiv:2411.04893 [quant-ph]} \BibitemShut {NoStop}%
\bibitem [{\citenamefont {Yu}\ \emph {et~al.}(2025)\citenamefont {Yu},
  \citenamefont {Li},\ and\ \citenamefont
  {Zhang}}]{yu2025symmetrybreakingdynamicsquantum}%
  \BibitemOpen
  \bibfield  {author} {\bibinfo {author} {\bibfnamefont {H.}~\bibnamefont
  {Yu}}, \bibinfo {author} {\bibfnamefont {Z.-X.}\ \bibnamefont {Li}},\ and\
  \bibinfo {author} {\bibfnamefont {S.-X.}\ \bibnamefont {Zhang}},\ }\href
  {https://arxiv.org/abs/2501.13459} {\bibinfo {title} {Symmetry breaking
  dynamics in quantum many-body systems}} (\bibinfo {year} {2025}),\ \Eprint
  {https://arxiv.org/abs/2501.13459} {arXiv:2501.13459 [quant-ph]} \BibitemShut
  {NoStop}%
\bibitem [{\citenamefont {Garratt}\ and\ \citenamefont
  {Altman}(2024)}]{garratt_probing_2024}%
  \BibitemOpen
  \bibfield  {author} {\bibinfo {author} {\bibfnamefont {S.~J.}\ \bibnamefont
  {Garratt}}\ and\ \bibinfo {author} {\bibfnamefont {E.}~\bibnamefont
  {Altman}},\ }\bibfield  {title} {\bibinfo {title} {Probing {Postmeasurement}
  {Entanglement} without {Postselection}},\ }\href
  {https://doi.org/10.1103/PRXQuantum.5.030311} {\bibfield  {journal} {\bibinfo
   {journal} {PRX Quantum}\ }\textbf {\bibinfo {volume} {5}},\ \bibinfo {pages}
  {030311} (\bibinfo {year} {2024})}\BibitemShut {NoStop}%
\bibitem [{\citenamefont {McGinley}(2024)}]{mcginley_postselection-free_2024}%
  \BibitemOpen
  \bibfield  {author} {\bibinfo {author} {\bibfnamefont {M.}~\bibnamefont
  {McGinley}},\ }\bibfield  {title} {\bibinfo {title} {Postselection-{Free}
  {Learning} of {Measurement}-{Induced} {Quantum} {Dynamics}},\ }\href
  {https://doi.org/10.1103/PRXQuantum.5.020347} {\bibfield  {journal} {\bibinfo
   {journal} {PRX Quantum}\ }\textbf {\bibinfo {volume} {5}},\ \bibinfo {pages}
  {020347} (\bibinfo {year} {2024})}\BibitemShut {NoStop}%
\bibitem [{\citenamefont {Hoke}\ \emph {et~al.}(2023)\citenamefont {Hoke},
  \citenamefont {Ippoliti}, \citenamefont {Rosenberg}, \citenamefont {Abanin},
  \citenamefont {Acharya}, \citenamefont {Andersen}, \citenamefont {Ansmann},
  \citenamefont {Arute}, \citenamefont {Arya}, \citenamefont {Asfaw},
  \citenamefont {Atalaya}, \citenamefont {Bardin}, \citenamefont {Bengtsson},
  \citenamefont {Bortoli}, \citenamefont {Bourassa}, \citenamefont {Bovaird}
  \emph {et~al.}}]{hoke_measurement-induced_2023}%
  \BibitemOpen
  \bibfield  {author} {\bibinfo {author} {\bibfnamefont {J.~C.}\ \bibnamefont
  {Hoke}}, \bibinfo {author} {\bibfnamefont {M.}~\bibnamefont {Ippoliti}},
  \bibinfo {author} {\bibfnamefont {E.}~\bibnamefont {Rosenberg}}, \bibinfo
  {author} {\bibfnamefont {D.}~\bibnamefont {Abanin}}, \bibinfo {author}
  {\bibfnamefont {R.}~\bibnamefont {Acharya}}, \bibinfo {author} {\bibfnamefont
  {T.~I.}\ \bibnamefont {Andersen}}, \bibinfo {author} {\bibfnamefont
  {M.}~\bibnamefont {Ansmann}}, \bibinfo {author} {\bibfnamefont
  {F.}~\bibnamefont {Arute}}, \bibinfo {author} {\bibfnamefont
  {K.}~\bibnamefont {Arya}}, \bibinfo {author} {\bibfnamefont {A.}~\bibnamefont
  {Asfaw}}, \bibinfo {author} {\bibfnamefont {J.}~\bibnamefont {Atalaya}},
  \bibinfo {author} {\bibfnamefont {J.~C.}\ \bibnamefont {Bardin}}, \bibinfo
  {author} {\bibfnamefont {A.}~\bibnamefont {Bengtsson}}, \bibinfo {author}
  {\bibfnamefont {G.}~\bibnamefont {Bortoli}}, \bibinfo {author} {\bibfnamefont
  {A.}~\bibnamefont {Bourassa}}, \bibinfo {author} {\bibfnamefont
  {J.}~\bibnamefont {Bovaird}}, \emph {et~al.},\ }\bibfield  {title} {\bibinfo
  {title} {Measurement-induced entanglement and teleportation on a noisy
  quantum processor},\ }\href {https://doi.org/10.1038/s41586-023-06505-7}
  {\bibfield  {journal} {\bibinfo  {journal} {Nature}\ }\textbf {\bibinfo
  {volume} {622}},\ \bibinfo {pages} {481} (\bibinfo {year}
  {2023})}\BibitemShut {NoStop}%
\bibitem [{\citenamefont {Alhambra}(2022)}]{alhambra_quantum_2022}%
  \BibitemOpen
  \bibfield  {author} {\bibinfo {author} {\bibfnamefont {A.~M.}\ \bibnamefont
  {Alhambra}},\ }\bibfield  {title} {\bibinfo {title} {Quantum many-body
  systems in thermal equilibrium},\ }\bibfield  {journal} {\bibinfo  {journal}
  {arXiv:2204.08349v1}\ }\href {https://doi.org/10.48550/arXiv.2204.08349}
  {10.48550/arXiv.2204.08349} (\bibinfo {year} {2022})\BibitemShut {NoStop}%
\bibitem [{\citenamefont {Jozsa}\ \emph {et~al.}(1994)\citenamefont {Jozsa},
  \citenamefont {Robb},\ and\ \citenamefont {Wootters}}]{jozsa_lower_1994}%
  \BibitemOpen
  \bibfield  {author} {\bibinfo {author} {\bibfnamefont {R.}~\bibnamefont
  {Jozsa}}, \bibinfo {author} {\bibfnamefont {D.}~\bibnamefont {Robb}},\ and\
  \bibinfo {author} {\bibfnamefont {W.~K.}\ \bibnamefont {Wootters}},\
  }\bibfield  {title} {\bibinfo {title} {Lower bound for accessible information
  in quantum mechanics},\ }\href {https://doi.org/10.1103/PhysRevA.49.668}
  {\bibfield  {journal} {\bibinfo  {journal} {Physical Review A}\ }\textbf
  {\bibinfo {volume} {49}},\ \bibinfo {pages} {668} (\bibinfo {year}
  {1994})}\BibitemShut {NoStop}%
\bibitem [{\citenamefont {Roberts}\ and\ \citenamefont
  {Yoshida}(2017)}]{roberts_chaos_2017}%
  \BibitemOpen
  \bibfield  {author} {\bibinfo {author} {\bibfnamefont {D.~A.}\ \bibnamefont
  {Roberts}}\ and\ \bibinfo {author} {\bibfnamefont {B.}~\bibnamefont
  {Yoshida}},\ }\bibfield  {title} {\bibinfo {title} {Chaos and complexity by
  design},\ }\href {https://doi.org/10.1007/JHEP04(2017)121} {\bibfield
  {journal} {\bibinfo  {journal} {Journal of High Energy Physics}\ }\textbf
  {\bibinfo {volume} {2017}},\ \bibinfo {pages} {121} (\bibinfo {year}
  {2017})}\BibitemShut {NoStop}%
\bibitem [{\citenamefont {Knill}\ \emph {et~al.}(2008)\citenamefont {Knill},
  \citenamefont {Leibfried}, \citenamefont {Reichle}, \citenamefont {Britton},
  \citenamefont {Blakestad}, \citenamefont {Jost}, \citenamefont {Langer},
  \citenamefont {Ozeri}, \citenamefont {Seidelin},\ and\ \citenamefont
  {Wineland}}]{knill_randomized_2008}%
  \BibitemOpen
  \bibfield  {author} {\bibinfo {author} {\bibfnamefont {E.}~\bibnamefont
  {Knill}}, \bibinfo {author} {\bibfnamefont {D.}~\bibnamefont {Leibfried}},
  \bibinfo {author} {\bibfnamefont {R.}~\bibnamefont {Reichle}}, \bibinfo
  {author} {\bibfnamefont {J.}~\bibnamefont {Britton}}, \bibinfo {author}
  {\bibfnamefont {R.~B.}\ \bibnamefont {Blakestad}}, \bibinfo {author}
  {\bibfnamefont {J.~D.}\ \bibnamefont {Jost}}, \bibinfo {author}
  {\bibfnamefont {C.}~\bibnamefont {Langer}}, \bibinfo {author} {\bibfnamefont
  {R.}~\bibnamefont {Ozeri}}, \bibinfo {author} {\bibfnamefont
  {S.}~\bibnamefont {Seidelin}},\ and\ \bibinfo {author} {\bibfnamefont
  {D.~J.}\ \bibnamefont {Wineland}},\ }\bibfield  {title} {\bibinfo {title}
  {Randomized benchmarking of quantum gates},\ }\href
  {https://doi.org/10.1103/PhysRevA.77.012307} {\bibfield  {journal} {\bibinfo
  {journal} {Physical Review A}\ }\textbf {\bibinfo {volume} {77}},\ \bibinfo
  {pages} {012307} (\bibinfo {year} {2008})}\BibitemShut {NoStop}%
\bibitem [{\citenamefont {Dankert}\ \emph {et~al.}(2009)\citenamefont
  {Dankert}, \citenamefont {Cleve}, \citenamefont {Emerson},\ and\
  \citenamefont {Livine}}]{dankert_exact_2009}%
  \BibitemOpen
  \bibfield  {author} {\bibinfo {author} {\bibfnamefont {C.}~\bibnamefont
  {Dankert}}, \bibinfo {author} {\bibfnamefont {R.}~\bibnamefont {Cleve}},
  \bibinfo {author} {\bibfnamefont {J.}~\bibnamefont {Emerson}},\ and\ \bibinfo
  {author} {\bibfnamefont {E.}~\bibnamefont {Livine}},\ }\bibfield  {title}
  {\bibinfo {title} {Exact and approximate unitary 2-designs and their
  application to fidelity estimation},\ }\href
  {https://doi.org/10.1103/PhysRevA.80.012304} {\bibfield  {journal} {\bibinfo
  {journal} {Physical Review A}\ }\textbf {\bibinfo {volume} {80}},\ \bibinfo
  {pages} {012304} (\bibinfo {year} {2009})}\BibitemShut {NoStop}%
\bibitem [{\citenamefont {Huang}\ \emph {et~al.}(2020)\citenamefont {Huang},
  \citenamefont {Kueng},\ and\ \citenamefont
  {Preskill}}]{huang_predicting_2020}%
  \BibitemOpen
  \bibfield  {author} {\bibinfo {author} {\bibfnamefont {H.-Y.}\ \bibnamefont
  {Huang}}, \bibinfo {author} {\bibfnamefont {R.}~\bibnamefont {Kueng}},\ and\
  \bibinfo {author} {\bibfnamefont {J.}~\bibnamefont {Preskill}},\ }\bibfield
  {title} {\bibinfo {title} {Predicting many properties of a quantum system
  from very few measurements},\ }\href
  {https://doi.org/10.1038/s41567-020-0932-7} {\bibfield  {journal} {\bibinfo
  {journal} {Nature Physics}\ }\textbf {\bibinfo {volume} {16}},\ \bibinfo
  {pages} {1050} (\bibinfo {year} {2020})}\BibitemShut {NoStop}%
\bibitem [{\citenamefont {Lancien}\ and\ \citenamefont
  {Majenz}(2020)}]{lancien_weak_2020}%
  \BibitemOpen
  \bibfield  {author} {\bibinfo {author} {\bibfnamefont {C.}~\bibnamefont
  {Lancien}}\ and\ \bibinfo {author} {\bibfnamefont {C.}~\bibnamefont
  {Majenz}},\ }\bibfield  {title} {\bibinfo {title} {Weak approximate unitary
  designs and applications to quantum encryption},\ }\href
  {https://doi.org/10.22331/q-2020-08-28-313} {\bibfield  {journal} {\bibinfo
  {journal} {Quantum}\ }\textbf {\bibinfo {volume} {4}},\ \bibinfo {pages}
  {313} (\bibinfo {year} {2020})}\BibitemShut {NoStop}%
\bibitem [{\citenamefont {Varikuti}\ and\ \citenamefont
  {Bandyopadhyay}(2024)}]{varikuti_unraveling_2024}%
  \BibitemOpen
  \bibfield  {author} {\bibinfo {author} {\bibfnamefont {N.~D.}\ \bibnamefont
  {Varikuti}}\ and\ \bibinfo {author} {\bibfnamefont {S.}~\bibnamefont
  {Bandyopadhyay}},\ }\bibfield  {title} {\bibinfo {title} {Unraveling the
  emergence of quantum state designs in systems with symmetry},\ }\bibfield
  {journal} {\bibinfo  {journal} {arXiv:2402.08949}\ }\href
  {https://doi.org/10.48550/arXiv.2402.08949} {10.48550/arXiv.2402.08949}
  (\bibinfo {year} {2024})\BibitemShut {NoStop}%
\bibitem [{\citenamefont {Goldstein}\ \emph
  {et~al.}(2006{\natexlab{a}})\citenamefont {Goldstein}, \citenamefont
  {Lebowitz}, \citenamefont {Tumulka},\ and\ \citenamefont
  {Zanghi}}]{goldstein_distribution_2006}%
  \BibitemOpen
  \bibfield  {author} {\bibinfo {author} {\bibfnamefont {S.}~\bibnamefont
  {Goldstein}}, \bibinfo {author} {\bibfnamefont {J.~L.}\ \bibnamefont
  {Lebowitz}}, \bibinfo {author} {\bibfnamefont {R.}~\bibnamefont {Tumulka}},\
  and\ \bibinfo {author} {\bibfnamefont {N.}~\bibnamefont {Zanghi}},\
  }\bibfield  {title} {\bibinfo {title} {On the {Distribution} of the {Wave}
  {Function} for {Systems} in {Thermal} {Equilibrium}},\ }\href
  {https://doi.org/10.1007/s10955-006-9210-z} {\bibfield  {journal} {\bibinfo
  {journal} {Journal of Statistical Physics}\ }\textbf {\bibinfo {volume}
  {125}},\ \bibinfo {pages} {1193} (\bibinfo {year}
  {2006}{\natexlab{a}})}\BibitemShut {NoStop}%
\bibitem [{\citenamefont {Goldstein}\ \emph {et~al.}(2016)\citenamefont
  {Goldstein}, \citenamefont {Lebowitz}, \citenamefont {Mastrodonato},
  \citenamefont {Tumulka},\ and\ \citenamefont
  {Zanghi}}]{goldstein_universal_2016}%
  \BibitemOpen
  \bibfield  {author} {\bibinfo {author} {\bibfnamefont {S.}~\bibnamefont
  {Goldstein}}, \bibinfo {author} {\bibfnamefont {J.~L.}\ \bibnamefont
  {Lebowitz}}, \bibinfo {author} {\bibfnamefont {C.}~\bibnamefont
  {Mastrodonato}}, \bibinfo {author} {\bibfnamefont {R.}~\bibnamefont
  {Tumulka}},\ and\ \bibinfo {author} {\bibfnamefont {N.}~\bibnamefont
  {Zanghi}},\ }\bibfield  {title} {\bibinfo {title} {Universal {Probability}
  {Distribution} for the {Wave} {Function} of a {Quantum} {System} {Entangled}
  with its {Environment}},\ }\href {https://doi.org/10.1007/s00220-015-2536-0}
  {\bibfield  {journal} {\bibinfo  {journal} {Communications in Mathematical
  Physics}\ }\textbf {\bibinfo {volume} {342}},\ \bibinfo {pages} {965}
  (\bibinfo {year} {2016})}\BibitemShut {NoStop}%
\bibitem [{\citenamefont {Marvian}(2022)}]{marvian_restrictions_2022}%
  \BibitemOpen
  \bibfield  {author} {\bibinfo {author} {\bibfnamefont {I.}~\bibnamefont
  {Marvian}},\ }\bibfield  {title} {\bibinfo {title} {Restrictions on
  realizable unitary operations imposed by symmetry and locality},\ }\href
  {https://doi.org/10.1038/s41567-021-01464-0} {\bibfield  {journal} {\bibinfo
  {journal} {Nature Physics}\ }\textbf {\bibinfo {volume} {18}},\ \bibinfo
  {pages} {283} (\bibinfo {year} {2022})}\BibitemShut {NoStop}%
\bibitem [{\citenamefont {Marvian}(2023)}]{marvian_theory_2023}%
  \BibitemOpen
  \bibfield  {author} {\bibinfo {author} {\bibfnamefont {I.}~\bibnamefont
  {Marvian}},\ }\bibfield  {title} {\bibinfo {title} {Theory of {Quantum}
  {Circuits} with {Abelian} {Symmetries}},\ }\href
  {https://arxiv.org/abs/2302.12466v2} {\bibfield  {journal} {\bibinfo
  {journal} {arXiv:2302.12466v2}\ } (\bibinfo {year} {2023})}\BibitemShut
  {NoStop}%
\bibitem [{\citenamefont {Ledoux}(2001)}]{ledoux_concentration_2001}%
  \BibitemOpen
  \bibfield  {author} {\bibinfo {author} {\bibfnamefont {M.}~\bibnamefont
  {Ledoux}},\ }\bibfield  {title} {\bibinfo {title} {The {Concentration} of
  {Measure} {Phenomenon}},\ }\bibfield  {journal} {\bibinfo  {journal}
  {Mathematical Surveys and Monographs}\ }\textbf {\bibinfo {volume} {89}},\
  \href {https://doi.org/https://doi.org/10.1090/surv/089}
  {https://doi.org/10.1090/surv/089} (\bibinfo {year} {2001})\BibitemShut
  {NoStop}%
\bibitem [{\citenamefont {Murthy}\ \emph {et~al.}(2023)\citenamefont {Murthy},
  \citenamefont {Babakhani}, \citenamefont {Iniguez}, \citenamefont
  {Srednicki},\ and\ \citenamefont {Yunger~Halpern}}]{murthy_non-abelian_2023}%
  \BibitemOpen
  \bibfield  {author} {\bibinfo {author} {\bibfnamefont {C.}~\bibnamefont
  {Murthy}}, \bibinfo {author} {\bibfnamefont {A.}~\bibnamefont {Babakhani}},
  \bibinfo {author} {\bibfnamefont {F.}~\bibnamefont {Iniguez}}, \bibinfo
  {author} {\bibfnamefont {M.}~\bibnamefont {Srednicki}},\ and\ \bibinfo
  {author} {\bibfnamefont {N.}~\bibnamefont {Yunger~Halpern}},\ }\bibfield
  {title} {\bibinfo {title} {Non-{Abelian} {Eigenstate} {Thermalization}
  {Hypothesis}},\ }\href {https://doi.org/10.1103/PhysRevLett.130.140402}
  {\bibfield  {journal} {\bibinfo  {journal} {Physical Review Letters}\
  }\textbf {\bibinfo {volume} {130}},\ \bibinfo {pages} {140402} (\bibinfo
  {year} {2023})}\BibitemShut {NoStop}%
\bibitem [{\citenamefont {Majidy}\ \emph
  {et~al.}(2023{\natexlab{a}})\citenamefont {Majidy}, \citenamefont {Lasek},
  \citenamefont {Huse},\ and\ \citenamefont
  {Yunger~Halpern}}]{majidy_non-abelian_2023}%
  \BibitemOpen
  \bibfield  {author} {\bibinfo {author} {\bibfnamefont {S.}~\bibnamefont
  {Majidy}}, \bibinfo {author} {\bibfnamefont {A.}~\bibnamefont {Lasek}},
  \bibinfo {author} {\bibfnamefont {D.~A.}\ \bibnamefont {Huse}},\ and\
  \bibinfo {author} {\bibfnamefont {N.}~\bibnamefont {Yunger~Halpern}},\
  }\bibfield  {title} {\bibinfo {title} {Non-{Abelian} symmetry can increase
  entanglement entropy},\ }\href {https://doi.org/10.1103/PhysRevB.107.045102}
  {\bibfield  {journal} {\bibinfo  {journal} {Physical Review B}\ }\textbf
  {\bibinfo {volume} {107}},\ \bibinfo {pages} {045102} (\bibinfo {year}
  {2023}{\natexlab{a}})}\BibitemShut {NoStop}%
\bibitem [{\citenamefont {Majidy}\ \emph
  {et~al.}(2023{\natexlab{b}})\citenamefont {Majidy}, \citenamefont {Braasch},
  \citenamefont {Lasek}, \citenamefont {Upadhyaya}, \citenamefont {Kalev},\
  and\ \citenamefont {Yunger~Halpern}}]{majidy_noncommuting_2023}%
  \BibitemOpen
  \bibfield  {author} {\bibinfo {author} {\bibfnamefont {S.}~\bibnamefont
  {Majidy}}, \bibinfo {author} {\bibfnamefont {W.~F.}\ \bibnamefont {Braasch}},
  \bibinfo {author} {\bibfnamefont {A.}~\bibnamefont {Lasek}}, \bibinfo
  {author} {\bibfnamefont {T.}~\bibnamefont {Upadhyaya}}, \bibinfo {author}
  {\bibfnamefont {A.}~\bibnamefont {Kalev}},\ and\ \bibinfo {author}
  {\bibfnamefont {N.}~\bibnamefont {Yunger~Halpern}},\ }\bibfield  {title}
  {\bibinfo {title} {Noncommuting conserved charges in quantum thermodynamics
  and beyond},\ }\href {https://doi.org/10.1038/s42254-023-00641-9} {\bibfield
  {journal} {\bibinfo  {journal} {Nature Reviews Physics}\ }\textbf {\bibinfo
  {volume} {5}},\ \bibinfo {pages} {689} (\bibinfo {year}
  {2023}{\natexlab{b}})}\BibitemShut {NoStop}%
\bibitem [{\citenamefont {Majidy}\ \emph
  {et~al.}(2023{\natexlab{c}})\citenamefont {Majidy}, \citenamefont {Agrawal},
  \citenamefont {Gopalakrishnan}, \citenamefont {Potter}, \citenamefont
  {Vasseur},\ and\ \citenamefont {Halpern}}]{majidy_critical_2023}%
  \BibitemOpen
  \bibfield  {author} {\bibinfo {author} {\bibfnamefont {S.}~\bibnamefont
  {Majidy}}, \bibinfo {author} {\bibfnamefont {U.}~\bibnamefont {Agrawal}},
  \bibinfo {author} {\bibfnamefont {S.}~\bibnamefont {Gopalakrishnan}},
  \bibinfo {author} {\bibfnamefont {A.~C.}\ \bibnamefont {Potter}}, \bibinfo
  {author} {\bibfnamefont {R.}~\bibnamefont {Vasseur}},\ and\ \bibinfo {author}
  {\bibfnamefont {N.~Y.}\ \bibnamefont {Halpern}},\ }\bibfield  {title}
  {\bibinfo {title} {Critical phase and spin sharpening in {SU}(2)-symmetric
  monitored quantum circuits},\ }\href
  {https://doi.org/10.1103/PhysRevB.108.054307} {\bibfield  {journal} {\bibinfo
   {journal} {Physical Review B}\ }\textbf {\bibinfo {volume} {108}},\ \bibinfo
  {pages} {054307} (\bibinfo {year} {2023}{\natexlab{c}})}\BibitemShut
  {NoStop}%
\bibitem [{\citenamefont {Feng}\ \emph {et~al.}(2024)\citenamefont {Feng},
  \citenamefont {Fishchenko}, \citenamefont {Gopalakrishnan},\ and\
  \citenamefont {Ippoliti}}]{feng_charge_2024}%
  \BibitemOpen
  \bibfield  {author} {\bibinfo {author} {\bibfnamefont {X.}~\bibnamefont
  {Feng}}, \bibinfo {author} {\bibfnamefont {N.}~\bibnamefont {Fishchenko}},
  \bibinfo {author} {\bibfnamefont {S.}~\bibnamefont {Gopalakrishnan}},\ and\
  \bibinfo {author} {\bibfnamefont {M.}~\bibnamefont {Ippoliti}},\ }\bibfield
  {title} {\bibinfo {title} {Charge and {Spin} {Sharpening} {Transitions} on
  {Dynamical} {Quantum} {Trees}},\ }\href {https://arxiv.org/abs/2405.13894v1}
  {\bibfield  {journal} {\bibinfo  {journal} {arXiv:2405.13894v1}\ } (\bibinfo
  {year} {2024})}\BibitemShut {NoStop}%
\bibitem [{\citenamefont {Tran}\ \emph {et~al.}(2023)\citenamefont {Tran},
  \citenamefont {Mark}, \citenamefont {Ho},\ and\ \citenamefont
  {Choi}}]{tran_measuring_2023}%
  \BibitemOpen
  \bibfield  {author} {\bibinfo {author} {\bibfnamefont {M.~C.}\ \bibnamefont
  {Tran}}, \bibinfo {author} {\bibfnamefont {D.~K.}\ \bibnamefont {Mark}},
  \bibinfo {author} {\bibfnamefont {W.~W.}\ \bibnamefont {Ho}},\ and\ \bibinfo
  {author} {\bibfnamefont {S.}~\bibnamefont {Choi}},\ }\bibfield  {title}
  {\bibinfo {title} {Measuring {Arbitrary} {Physical} {Properties} in {Analog}
  {Quantum} {Simulation}},\ }\href {https://doi.org/10.1103/PhysRevX.13.011049}
  {\bibfield  {journal} {\bibinfo  {journal} {Physical Review X}\ }\textbf
  {\bibinfo {volume} {13}},\ \bibinfo {pages} {011049} (\bibinfo {year}
  {2023})}\BibitemShut {NoStop}%
\bibitem [{\citenamefont {McGinley}\ and\ \citenamefont
  {Fava}(2023)}]{mcginley_shadow_2023}%
  \BibitemOpen
  \bibfield  {author} {\bibinfo {author} {\bibfnamefont {M.}~\bibnamefont
  {McGinley}}\ and\ \bibinfo {author} {\bibfnamefont {M.}~\bibnamefont
  {Fava}},\ }\bibfield  {title} {\bibinfo {title} {Shadow {Tomography} from
  {Emergent} {State} {Designs} in {Analog} {Quantum} {Simulators}},\ }\href
  {https://doi.org/10.1103/PhysRevLett.131.160601} {\bibfield  {journal}
  {\bibinfo  {journal} {Physical Review Letters}\ }\textbf {\bibinfo {volume}
  {131}},\ \bibinfo {pages} {160601} (\bibinfo {year} {2023})}\BibitemShut
  {NoStop}%
\bibitem [{\citenamefont {Hearth}\ \emph {et~al.}(2024)\citenamefont {Hearth},
  \citenamefont {Flynn}, \citenamefont {Chandran},\ and\ \citenamefont
  {Laumann}}]{hearth_efficient_2024}%
  \BibitemOpen
  \bibfield  {author} {\bibinfo {author} {\bibfnamefont {S.~N.}\ \bibnamefont
  {Hearth}}, \bibinfo {author} {\bibfnamefont {M.~O.}\ \bibnamefont {Flynn}},
  \bibinfo {author} {\bibfnamefont {A.}~\bibnamefont {Chandran}},\ and\
  \bibinfo {author} {\bibfnamefont {C.~R.}\ \bibnamefont {Laumann}},\
  }\bibfield  {title} {\bibinfo {title} {Efficient {Local} {Classical} {Shadow}
  {Tomography} with {Number} {Conservation}},\ }\href
  {https://doi.org/10.1103/PhysRevLett.133.060802} {\bibfield  {journal}
  {\bibinfo  {journal} {Physical Review Letters}\ }\textbf {\bibinfo {volume}
  {133}},\ \bibinfo {pages} {060802} (\bibinfo {year} {2024})}\BibitemShut
  {NoStop}%
\bibitem [{\citenamefont {Baldwin}\ \emph {et~al.}(2020)\citenamefont
  {Baldwin}, \citenamefont {Bjork}, \citenamefont {Gaebler}, \citenamefont
  {Hayes},\ and\ \citenamefont {Stack}}]{baldwin_subspace_2020}%
  \BibitemOpen
  \bibfield  {author} {\bibinfo {author} {\bibfnamefont {C.~H.}\ \bibnamefont
  {Baldwin}}, \bibinfo {author} {\bibfnamefont {B.~J.}\ \bibnamefont {Bjork}},
  \bibinfo {author} {\bibfnamefont {J.~P.}\ \bibnamefont {Gaebler}}, \bibinfo
  {author} {\bibfnamefont {D.}~\bibnamefont {Hayes}},\ and\ \bibinfo {author}
  {\bibfnamefont {D.}~\bibnamefont {Stack}},\ }\bibfield  {title} {\bibinfo
  {title} {Subspace benchmarking high-fidelity entangling operations with
  trapped ions},\ }\href {https://doi.org/10.1103/PhysRevResearch.2.013317}
  {\bibfield  {journal} {\bibinfo  {journal} {Physical Review Research}\
  }\textbf {\bibinfo {volume} {2}},\ \bibinfo {pages} {013317} (\bibinfo {year}
  {2020})}\BibitemShut {NoStop}%
\bibitem [{\citenamefont {Goldstein}\ \emph
  {et~al.}(2006{\natexlab{b}})\citenamefont {Goldstein}, \citenamefont
  {Lebowitz}, \citenamefont {Tumulka},\ and\ \citenamefont
  {Zanghi}}]{goldstein_canonical_2006}%
  \BibitemOpen
  \bibfield  {author} {\bibinfo {author} {\bibfnamefont {S.}~\bibnamefont
  {Goldstein}}, \bibinfo {author} {\bibfnamefont {J.~L.}\ \bibnamefont
  {Lebowitz}}, \bibinfo {author} {\bibfnamefont {R.}~\bibnamefont {Tumulka}},\
  and\ \bibinfo {author} {\bibfnamefont {N.}~\bibnamefont {Zanghi}},\
  }\bibfield  {title} {\bibinfo {title} {Canonical {Typicality}},\ }\href
  {https://doi.org/10.1103/PhysRevLett.96.050403} {\bibfield  {journal}
  {\bibinfo  {journal} {Physical Review Letters}\ }\textbf {\bibinfo {volume}
  {96}},\ \bibinfo {pages} {050403} (\bibinfo {year}
  {2006}{\natexlab{b}})}\BibitemShut {NoStop}%
\bibitem [{\citenamefont {Mi}\ \emph {et~al.}(2022)\citenamefont {Mi},
  \citenamefont {Ippoliti}, \citenamefont {Quintana}, \citenamefont {Greene},
  \citenamefont {Chen}, \citenamefont {Gross}, \citenamefont {Arute},
  \citenamefont {Arya}, \citenamefont {Atalaya}, \citenamefont {Babbush},
  \citenamefont {Bardin}, \citenamefont {Basso}, \citenamefont {Bengtsson},
  \citenamefont {Bilmes}, \citenamefont {Bourassa}, \citenamefont {Brill} \emph
  {et~al.}}]{mi_time-crystalline_2022}%
  \BibitemOpen
  \bibfield  {author} {\bibinfo {author} {\bibfnamefont {X.}~\bibnamefont
  {Mi}}, \bibinfo {author} {\bibfnamefont {M.}~\bibnamefont {Ippoliti}},
  \bibinfo {author} {\bibfnamefont {C.}~\bibnamefont {Quintana}}, \bibinfo
  {author} {\bibfnamefont {A.}~\bibnamefont {Greene}}, \bibinfo {author}
  {\bibfnamefont {Z.}~\bibnamefont {Chen}}, \bibinfo {author} {\bibfnamefont
  {J.}~\bibnamefont {Gross}}, \bibinfo {author} {\bibfnamefont
  {F.}~\bibnamefont {Arute}}, \bibinfo {author} {\bibfnamefont
  {K.}~\bibnamefont {Arya}}, \bibinfo {author} {\bibfnamefont {J.}~\bibnamefont
  {Atalaya}}, \bibinfo {author} {\bibfnamefont {R.}~\bibnamefont {Babbush}},
  \bibinfo {author} {\bibfnamefont {J.~C.}\ \bibnamefont {Bardin}}, \bibinfo
  {author} {\bibfnamefont {J.}~\bibnamefont {Basso}}, \bibinfo {author}
  {\bibfnamefont {A.}~\bibnamefont {Bengtsson}}, \bibinfo {author}
  {\bibfnamefont {A.}~\bibnamefont {Bilmes}}, \bibinfo {author} {\bibfnamefont
  {A.}~\bibnamefont {Bourassa}}, \bibinfo {author} {\bibfnamefont
  {L.}~\bibnamefont {Brill}}, \emph {et~al.},\ }\bibfield  {title} {\bibinfo
  {title} {Time-crystalline eigenstate order on a quantum processor},\ }\href
  {https://doi.org/10.1038/s41586-021-04257-w} {\bibfield  {journal} {\bibinfo
  {journal} {Nature}\ }\textbf {\bibinfo {volume} {601}},\ \bibinfo {pages}
  {531} (\bibinfo {year} {2022})}\BibitemShut {NoStop}%
\bibitem [{\citenamefont {Richter}\ and\ \citenamefont
  {Pal}(2021)}]{richter_simulating_2021}%
  \BibitemOpen
  \bibfield  {author} {\bibinfo {author} {\bibfnamefont {J.}~\bibnamefont
  {Richter}}\ and\ \bibinfo {author} {\bibfnamefont {A.}~\bibnamefont {Pal}},\
  }\bibfield  {title} {\bibinfo {title} {Simulating {Hydrodynamics} on {Noisy}
  {Intermediate}-{Scale} {Quantum} {Devices} with {Random} {Circuits}},\ }\href
  {https://doi.org/10.1103/PhysRevLett.126.230501} {\bibfield  {journal}
  {\bibinfo  {journal} {Physical Review Letters}\ }\textbf {\bibinfo {volume}
  {126}},\ \bibinfo {pages} {230501} (\bibinfo {year} {2021})}\BibitemShut
  {NoStop}%
\bibitem [{\citenamefont {Andersen}\ \emph {et~al.}(2024)\citenamefont
  {Andersen}, \citenamefont {Astrakhantsev}, \citenamefont {Karamlou},
  \citenamefont {Berndtsson}, \citenamefont {Motruk}, \citenamefont {Szasz},
  \citenamefont {Gross}, \citenamefont {Schuckert}, \citenamefont {Westerhout},
  \citenamefont {Zhang}, \citenamefont {Forati}, \citenamefont {Rossi},
  \citenamefont {Kobrin}, \citenamefont {Di~Paolo}, \citenamefont {Klots},
  \citenamefont {Drozdov} \emph {et~al.}}]{andersen_thermalization_2024}%
  \BibitemOpen
  \bibfield  {author} {\bibinfo {author} {\bibfnamefont {T.~I.}\ \bibnamefont
  {Andersen}}, \bibinfo {author} {\bibfnamefont {N.}~\bibnamefont
  {Astrakhantsev}}, \bibinfo {author} {\bibfnamefont {A.~H.}\ \bibnamefont
  {Karamlou}}, \bibinfo {author} {\bibfnamefont {J.}~\bibnamefont
  {Berndtsson}}, \bibinfo {author} {\bibfnamefont {J.}~\bibnamefont {Motruk}},
  \bibinfo {author} {\bibfnamefont {A.}~\bibnamefont {Szasz}}, \bibinfo
  {author} {\bibfnamefont {J.~A.}\ \bibnamefont {Gross}}, \bibinfo {author}
  {\bibfnamefont {A.}~\bibnamefont {Schuckert}}, \bibinfo {author}
  {\bibfnamefont {T.}~\bibnamefont {Westerhout}}, \bibinfo {author}
  {\bibfnamefont {Y.}~\bibnamefont {Zhang}}, \bibinfo {author} {\bibfnamefont
  {E.}~\bibnamefont {Forati}}, \bibinfo {author} {\bibfnamefont
  {D.}~\bibnamefont {Rossi}}, \bibinfo {author} {\bibfnamefont
  {B.}~\bibnamefont {Kobrin}}, \bibinfo {author} {\bibfnamefont
  {A.}~\bibnamefont {Di~Paolo}}, \bibinfo {author} {\bibfnamefont {A.~R.}\
  \bibnamefont {Klots}}, \bibinfo {author} {\bibfnamefont {I.}~\bibnamefont
  {Drozdov}}, \emph {et~al.},\ }\bibfield  {title} {\bibinfo {title}
  {Thermalization and {Criticality} on an {Analog}-{Digital} {Quantum}
  {Simulator}},\ }\bibfield  {journal} {\bibinfo  {journal} {arXiv:2405.17385}\
  }\href {https://doi.org/10.48550/arXiv.2405.17385}
  {10.48550/arXiv.2405.17385} (\bibinfo {year} {2024})\BibitemShut {NoStop}%
\end{thebibliography}%

\clearpage


\widetext

\appendix

\section{Proof of Theorem~\ref{thm:1} \label{app:thm1}}
Here we prove Theorem~\ref{thm:1} stated in the main text, which is that  with high probability, the projected ensemble $\mathcal{E}_A$ of a Haar random $U(1)$-symmetric state, constructed using charge-revealing computational basis ($z$) measurements, is the direct sum ensemble.  We restate Theorem~\ref{thm:1} in slightly more formal fashion:

{\bf Theorem~\ref{thm:1}.}
{\it Let $|\Psi\rangle$ be a Haar random state on $N$ qubits with definite charge $Q_0$, so that the state lives in a Hilbert space with dimension $d_{Q_0} = \binom{N}{Q_0}$.
Construct  the projected ensemble $\mathcal{E}_A$ on $N_A$ qubits, using  measurements of $B$ in the computational $(z)$ basis. 
Then as long as 
\begin{align}
    d_{Q_0} & \geq \frac{18 \pi^3 (2k-1)^2 (N_A+1)^2 d_A^{4k}}{\epsilon^2} \left( \log(2/\delta) + \log\left( 2C_k d_A^{2k} N_A^{1-k} \right) \right),
\end{align}
($C_k$ is  a numerical constant depending on $k$ but not depending on $N_A$) 
with probability at least  $1-\delta$ the trace difference between the $k$-th moments of the projected ensemble and the `direct sum (DS)' ensemble,  obeys 
\begin{align}
    \frac{1}{2} \| \rho^{(k)}_{\mathcal E} - \rho^{(k)}_{\text{DS},Q_0} \| \leq \epsilon,
\end{align}
where
\begin{align}
    \rho^{(k)}_{\text{DS},Q_0} = \bigoplus \pi(Q_A|Q_0) \rho^{(k)}_{{\rm Haar},A,Q_A}.
\end{align}
Above, $\rho^{(k)}_{{\rm Haar},A,Q_A}$ is the $k$-th moment of the Haar ensemble on states on $A$ with fixed charge $Q_A$,   
$d_{Q_A} = \binom{N_A}{Q_A}$ is the dimension of charge sector $Q_A$ on $N_A$ qubits, and  
 the probability $\pi(Q_A|Q_0) = d_{Q_A} d_{Q_B}/d_{Q_0}$, where $d_{Q_B} = \binom{N-N_A}{Q_0-Q_A}$. 
 \\\\
 Letting $\sigma \equiv Q_0/N \in [0,1]$, the condition on $d_{Q_0}$ can be satisfied with
\begin{align}
    N_B H(\sigma) = \Omega\left(k N_A + \log\left({\epsilon}^{-1} \right) + \log \log\left({\delta}^{-1} \right) \right)
\end{align}
where $H(\sigma) = -\sigma \log \sigma - (1-\sigma)\log(1-\sigma)$.  
}

    {\it Proof.} 
    We break the proof in two parts. We first show that the $k^{th}$-moment of  theprojected ensemble $\mathcal{E}_{A}$ averaged over all Haar random $U(1)$-symmetric generator states $|\Phi\rangle$ is exactly the  $k^{th}$-moment of the direct sum ensemble, i.e.
    \begin{equation}
    \label{eq:avg_haar_gen}
        \mathop{\mathbb{E}}_{\Phi \sim \text{Haar}(\mathcal{H}_{Q_0})} \left[\sum_z \frac{\left(|\Tilde{\psi}_z\rangle\langle\Tilde{\psi}_z|\right)^{\otimes k}}{\langle\Tilde{\psi}_z|\Tilde{\psi}_z\rangle^{k-1}}\right] = \mathop{\bigoplus}_{Q_A} \pi(Q_A|Q_0)\rho^{(k)}_{{\rm Haar},A,Q_A} = \rho^{(k)}_{\text{DS},Q_0}.
    \end{equation}
    To do so we break the summation $\sum_z [\ldots] = \sum_{Q_B}\sum_{z_{Q_B}}[\ldots]$ into sums over charge sectors $Q_B$ and outcome strings $z_{Q_B}$ of fixed charge $Q_B$ and define the normalized state $|\psi_{z_{Q_B}}\rangle = |\Tilde{\psi}_{z_{Q_B}}\rangle / \sqrt{p(z_{Q_B})}$ with $p(z_{Q_B}) = \langle\Tilde{\psi}_{z_{Q_B}}|\Tilde{\psi}_{z_{Q_B}}\rangle$. This allows us to write the LHS as
    \begin{align}
        \sum_{Q_B} \sum_{z_{Q_B}} \mathop{\mathbb{E}}_{\Phi \sim \text{Haar}(\mathcal{H}_{Q_0})} \left[  \left(|\psi_{z_{Q_B} }\rangle \langle \psi_{z_{Q_B}}| \right)^{\otimes k} p(z_{Q_B})  \right].
    \end{align}
Then we note that we can write $|\Phi\rangle = U |Q_0\rangle$ where $U$ is a charge-conserving unitary and $|Q_0\rangle$ is any fixed, reference state with charge $Q_0$, and the Haar averaging of the states comes from Haar averaging the unitaries $U$. 
Because of the left invariance of the Haar measure, we can send $U \to (U_A \otimes \mathbb{I}_B) U$ where $U_A$ is {\it any} charge-conserving unitary on $A$, and all expected values remain invariant. But under this transformation  $p(z_{Q_B})$ is invariant while $|\psi_{z_{Q_B}}\rangle \langle \psi_{z_{Q_B}}| \to U_A|\psi_{z_{Q_B}}\rangle \langle \psi_{z_{Q_B}}| U_A^\dagger  $, so the LHS equals
\begin{align}
        \sum_{Q_B} \sum_{z_{Q_B}} \mathop{\mathbb{E}}_{\Phi \sim \text{Haar}(\mathcal{H}_{Q_0})} \left[  \left(U_A |\psi_{z_{Q_B} }\rangle \langle \psi_{z_{Q_B}}| U_A^\dagger \right)^{\otimes k} p(z_{Q_B})  \right].
    \end{align}
 We are also free to average over $U_A$ in any fashion without changing the expected value; a convenient averaging is to also {\it Haar average} over $U_A$. Then we get 
 \begin{align}
        & \sum_{Q_B} \sum_{z_{Q_B}} \mathop{\mathbb{E}}_{\Phi \sim \text{Haar}(\mathcal{H}_{Q_0})} \left[  \left(U_A |\psi_{z_{Q_B} }\rangle \langle \psi_{z_{Q_B}}| U_A^\dagger \right)^{\otimes k} p(z_{Q_B})  \right] \nonumber \\
        = & \sum_{Q_B} \sum_{z_{Q_B}} \mathop{\mathbb{E}}_{\Phi \sim \text{Haar}(\mathcal{H}_{Q_0})} \left[ \mathbb{E}_{U_A \sim \text{Haar}(\mathcal{H}_{Q_A})}\left[  \left(U_A |\psi_{z_{Q_B} }\rangle \langle \psi_{z_{Q_B}}| U_A^\dagger  \right)^{\otimes k} \right] p(z_{Q_B})  \right] \nonumber \\
        = & \sum_{Q_B} \sum_{z_{Q_B}} \mathop{\mathbb{E}}_{\Phi \sim \text{Haar}(\mathcal{H}_{Q_0})} \left[ \rho^{(k)}_{\text{Haar},A,Q_A} p(z_{Q_B})  \right] \nonumber \\
        = & \sum_{Q_B}  \rho^{(k)}_{\text{Haar},A,Q_A} \sum_{z_{Q_B}} \mathop{\mathbb{E}}_{\Phi \sim \text{Haar}(\mathcal{H}_{Q_0})} \left[  p(z_{Q_B})  \right] \nonumber \\
        = & \sum_{Q_B}  \rho^{(k)}_{\text{Haar},A,Q_A} \left( \frac{d_{Q_A} d_{Q_B}}{d_{Q_0}}\right) \nonumber \\
        = & \bigoplus_{Q_A} \pi(Q_A|Q_0)  \rho^{(k)}_{\text{Haar},A,Q_A}.
    \end{align}
    Above, we used that the Haar average over $U_A$ of $k$ copies of a pure state with definite charge $Q_A$ yields $\rho^{(k)}_{\text{Haar},A,Q_A}$, and also  that the sum over $Q_B$ is equivalent to the sum over $Q_A$.
    
    Having shown the assertion \ref{eq:avg_haar_gen}, we proceed with the second part by bounding the deviation of $k^{th}$ moment of the ensemble $\mathcal{E}_{A}$ obtained from a single instance of a Haar random symmetric $|\Phi\rangle$, from the average behavior. To that end, we will use a standard  concentration of measure lemma, which is~\cite{ledoux_concentration_2001}
    \begin{lemma}[Levy]\label{lemma:levy}
            Let $f: \mathbb{S}^{2d - 1} \rightarrow \mathbb{R}$ be Lipschitz continuous i.e.~satisfying $|f(v) - f(w)| \leq \eta \norm{v-w}_2$ for some Liftshitz constant $\eta < \infty$. Then, for any $\delta \geq 0$, we have 
        \begin{equation}
            \text{Prob}_{\Phi\sim \text{Haar}(\mathcal{H})} [|f(\Phi) - \mathbb{E}_{\Psi\sim \text{Haar}(\mathcal{H})}[f(\Psi)]| \geq \delta] \leq 2\exp \left( -\frac{2d\delta^2}{9\pi^3 \eta^2} \right).
        \end{equation}
    \end{lemma}
    
 Define the function $f$,
    \begin{equation}
        f(|\Phi\rangle) = \sum_z \frac{(|\Tilde{\psi}_z\rangle\langle\Tilde{\psi}_z|)^{\otimes k}}{\langle\Tilde{\psi}_z|\Tilde{\psi}_z\rangle^{k-1}}.
    \end{equation}
  We can break the sum over outcomes $z$ again as
    \begin{align}
        f(|\Phi\rangle) = \sum_{Q_B} \sum_{z_{Q_B}} \frac{(|\Tilde{\psi}_{z_{Q_B}} \rangle\langle\Tilde{\psi}_{z_{Q_B}} |)^{\otimes k}}{\langle\Tilde{\psi}_{z_{Q_B}} |\Tilde{\psi}_{z_{Q_B}} \rangle^{k-1}} =: \sum_{Q_B} f_{Q_B} (|\Phi\rangle).
    \end{align}
    Thus consider the function $f_{Q_B, ij} (|\Phi\rangle): \mathbb{S}^{2d_{Q_0} - 1} \rightarrow \mathbb{R}$ as
    \begin{align}
        f_{Q_B, ij} (|\Phi\rangle) = \langle i| \left( \sum_{z_{Q_B}} \frac{(|\Tilde{\psi}_{z_{Q_B}}\rangle\langle\Tilde{\psi}_{z_{Q_B}}|)^{\otimes k}}{\langle\Tilde{\psi}_{z_{Q_B}}|\Tilde{\psi}_{z_{Q_B}}\rangle^{k-1}} \right) |j\rangle
    \end{align}
    where $|i\rangle = \otimes_{l=1}^k |i^{(l)}\rangle$ and $|i^{(l)}\rangle$ is a basis element of $\mathcal{H}^{Q_A}$. From Lemma 2 of Ref.~\cite{cotler_emergent_2023}, we have that a Lipschitz constant for $f_{Q_B, ij}$ is $\eta = 2(2k-1)$.

    Now we apply Levy's lemma to $f_{Q_B, ij}$ which gives
    \begin{align}
        \text{Prob}_{\Phi \sim \text{Haar}(\mathcal{H}_{Q_0})} [|f_{Q_B, ij}(\Phi) - \mathbb{E}_{\Psi_Q \sim \text{Haar}(\mathcal{H}_{Q_0})}[f_{Q_B, ij}(\Psi)]| \geq \epsilon] \leq 2\exp \left( -\frac{2d_{Q_0} \epsilon^2}{9\pi^3 4(2k-1)^2} \right).
    \end{align}
    A union bound on entries $(i,j)$ and re-scaling $\epsilon \rightarrow \epsilon/d_{Q_A}^{2k}$ gives
    \begin{align}
        \mathop{\text{Prob}}_{\Phi \sim \text{Haar}(\mathcal{H}_{Q_0})} &\left[ \left|f_{Q_B, ij}(\Phi) - \mathop{\mathbb{E}}_{\Psi \sim \text{Haar}(\mathcal{H}_{Q_0})}[f_{Q_B, ij}(\Psi)] \right| \geq \frac{\epsilon}{d_{Q_A}^{2k}}, \text{ for some } i,j \right] \nonumber \\
        &\leq 2d_{Q_A}^{2k} \exp \left( -\frac{d_{Q_0} \epsilon^2}{18\pi^3 (2k-1)^2 d_{Q_A}^{4k}} \right).
    \end{align}
 Define the entrywise (e) norm as $\|{A}\|_{e,1} \equiv  \sum_{i,j} |A_{ij}|$ and note that,
 \begin{align}
     \text{Prob} [|A_{ij}| \geq \epsilon/N, \text{ for some } (i,j)] &= 1 - \text{Prob} [|A_{ij}| < \epsilon/N, \text{ for all } (i,j)] \nonumber \\
     &\geq 1 - \text{Prob} [ \norm{A}_{e,1}  < \epsilon] \nonumber \\
     &= \text{Prob} [ \norm{A}_{e,1}  \geq \epsilon],
 \end{align}
    which gives
    \begin{align}
        \mathop{\text{Prob}}_{\Phi \sim \text{Haar}(\mathcal{H}_{Q_0})} \left[ \norm{f_{Q_B}(\Phi) - \mathop{\mathbb{E}}_{\Psi \sim \text{Haar}(\mathcal{H}_{Q_0})}[f_{Q_B}(\Psi)]}_{e,1}  \geq \epsilon \right] 
        \leq 2d_{Q_A}^{2k} \exp \left( -\frac{d_{Q_0} \epsilon^2}{18\pi^3 (2k-1)^2 d_{Q_A}^{4k}}. \right)
    \end{align}
    Since $\norm{A}_{e,1} \geq \norm{A}_{1}$, we have
    \begin{align}
        \text{Prob} [\norm{A}_1 \geq \epsilon] \leq \text{Prob} [\norm{A}_{e,1} \geq \epsilon]
    \end{align}
    using which we get (substituting the functional form of $f_{Q_B}$)
    \begin{align}
        \mathop{\text{Prob}}_{\Phi \sim \text{Haar}(\mathcal{H}_{Q_0})} \left[ \norm{ \sum_{z_{Q_B}} \frac{(|\Tilde{\psi}_{z_{Q_B}} \rangle\langle\Tilde{\psi}_{z_{Q_B}} |)^{\otimes k}}{\langle\Tilde{\psi}_{z_{Q_B}} |\Tilde{\psi}_{z_{Q_B}} \rangle^{k-1}} - \pi(Q_A|Q_0)\rho^{(k)}_{\text{Haar},A,Q_A} }_{1}  \geq \epsilon \right] 
        \leq 2d_{Q_A}^{2k} \exp \left( -\frac{d_{Q_0} \epsilon^2}{18\pi^3 (2k-1)^2 d_{Q_A}^{4k}} \right).
    \end{align}
    Next, we again apply a union bound for the different charge sectors and re-scale $\epsilon \rightarrow \epsilon/(N_A+1)$. Then
    \begin{align}
        \mathop{\text{Prob}}_{\Phi \sim \text{Haar}(\mathcal{H}_{Q_0})} &\left[ \norm{ \sum_{z_{Q_B}} \frac{(|\Tilde{\psi}_{z_{Q_B}} \rangle\langle\Tilde{\psi}_{z_{Q_B}} |)^{\otimes k}}{\langle\Tilde{\psi}_{z_{Q_B}} |\Tilde{\psi}_{z_{Q_B}} \rangle^{k-1}} - \pi(Q_A|Q_0)\rho^{(k)}_{\text{Haar},A,Q_A} }_{1}  \geq \frac{\epsilon}{(N_A+1)}, \text{ for some charge sector } Q_A \right] \nonumber  \\
        &\leq \sum_{Q_A} 2d_{Q_A}^{2k} \exp \left( -\frac{d_{Q_0} \epsilon^2}{18\pi^3 (N_A+1)^2 (2k-1)^2 d_{Q_A}^{4k}} \right).
    \end{align}
    For combining probabilities from different charge sectors we note that,
    \begin{align}
        \text{Prob}\left[\Delta^{(k)}_{Q_A} \geq \frac{\epsilon}{(N_A+1)}, \text{ for some sector } Q_A \right] &= 1- \text{Prob}\left[\Delta^{(k)}_{Q_A} < \frac{\epsilon}{(N_A+1)}, \text{ for all  sectors } Q_A \right] \nonumber \\
        &\geq  1- \text{Prob}\left[\Delta^{(k)} < \epsilon \right] \nonumber  \\
        &= \text{Prob}\left[\Delta^{(k)} \geq \epsilon \right].
    \end{align}
    Thus we finally get
    \begin{align}
        \mathop{\text{Prob}}_{\Phi \sim \text{Haar}(\mathcal{H}_{Q_0})} &\left[ \norm{ \sum_{Q_B}\sum_{z_{Q_B}} \frac{(|\Tilde{\psi}_{z_{Q_B}} \rangle\langle\Tilde{\psi}_{z_{Q_B}} |)^{\otimes k}}{\langle\Tilde{\psi}_{z_{Q_B}} |\Tilde{\psi}_{z_{Q_B}} \rangle^{k-1}} - \mathop{\bigoplus}_{Q_A} \pi(Q_A|Q_0)\rho^{(k)}_{\text{Haar},A,Q_A} }_{1}  \geq \epsilon \right] \nonumber \\
        &\leq \sum_{Q_A} 2d_{Q_A}^{2k} \exp \left( -\frac{d_{Q_0} \epsilon^2}{18\pi^3 (N_A+1)^2 (2k-1)^2 d_{Q_A}^{4k}} \right).
    \end{align}
    We can simplify the above expression, noting that $d_{Q_A} \leq d_A = 2^{N_A}$. Also, $d_{Q_A} = \binom{N_A}{Q_A}$ and so $\sum_{Q_A} d_{Q_A}^{2k} \leq C_k \sum_{Q_A} 2^{2N_A k}/N_A^k \leq C_k 2^{2N_A k}/N_A^{k-1}$ using known bounds on the central binomial coefficient which dominates all other binomial coefficients. $C_k$ is a constant depending on $k$ but not $N_A$.  
    
    So we get
    \begin{align}
        \mathop{\text{Prob}}_{\Phi \sim \text{Haar}(\mathcal{H}_{Q_0})} \left[ \norm{ \sum_z \frac{\left(|\Tilde{\psi}_z\rangle\langle\Tilde{\psi}_z|\right)^{\otimes k}}{\langle\Tilde{\psi}_z|\Tilde{\psi}_z\rangle^{k-1}} - \mathop{\bigoplus}_{Q_A} \pi(Q_A|Q_0) \rho^{(k)}_{{\rm Haar},A,Q_A} }_{1}  \geq \epsilon \right]  \nonumber \\
        \leq 2 C_k \frac{2^{2k N_A}}{N_A^{k-1}} \exp \left( -\frac{d_{Q_0} \epsilon^2}{18\pi^3 (N_A +1)^2 (2k-1)^2 d_{A}^{4k}} \right).
    \end{align}
    Thus, if
    \begin{equation}
        d_{Q_0} \geq \frac{18\pi^3 (N_A +1)^2 (2k-1)^2 d_{A}^{4k}}{\epsilon^2}\left(\log{\left(\frac{2}{\delta}\right)} + \log{\left( 2 C_k d_A^{2k} N_A^{1-k} \right)} \right)
    \end{equation}
    then  the $k^{th}$ moment of the projected ensemble $\mathcal{E}_{A}$ is $\epsilon$-close to $\mathop{\bigoplus}_{Q_A} p(Q_A)\rho^{(k)}_{{\rm Haar},A,Q_A}$ with probability at least $1-\delta$, as claimed.
    
    Lastly, using Stirling's approximation
    \begin{align}
        \log{d_{Q_0}} &= \log\left[\binom{N}{Q}\right]  \nonumber \\
        & \approx  N\log N - N - \left( (\sigma N)\log(\sigma N) - (\sigma N) + (N-\sigma N)\log (N-\sigma N) - (N-\sigma N) \right) \nonumber \\
        &= N \left( -\sigma \log (\sigma) - (1-\sigma) \log (1-\sigma) \right) \nonumber \\
        & = N H(\sigma).
    \end{align}
    The  requirement  on $d_{Q_0}$ can be seen to be satisfied if the number of bath qubits obeys
 \begin{align}
    N_B H(\sigma) = \Omega\left(k N_A + \log\left({\epsilon}^{-1} \right) + \log \log\left({\delta}^{-1} \right) \right).
\end{align}
$\blacksquare$

\section{Replica calculation of moments of the projected ensemble}\label{app:ansatz}

In this appendix we present analytical calculations for the $k$-th moment operators of the projected ensemble at late times under chaotic $U(1)$-symmetric dynamics.
First we derive a general, implicit form of the moment operator $\rho^{(k)}$ averaged over random $U(1)$-symmetric transformations of the generator state in Sec.~\ref{app:general_form_rhok}. Then we make progress, within a `replica limit' approach, for two special cases of interest:
(i) when the the measurement basis is `maximally charge-revealing' ($z$ basis) in Sec.~\ref{app:charge_revealing}, and 
(ii) when the generator state is symmetric and the measurement basis is `charge non-revealing' ($x-y$ plane basis) in Sec.~\ref{app:charge_non_revealing}. From the expressions of the averaged moment operators $\rho^{(k)}$ we are then able to reverse-engineer the form of the projected ensemble itself. 
Our approach is tractable only in the limit of $N, N_A\to\infty$ taken {\it before} the replica limit. We estimate the approximation error induced by this incorrect order of limits in Sec.~\ref{app:approximation_error}; the leading error term suggests our results should be accurate for $N_A \gg \ln(k)$. 
Numerical evidence presented in Sec.~\ref{sec:numerics} shows that, in practice, our results are correct in the thermodynamic limit $N\to \infty$ even for the smallest value of $N_A = 2$. 

\subsection{Setup}
We consider a generator ensemble $\mathcal{F}$ of many-body pure state $\ket{\Psi} = \sum_{Q=0}^N \sqrt{p_{\rm in} (Q)} \ket{\Phi_Q}$, with $p_{\rm in} (Q)$ the distribution of charge. We perform projective measurements on the bath in an as-of-yet unspecified basis $\{\ket{\nu }_B\}_{\nu =1}^{d_B}$ of orthonormal pure states (we will later take these to be the $X$ or $Z$ basis). The $k$-th moment of the projected ensemble then reads 
\begin{equation}
    \rho^{(k)}_{\mathcal E(\Psi)} = \sum_{\nu =1}^{N_B} p(\nu) (\ketbra{\psi_\nu })^{\otimes k}, 
    \qquad 
    p(\nu ) = \bra{\Psi} (\mathbb{I}_A\otimes \ketbra{\nu }_B) \ket{\Psi},
    \qquad
    \ket{\psi_\nu }_A = {}_B\!\braket{\nu }{\Psi}_{AB} / \sqrt{p(\nu )}.
\end{equation}
Our ansatz is that general dynamics starting from a state with charge distribution $p(Q)$ should converge to a projected ensemble whose moments are given by 
\begin{equation}
    \rho^{(k)}_{\rm target} \equiv \mathop{\mathbb{E}}_{\{ \Phi_Q\sim \text{Haar}(\mathcal{H}_Q) \} } \left[ \rho^{(k)}_{\mathcal{E}(\Psi)} \right].
    \label{eq:app_pe_ansatz}
\end{equation}
This is the average moment operator across all states $\ket{\Psi}$ with the given charge distribution, where the state in each charge sector is taken to be Haar-random; the content of our conjecture is that this should be the moment operator not just on average, but for {\it almost all individual instances} of $\ket{\Psi} \in \mathcal{F}$, in a suitable thermodynamic limit. 

\subsection{General structure of the moment operators \label{app:general_form_rhok}}

Here we turn Eq.~\eqref{eq:app_pe_ansatz} into a universal form that explicitly depends only on the charge distribution $p_{\rm in} (Q)$ and on the measurement basis $\{ \ket{\nu} \}$. 

Given the generator state $\ket{\Psi}$, we may in general define a joint distribution over measurement outcomes $\nu$ and values of the charge on subsystem $A$, $Q_A$, by 
\begin{equation}
    p(Q_A, \nu) = \bra{\Psi} (\hat{\Pi}_{Q_A} \otimes \ketbra{\nu }) \ket{\Psi} = p(\nu) \bra{\psi_\nu } \hat{\Pi}_{Q_A} \ket{\psi_\nu }. \label{eq:app_joint_dist}
\end{equation}
Note that here $p(\nu)$ depends on the generator state $\ket{\Psi}$, i.e., it is not yet averaged over $\mathcal{F}$. 
This immediately implies the following decomposition for the projected states,
\begin{equation}
    \ket{\psi_\nu} = \sum_{Q_A} \sqrt{p(Q_A|\nu)} \ket{\phi_{\nu,Q_A}},
\end{equation}
where the states $\ket{\phi_{\nu,Q_A}}$ belong to charge sector $Q_A$ of subsystem $A$, and $p(Q_A|\nu) = p(Q_A,\nu) / p(\nu)$ is the conditional distribution of charge $Q_A$ given outcome $\nu$ on the bath, derived from the joint distribution in Eq.~\eqref{eq:app_joint_dist}. 

Plugging this expansion of the projected states $\ket{\psi_\nu}$ into the expression for the moment operator yields
\begin{align}
    \rho^{(k)}_{\mathcal E} 
    & = \sum_\nu p(\nu) \left( \sum_{Q_A} \sqrt{p(Q_A|\nu)} \ket{\phi_{\nu,Q_A}} \right)^{\otimes k} \left( \sum_{Q_A} \sqrt{p(Q_A|\nu)} \bra{\phi_{\nu,Q_A}} \right)^{\otimes k} \nonumber \\
    & = \sum_\nu p(\nu) \sum_{\mathbf{Q}_A, \mathbf{Q}_A'}
    \prod_{i=1}^k \sqrt{p(Q_{A,i} |\nu) p(Q_{A,i}'|\nu)} \bigotimes_{i=1}^k \ketbra{\phi_{\nu,Q_{A,i}}}{\phi_{\nu,Q_{A,i}'}},
\end{align}
where $\mathbf{Q}_A,\mathbf{Q}_A'\in \{0,\dots N_A\}^k$ are strings of values of the charge in each Hilbert space replica.

Our ansatz Eq.~\eqref{eq:app_pe_ansatz} is that the target ensemble should be invariant under $U(1)$-symmetric unitary transformations. A special case of such transformations is one that acts nontrivially only on $A$: $U = V_A \otimes \mathbb{I}_B$, where $V$ conserves the value of $Q_A$. Given unitary invariance of the Haar measure, we can write 
\begin{equation}
    \mathop{\mathbb{E}}_{U \sim U(1)\text{-Haar}(\mathcal{H}_{AB})} (\cdots)
    = \mathop{\mathbb{E}}_{U \sim U(1)\text{-Haar}(\mathcal{H}_{AB})} \left[
    \mathop{\mathbb{E}}_{V \sim U(1)\text{-Haar}(\mathcal{H}_{A})} (\cdots) \right],
\end{equation}
with `$U(1)$-Haar' denoting the Haar measure over unitary matrices that preserve the $U(1)$-symmetry (this is a compact Lie group and so has a Haar measure).
Then we can carry out the average over $V$ (a random $U(1)$-symmetric unitary on $A$ only) first. 
This has the effect of replacing the vectors $\ket{\phi_{b,Q_A}}$ by Haar-random vectors in their respective sector, which we denote by $\ket{\phi_{Q_A}}$:
\begin{align}
    \rho^{(k)}_{\rm target} 
    & = \sum_\nu \sum_{\mathbf{Q}_A, \mathbf{Q}_A'}
    \mathop{\mathbb{E}}_{\{ \Phi_Q \sim \text{Haar}(\mathcal{H}_{Q}) \} } \left[
    p(\nu) \prod_{i=1}^k \sqrt{p(Q_{A,i} |\nu) p(Q_{A,i}'|\nu)} 
    \right] 
    \mathop{\mathbb{E}}_{\{ \phi_{Q_A} \sim \text{Haar}(\mathcal{H}_{Q_A}) \} } \left[
    \bigotimes_{i=1}^k \ketbra{\phi_{Q_{A,i}}}{\phi_{Q_{A,i}'}} \right]
    \label{eq:app_targetrho_factor}
\end{align}
The average thus breaks up into a numerical factor and an operator factor; this is analogous to the observation used in proving Theorem~\ref{thm:1} on the statistical independence between $p(z)$ and $\ket{\psi_z}$.

In carrying out the Haar average over states $\{ \ket{\phi_{Q_A}}\}$, we note that the average vanishes due to random dephasing unless the charge strings $\mathbf{Q}_A$, $\mathbf{Q}_A'$ have the same content. 
We can capture this constraint by introducing the notion of a ``type'':
\begin{equation}
    {\rm type}(\mathbf Q) = \vec{T} = (T_0, T_1,\dots T_N):
    \qquad 
    T_q = |\{i: Q_i=q \}|.
    \label{eq:app_type_def}
\end{equation}
In words, the type of $\mathbf Q$ is the list of multiplicities of elements in $\mathbf Q$ ($T_q$ is the number of times that the value $q$ appears in the sequence $\mathbf{Q}$). We denote the set of possible types for lists $\mathbf{Q}$ of length $k$, where each entry is valued in $\{0,\dots N\}$, as $\mathcal{T}(k,N)$.
Then, to avoid a cancellation in Eq.~\eqref{eq:app_targetrho_factor}, we must have ${\rm type} (\mathbf Q) = {\rm type}(\mathbf Q')$. 
The sum over $\mathbf{Q}, \mathbf{Q}'$ can thus be replaced by a sum over ``types'' $\vec{T} \in \mathcal{T}(k,N)$ and a separate sum over permutations $\sigma,\tau\in S_k$. The idea is that an ordered list $\mathbf{Q}$ is fully specified by two things: its unsorted content (the type $\vec T$) and its ordering (a permutation $\sigma \in S_k$ of its elements). 
This decomposition, however, has some redundancy---specifically whenever $\mathbf{Q}$ is degenerate, i.e. when $T_q > 1$ for some $q$. In general, by summing over $\sigma\in S_k$ we count the same state a number of times $\prod_q (T_q!)$.

With this in mind, we can rewrite   Eq.~\eqref{eq:app_targetrho_factor} as 
\begin{align}
    \rho^{(k)}_{\rm target} 
    & = \sum_{\nu} \sum_{\vec{T} \in \mathcal{T}(k,N)} 
    \mathop{\mathbb{E}}_{\{ \Phi_Q \sim \text{Haar}(\mathcal{H}_{Q}) \} } \left[
    p(\nu) \prod_{Q_A=0}^{N_A} p(Q_A |\nu)^{T_{Q_A}}
    \right] 
    \prod_{Q_A=0}^{N_A} \frac{1}{(T_{Q_A}!)^2}
    \sum_{\sigma,\tau \in S_k} 
    \hat{\sigma} \left( \bigotimes_{Q_A=0}^{N_A} \rho_{{\rm Haar},A,Q_A}^{(T_{Q_A})} \right) \hat{\tau} \nonumber \\
    & = \sum_{\nu} \sum_{\vec{T} \in \mathcal{T}(k,N)} 
    \mathop{\mathbb{E}}_{\{ \Phi_Q \sim \text{Haar}(\mathcal{H}_{Q}) \} } \left[
    p(\nu) \prod_{Q_A=0}^{N_A} p(Q_A | \nu)^{T_{Q_A}}
    \right] 
    \binom{k}{\vec{T}}^2 
    \hat{\Pi}_{\rm sym}^{(k)} \left( \bigotimes_{Q_A=0}^{N_A} \rho_{{\rm Haar},A,Q_A}^{(T_{Q_A})} \right) \hat{\Pi}_{\rm sym}^{(k)}.
    \label{eq:app_rhotarget_moment}
\end{align}
Here, $\hat{\sigma}$ is the replica permutation operator associated to permutation $\sigma \in S_k$, $\rho_{{\rm Haar},A,Q_A}^{(m)}$ is the $m$-th moment operator of the Haar ensemble on charge sector $Q_A$ in subsystem $A$, and in the second line we have introduced the projector on the symmetric subspace and the multinomial coefficient,
\begin{equation}
    \hat{\Pi}_{\rm sym} = \frac{1}{k!} \sum_{\sigma \in S_k} \hat{\sigma}, 
    \qquad 
    \binom{k}{\vec{T}} = \frac{k!}{T_0! \cdots T_N!}
\end{equation}
(note that $\sum_{Q_A=0}^{N_A} T_{Q_A} = k$, so $\vec{T}$ is a partition of $k$). 

It remains to calculate the numerical coefficients
\begin{align}
    f_p(\vec T, \nu)  
    & = \mathop{\mathbb{E}}_{\{ \Phi_Q\sim {\rm Haar}(\mathcal{H}_Q) \} } 
    \left[ 
    p(\nu) \prod_{Q_A} p(Q_A|\nu)^{T_{Q_A}}
    \right].
    \label{eq:fp_def}
\end{align}
This calculation is not straightfoward due to the nontrivial dependence on the initial charge-sector states $\{\ket{\Phi_Q}\}$ to be averaged over. However, it can be carried out in an approximate way in several cases of interest, as we will show next.

\subsection{Charge-revealing measurements on general states}\label{app:charge_revealing}

Here we focus on the case of $Z$-basis measurements ($\nu\mapsto z$) on arbitrary initial states, i.e., we place no restrictions on $p(Q)$. We will compute the $f_p(\vec{T},z)$ coefficients from Eq.~\eqref{eq:fp_def} within a replica limit approach, then perform an approximation of large Hilbert space dimension to express the resulting moment operator in a simpler way, and deduce a form of the projected ensemble compatible with such moment operators. The result is the generalized Scrooge ensemble (GSE), Eq.~\eqref{eq:gen_scrooge_compressed}. 

\subsubsection{Calculation of $f_p$ coefficients}
We have 
\begin{align}
    f_p(\vec T, z)  
    & = \mathop{\mathbb{E}}_{\{ \Phi_Q\sim {\rm Haar}(\mathcal{H}_Q) \} } 
    \left[ p(z) \prod_{Q_A} p(Q_A|z)^{T_{Q_A}} \right] 
    = \mathop{\mathbb{E}}_{\{ \Phi_Q\sim {\rm Haar}(\mathcal{H}_Q) \} } 
    \left[ \left(\sum_{Q_A} p(Q_A,z) \right)^{1-k} \prod_{Q_A} p(Q_A,z)^{T_{Q_A}} \right],
\end{align}
where we expressed everything in terms of the joint distribution $p(Q_A,z)$, Eq.~\eqref{eq:app_joint_dist}.
To make progress, we adopt a replica limit approach: we set $1-k\mapsto n$, compute the integral for positive integer $n$, then take the limit $n\to 1-k$ in the result. 

By expanding the $n$-th power of the multinomial, we obtain
\begin{align}
    f_p^{(n)}(\vec{T},z)
    & = \sum_{\vec{T}'\in \mathcal{T}(n,N)} \binom{n}{\vec{T}'}  
    \mathop{\mathbb{E}}_{\{ \Phi_Q\sim {\rm Haar}(\mathcal{H}_Q) \} } 
    \prod_{Q_A} p(Q_A,z)^{T_{Q_A} + T'_{Q_A} }.
\end{align}
Now we note that 
\begin{equation} 
p(Q_A,z) = \bra{\Psi} (\hat{\Pi}_{Q_A} \otimes \ketbra{z}) \ket{\Psi} 
= p_{\rm in} (Q_A+Q_B(z)) \bra{\Phi_{Q_A+Q_B(z)}} (\hat{\Pi}_{Q_A} \otimes \ketbra{z})  \ket{\Phi_{Q_A+Q_B(z)}},
\end{equation}
i.e., the probability $p(Q_A,z)$ depends only on one charge component of $\ket{\Psi}$, namely $Q = Q_A + Q_B(z)$, with $Q_B(z) = \sum_{i=1}^{N_B} z_i$ the total charge of bit string $z$ on $B$. 
This allows us to factor the average as 
\begin{align}
    f_p^{(n)}(\vec{T},z)
    & = \sum_{\vec{T}'\in \mathcal{T}(n,N)} \binom{n}{\vec{T}'}  
    \prod_{Q_A} \mathop{\mathbb{E}}_{\Phi_{Q_A+Q_B(z)}} \left[ p(Q_A,z)^{T_{Q_A} + T'_{Q_A}} \right].
\end{align}

The calculation thus reduces to the Haar average 
\begin{equation}
\mathop{\mathbb{E}}_{\Phi_Q\sim {\rm Haar}(\mathcal{H}_Q)} \bra{\Phi_Q} (\mathbb{I}_A \otimes \ketbra{z}_B) \ket{\Phi_Q}^\ell
= \Tr \left( (\mathbb{I}_A \otimes \ketbra{z}_B)^{\otimes \ell} \rho_{{\rm Haar},AB,Q}^{(\ell)} \right)
\end{equation}
with $\rho_{{\rm Haar},AB,Q}^{(\ell)}$ the moment of the Haar ensemble on the charge sector $Q$ for the whole system $AB$, for arbitrary integer $\ell$ (to be set to $T_Q + T'_Q$ later). Note the projector $\hat{\Pi}_{Q_A}$ on $A$ is redundant as both $Q$ and $Q_B$ are fixed. 
Writing out $\rho_{{\rm Haar},AB,Q}^{(\ell)}$ in terms of permutations, we have
\begin{align}
    \mathop{\mathbb{E}}_{\Phi_Q\sim {\rm Haar}(\mathcal{H}_Q)} \bra{\Phi_Q} (\mathbb{I}_A \otimes \ketbra{z}_B) \ket{\Phi_Q}^\ell    
    & = \frac{[\binom{N}{Q}-1]!}{[\binom{N}{Q}-1+\ell]!} \sum_{\sigma \in S_k} \Tr \left( (\mathbb{I}_A \otimes \ketbra{z}_B)^{\otimes \ell} \hat{\sigma}_{AB,Q} \right).
\end{align}
with $\hat{\sigma}_{AB,Q}$ the permutation operator acting on charge sector $Q$ of the total system $AB$. 
We have $\Tr \left( (\mathbb{I}_A \otimes \ketbra{z}_B)^{\otimes \ell} \hat{\sigma}_{AB,Q} \right) = \Tr_A \bra{z}^{\otimes \ell}_B \hat{\sigma}_{AB,Q} \ket{z}_B^{\otimes \ell} = \hat{\sigma}_{A,Q_A}$, i.e. the replica permutation acting on charge sector $Q_A = Q-Q_B(z)$ on subsystem $A$ only. This allows us to resum the expression exactly, obtaining
\begin{align}
    \mathop{\mathbb{E}}_{\Phi_Q\sim {\rm Haar}(\mathcal{H}_Q)} \bra{\Phi_Q} (\mathbb{I}_A \otimes \ketbra{z}_B) \ket{\Phi_Q}^\ell    
    & = \frac{[\binom{N}{Q}-1]!}{[\binom{N}{Q}-1+\ell]!}  \frac{[\binom{N_A}{Q_A}-1+\ell]!}{[\binom{N_A}{Q_A}-1]!},
\end{align}
with $Q = Q_A + Q_B(z)$. 
We conclude that 
\begin{align}
    f_p^{(n)}(\vec{T},z)
    & = \sum_{\vec{T}'\in \mathcal{T}(n,N)} \binom{n}{\vec{T}'} 
    \prod_{Q_A=0}^{N_A} \left[ 
    p_{\rm in} (Q)^{T_{Q_A} + T'_{Q_A}} 
    \frac{[\binom{N}{Q}-1]!}{[\binom{N_A}{Q_A}-1]!} 
    \frac{[\binom{N_A}{Q_A}-1+T_{Q_A}+T'_{Q_A}]!}{[\binom{N}{Q}-1+T_{Q_A} + T'_{Q_A}]!} \right]_{Q = Q_A + Q_B(z)} . \label{eq:app_fpn_exact}
\end{align}

\subsubsection{Approximation and replica limit}

Unfortunately, the replica limit $n\to 1-k$ in Eq.~\eqref{eq:app_fpn_exact} does not appear to be tractable. To make progress, we make a further approximation: we assume that $\binom{N}{Q}, \binom{N_A}{Q_A} \gg 1$, and approximate the ratios of factorials as 
\begin{equation}    
\frac{[\binom{N}{Q}-1]!}{[\binom{N}{Q}-1+\ell]!} \frac{[\binom{N_A}{Q_A}-1+\ell]!}{[\binom{N_A}{Q_A}-1]!} 
\approx \left[ \binom{N_A}{Q_A} \binom{N}{Q}^{-1} \right]^\ell. \label{eq:app_factorial_approx}
\end{equation}
This approximation neglects terms of order $\ell^2 / \binom{N_A}{Q_A}$, $\ell^2 / \binom{N}{Q}$. 
Since the replica limit $n \to 1-k$ in principle depends on all integer values of the replica number $n$, and $\ell = T_{Q_A} + T_{Q_A}'$ can take on values up to the total number of replicas $n+k$ (which is unbounded), this approximation is not rigorously justified. 
In effect, we are taking the $N,N_A\to\infty$ limit before the replica limit $n\to 1-k$, where in reality the opposite order should be followed.
For this reason, the following analysis is non-rigorous and serves only to produce a conjecture for the form of the target ensemble, which has to then be checked numerically or through other means.

Within the approximation in Eq.~\eqref{eq:app_factorial_approx}, we get
\begin{align}
    f_p^{(n)}(\vec T, z)
    & \simeq \sum_{\vec T' \in \mathcal{T}(n,N)} \binom{n}{\vec T'} \prod_{Q_A} \left[ p_{\rm in} (Q) \binom{N_A}{Q_A} \binom{N}{Q}^{-1} \right]^{T_{Q_A} + T'_{Q_A}} \nonumber \\
    & = \left[ \sum_{Q_A} p_{\rm in} (Q)  \binom{N_A}{Q_A} \binom{N}{Q}^{-1} \right]^n 
    \prod_{Q_A} \left[ p_{\rm in} (Q)  \binom{N_A}{Q_A} \binom{N}{Q}^{-1} \right]^{T_{Q_A}},
\end{align}
using $Q$ as shorthand for $Q_A + Q_B(z)$.
Now we can straightforwardly take the $n \to 1-k$ replica limit:
\begin{align}
    f_p(\vec T, z)
    & \simeq  \left( \sum_{Q_A} p_{\rm in} (Q)  \binom{N_A}{Q_A} \binom{N}{Q}^{-1} \right)
    \prod_{Q_A} \left[ \frac{ p_{\rm in} (Q)  \binom{N_A}{Q_A} \binom{N}{Q}^{-1}}{\sum_{Q_A'} p_{\rm in} (Q') \binom{N_A}{Q_A'} \binom{N}{Q'}^{-1} } \right]^{T_{Q_A}},
\end{align}
Finally, recalling the the probability distributions $p(\nu)$ from Eq.~\eqref{eq:maxrev_pnu} and $p(Q_A|\nu)$ from Eq.~\eqref{eq:maxrev_pqanu}, we obtain
we can rewrite the expression as
\begin{align}
f_p (\vec{T},z) 
    & \simeq p(z) 
    \prod_{Q_A} p(Q_A|z)^{T_{Q_A}}
    = \binom{N_B}{Q_B(z) }^{-1} p(Q_B(z) ) \prod_{Q_A} p(Q_A|z)^{T_{Q_A}}
    \label{eq:app_zbasis_fp}. 
\end{align}

\subsubsection{Reverse-engineering the ensemble from the moment operators}\label{recipe}

Plugging the approximate result Eq.~\eqref{eq:app_zbasis_fp} back into Eq.~\eqref{eq:app_rhotarget_moment} gives
\begin{equation}
    \rho_{\rm target}^{(k)} \simeq 
    \sum_{Q_B} p(Q_B) \hat{\Pi}_{\rm sym}^{(k)} 
    \left( \sum_{\vec{T} \in \mathcal{T}(k,N) } \binom{k}{\vec T}^2 \bigotimes_{Q_A=0}^{N_A} p(Q_A|z)^{T_{Q_A}} \rho_{{\rm Haar},A,Q_A}^{(T_{Q_A})} \right) \hat{\Pi}_{\rm sym}^{(k)}
    \label{eq:simplified_moment}
\end{equation}
(note that Eq.~\eqref{eq:app_zbasis_fp} depends on $z$ only through $Q_B(z)$, so the sum over $z$ cancels out the $\binom{N_B}{Q_B}^{-1}$ prefactor, and $z$ in $p(Q_A|z)$ is any bitstring of charge $Q_B$). 
From these moment operators, it is possible to ``reverse-engineer'' the form of the ensemble. 
Two key observations can guide us:
\begin{enumerate}
    \item The moment operator is in the form 
    $\rho^{(k)}_{\rm target} = \sum_{Q_B} p(Q_B) \rho_{\text{target},Q_B}^{(k)}$, which implies the target ensemble $\mathcal{E}_{\rm target}$ is a stochastic mixture of multiple ensembles $\mathcal{E}_{\text{target},Q_B}$, each one weighted by a probability $p(Q_B)$ (in other words: to sample a state from $\mathcal{E}_{\rm target}$, first sample $Q_B \sim p(Q_B)$, then sample a state from $\mathcal{E}_{\text{target},Q_B}$).
    \item For each value of $Q_B$, the moment operator $\rho_{\text{target},Q_B}^{(k)}$ is a product of Haar moment operators in different charge sectors, with each sector $Q_A$ weighted by a factor of $p(Q_A|z)$. 
\end{enumerate}
Building on these observations, it is possible to formulate a recipe to sample from a target ensemble $\mathcal{E}_{\rm target}$ whose moment operators are as in Eq.~\eqref{eq:simplified_moment}:
\begin{itemize}
    \item[(i)] Draw a value of $Q_B \sim p(Q_B)$;
    \item[(ii)] Draw a collection of Haar-random states $\{\ket{\phi_{Q_A}}\}_{Q_A}$ in each charge sector $Q_A$ of the subsystem;
    \item[(iii)] Output the state 
    \begin{equation}
        \ket{\psi} = \sum_{Q_A} \sqrt{p(Q_A|z)} \ket{\phi_{Q_A}} .
        \label{eq:app_ansatz_ensemble}
    \end{equation}
\end{itemize}

Finally, since our derivation is valid only in the limit of large Hilbert space dimensions $\binom{N_A}{Q_A} \gg 1$, the above recipe can be further simplified and reduced to the GSE, Eq.~\eqref{eq:gen_scrooge_compressed}. 
The output state in Eq.~\eqref{eq:app_ansatz_ensemble} equals 
\begin{align}
\ket{\psi} 
& = \sum_{Q_A} \sqrt{p(Q_A|z) } \ket{\phi_{Q_A}}
= \frac{ \rho_z^{1/2} \ket{\xi} }{\bra{\xi} \rho_z \ket{\xi}^{1/2}},
\end{align}
where
\begin{align}
\ket{\xi} & = 2^{-N_A/2} \sum_{Q_A} \binom{N_A}{Q_A}^{1/2} \ket{\phi_{Q_A}}, \label{eq:app_xi} \\
\rho_z & = \sum_{Q_A=0}^{N_A} p(Q_A|z) \frac{\hat{\Pi}_{Q_A}}{\Tr(\hat{\Pi}_{Q_A})}.  \label{eq:app_rhobar}
\end{align}
In words, $\ket{\xi}$ is a random state with {\it fixed} charge distribution $p(Q_A) = 2^{-N_A} \binom{N_A}{Q_A}$, and $\ket{\psi}$ is a ``$\rho$-distortion'' of $\ket{\xi}$ by the density matrix $\rho_z$.
But in the limit of large charge sectors $\binom{N_A}{Q_A} \gg 1$, the constraint on the charge distribution of $\ket{\xi}$ becomes unimportant: expressing a Haar-random state on $A$ as $\sum_{Q_A} \sqrt{p(Q_A)} \ket{\phi_{Q_A}}$, one would find $p(Q_A) \approx 2^{-N_A} \binom{N_A}{Q_A}$ almost surely. Specifically, we have 
\begin{equation}
\mathop{\mathbb{E}}_{\psi \sim \text{Haar}(\mathcal{H}_A)} \left[\left(p(Q_A)-2^{-N_A}\binom{N_A}{Q_A}\right)^2 \right]
= 2^{-2N_A} \binom{N_A}{Q_A}^2 \left[ \frac{1+ \binom{N_A}{Q_A}^{-1}}{1+2^{-N_A}} -1\right],
\end{equation}
corresponding to a relative error on $p(Q_A)$ of order $\binom{N_A}{Q_A}^{-1/2}$, which we assumed small. 

Thus the state $\ket{\xi}$, Eq.~\eqref{eq:app_xi}, becomes indistinguishable from an unconstrained Haar-random state on the whole Hilbert space in the limit of large $N_A$. 
Since $\rho$-distortion of the Haar ensemble gives the Scrooge ensemble, what we just described is a stochastic mixture of Scrooge ensembles, i.e. a GSE (note also that the factor of $2^{N_A} \bra{\psi}\rho\ket{\psi}$, which modifies the measure in the definition of the Scrooge ensemble, concentrates around 1 exponentially with increasing $N_A$ and so can be neglected in this same limit). The GSE is specified by the distribution $p(z)$ and the density matrices $\{\rho_z \}$, Eq.~\eqref{eq:app_rhobar}, which depend only on the equivalence class $[z] = Q_B(z)$. 


\subsection{Charge-non-revealing measurements on symmetric states}\label{app:charge_non_revealing}

Here we focus on the case of $X$-basis measurements ($\nu \mapsto x$) on initial states with well-defined charge, i.e., we assume $p(Q) = \delta_{Q,Q_0}$ for some fixed value $Q_0$. 
We will follow the same strategy as in the previous case, and will obtain as a result a standard (non-generalized) Scrooge ensemble determined uniquely by the value of $Q_0$. In the special case of $Q_0 = N/2$ (charge neutrality), this will reduce to the Haar ensemble. 

\subsubsection{Calculation of $f_p$ coefficients}

In this case, the general form of the projected ensemble's moment operators, Eq.~\eqref{eq:app_rhotarget_moment} depends on coefficients 
\begin{align}
    f_p(\vec{T},x)
    & = \mathop{\mathbb{E}}_{\Phi_{Q_0} \sim {\rm Haar}(\mathcal{H}_{Q_0}) } 
    \left[ p(x) \prod_{Q_A} p(Q_A|x)^{T_{Q_A}} \right] 
    = \mathop{\mathbb{E}}_{\Phi_{Q_0} \sim {\rm Haar}(\mathcal{H}_{Q_0}) } 
    \left[ p(+) \prod_{Q_A} p(Q_A|+)^{T_{Q_A}} \right].
\end{align}
Here, using the invariance of the measure under $U(1)$-symmetric unitaries, we have absorbed a unitary $\bigotimes_i Z_i^{x_i}$ to map any measurement outcome $\ket{x}$ to $\ket{+}^{\otimes N_B}$. In this derivation we will drop the tensor power from the notation and just write $\ket{+}$ in lieu of $\ket{+}^{\otimes N_B}$. 
Following the same replica approach used for the case of $Z$ measurements before, we can define
\begin{align}
    f_p^{(n)} (\vec{T},+)
    & = \mathop{\mathbb{E}}_{\Phi_{Q_0} \sim {\rm Haar}(\mathcal{H}_{Q_0}) } 
    \left[ \left(\sum_{Q_A} p(Q_A,+) \right)^n \prod_{Q_A} p(Q_A,+)^{T_{Q_A}} \right] \nonumber \\
    & = \sum_{\vec{T}' \in \mathcal{T}(n,N)} \binom{n}{\vec{T}'} \mathop{\mathbb{E}}_{\Phi_{Q_0} \sim {\rm Haar}(\mathcal{H}_{Q_0}) }  \left[ \prod_{Q_A} p(Q_A,+)^{T_{Q_A} + T_{Q_A}'}\right]
\end{align}

The argument of the Haar average can now be written as
\begin{equation}
    \prod_{Q_A=0}^{N_A} p(Q_A,+)^{\ell(Q_A)} 
    = \bra{\Phi_{Q_0}}^{\otimes r} \left( \bigotimes_{Q_A=0}^{N_A} \hat{\Pi}_{Q_A}^{\otimes \ell(Q_A)} \otimes \ketbra{+}^{\otimes r} \right) \ket{\Phi_{Q_0}}^{\otimes r},
\end{equation}
with $r = n+k$ the total number of replicas and $\ell(Q_A) = T_{Q_A} + T'_{Q_A}$. Thus the average yields 
\begin{align}
    \mathop{\mathbb{E}}_{\Phi_{Q_0} \sim {\rm Haar}(\mathcal{H}_{Q_0}) }  \left[ \prod_{Q_A} p(Q_A,+)^{\ell(Q_A)}\right]
    & = \frac{[\binom{N}{Q_0}-1]!}{[\binom{N}{Q_0}-1+r]!} \sum_{\sigma\in S_r} 
    {\rm Tr} \left[ \hat{\sigma}_{AB,Q_0} 
    \left( \bigotimes_{Q_A=0}^{N_A} \hat{\Pi}_{Q_A}^{\otimes \ell(Q_A)} \otimes \ketbra{+}^{\otimes r} \right) \right],
\end{align}
with $\hat{\sigma}_{AB,Q_0}$ the replica permutation operator acting on charge sector $Q_0$ of the whole system $AB$. This equals $\hat{\sigma}_{AB} \hat{\Pi}_{Q_0}^{\otimes r}$, with $\hat{\sigma}_{AB}$ the replica permutation acting on the whole Hilbert space of $AB$ (not just sector $Q_0$); in turn, this factors across the Hilbert spaces of the two subsystems, so in all
$\hat{\sigma}_{AB,Q_0} = (\hat{\sigma}_A \otimes \hat{\sigma}_B) \hat{\Pi}_{Q_0}^{\otimes r}$. 
Thus we have 
\begin{align}
    {\rm Tr} \left[ \hat{\sigma}_{AB,Q_0} 
    \left( \bigotimes_{Q_A=0}^{N_A} \hat{\Pi}_{Q_A}^{\otimes \ell(Q_A)} \otimes \ketbra{+}^{\otimes r} \right) \right]
    & = {\rm Tr} \left[ (\hat{\sigma}_A\otimes\hat{\sigma}_B)
    \left( \bigotimes_{Q_A=0}^{N_A} (\hat{\Pi}_{Q_A} \otimes \hat{\Pi}_{Q_B})^{\otimes \ell(Q_A)} \right) (\mathbb{I}_A\otimes \ketbra{+})^{\otimes r} \right] \nonumber \\
    & = {\rm Tr}_A \left[\hat{\sigma}_A \bigotimes_{Q_A} \hat{\Pi}_{Q_A}^{\otimes \ell(Q_A)} \right]
    \bra{+}^{\otimes r} \hat{\sigma}_B \bigotimes_{Q_A} \hat{\Pi}_{Q_0-Q_A}^{\otimes \ell(Q_A)} \ket{+}^{\otimes r} \nonumber \\
    & = {\rm Tr}_A \left[\hat{\sigma}_A \bigotimes_{Q_A} \hat{\Pi}_{Q_A}^{\otimes \ell(Q_A)} \right]
    \prod_{Q_A} \bra{+} \hat{\Pi}_{Q_0-Q_A} \ket{+}^{\ell(Q_A)} \label{eq:app_xmeas_fp_trace}
\end{align}
where we used the fact that $ \hat{\Pi}_{Q_0} (\hat{\Pi}_{Q_A} \otimes \mathbb{I}_B) = \hat{\Pi}_{Q_A} \otimes \hat{\Pi}_{Q_B}$ with $Q_B = Q_0 - Q_A$, and that $\ket{+}^{\otimes r}$ is invariant under replica permutations. 

Now we will analyze the two terms in Eq.~\eqref{eq:app_xmeas_fp_trace}. 
Starting with the second term, we have 
\begin{equation}
    \bra{+} \hat{\Pi}_{Q_B} \ket{+}
    = \sum_{z:\, Q_B(z)=Q_B} |\braket{+}{z}|^2 = 2^{-N_B} \binom{N_B}{Q_B},
\end{equation}
which is independent of the permutation $\sigma$. 
For the first term, we note that it vanishes if $\sigma$ connects two replicas with a different value of $Q_A$, since in that case one gets a product of orthogonal projectors $\hat{\Pi}_{Q_A} \hat{\Pi}_{Q_A'} = 0$ if $Q_A\neq Q_A'$. This restricts $\sigma$ to be a product of permutations that only mix replicas with the same value of $Q_A$: $\sigma \in S_{\ell(0)} \times S_{\ell(1)} \times \cdots \times S_{\ell(N_A)}$ (for cases where $\ell(Q_A) = 0$ we take $S_0$ to be the trivial group). With this restriction on $\sigma$, we obtain
\begin{align}
    \sum_{\sigma\in S_r} {\rm Tr}_A \left[\hat{\sigma}_A \bigotimes_{Q_A} \hat{\Pi}_{Q_A}^{\otimes \ell(Q_A)} \right]
    & = \prod_{Q_A} \sum_{\sigma\in S_{\ell(Q_A)}} {\rm Tr}_A \left[\hat{\sigma}_A \hat{\Pi}_{Q_A}^{\otimes \ell(Q_A)} \right] \nonumber \\
    & = \prod_{Q_A} \sum_{\sigma\in S_{\ell(Q_A)}} {\rm Tr}_A (\hat{\sigma}_{A,Q_A}) \nonumber \\
    & = \prod_{Q_A} \frac{[\binom{N_A}{Q_A}-1+\ell(Q_A)]!}{[\binom{N_A}{Q_A}-1]!} {\rm Tr}(\rho^{(\ell(Q_A))}_{{\rm Haar},A,Q_A}) \nonumber \\
    & = \prod_{Q_A} \frac{[\binom{N_A}{Q_A}-1+\ell(Q_A)]!}{[\binom{N_A}{Q_A}-1]!},
\end{align}
where we carried out the sum over permutations by composing the $\ell(Q_A)$-th moment operator for the Haar ensemble in each charge sector.

Plugging these results back in the expression for $f_p^{(n)}(\vec{T},+)$ yields
\begin{align}
f_p^{(n)} (\vec{T},+)
    & = \frac{[\binom{N}{Q_0}-1]!}{[\binom{N}{Q_0}-1+r]!} 2^{-rN_B}
    \sum_{\vec{T}' \in \mathcal{T}(n,N)} \binom{n}{\vec{T}'} 
    \prod_{Q_A=0}^{N_A} \binom{N_B}{Q_0-Q_A}^{T_{Q_A} + T'_{Q_A}} \frac{[\binom{N_A}{Q_A}-1+T_{Q_A} + T'_{Q_A}]!}{[\binom{N_A}{Q_A}-1]!}.
    \label{eq:app_fpn_exact_x}
\end{align}

\subsubsection{Approximation and replica limit}

Again, Eq.~\eqref{eq:app_fpn_exact_x} is not directly amenable to a replica limit $n\to 1-k$. However, if we take the limit of large Hilbert space dimension $\binom{N_A}{Q_A} \to \infty$ before the replica limit, we can approximate 
\begin{equation}
    \frac{[\binom{N_A}{Q_A}-1+T_{Q_A} + T'_{Q_A}]!}{[\binom{N_A}{Q_A}-1]!}
    \simeq \binom{N_A}{Q_A}^{T_{Q_A} + T_{Q_A}'}.
\end{equation}
This allows an exact (within the large-$N_A$ limit) resummation of Eq.~\eqref{eq:app_fpn_exact_x}, 
\begin{align}
f_p^{(n)} (\vec{T},+)
    & \simeq \frac{[\binom{N}{Q_0}-1]!}{[\binom{N}{Q_0}-1+r]!} 2^{-rN_B}
    \sum_{\vec{T}' \in \mathcal{T}(n,N)} \binom{n}{\vec{T}'} 
    \prod_{Q_A=0}^{N_A} \left[ \binom{N_B}{Q_0-Q_A} \binom{N_A}{Q_A} \right]^{T_{Q_A} + T'_{Q_A}} \nonumber \\
    & = \frac{[\binom{N}{Q_0}-1]!}{[\binom{N}{Q_0}-1+r]!} 2^{-rN_B} \left(\sum_{Q_A} \binom{N_B}{Q_0-Q_A} \binom{N_A}{Q_A} \right)^n \prod_{Q_A=0}^{N_A} \left[ \binom{N_B}{Q_0-Q_A} \binom{N_A}{Q_A} \right]^{T_{Q_A}}.
\end{align}
We can now take $n\to 1-k$ (and $r=n+k\to 1$) and, using the identity $\sum_{Q_A} \binom{N_A}{Q_A} \binom{N-N_A}{Q_0-N_A} = \binom{N}{Q_0}$, we obtain 
\begin{align}
f_p (\vec{T},+)
    & \simeq 2^{-N_B} 
    \prod_{Q_A=0}^{N_A} \left[ \binom{N_B}{Q_0-Q_A} \binom{N_A}{Q_A} \binom{N}{Q_0}^{-1} \right]^{T_{Q_A}}
    = 2^{-N_B} \prod_{Q_A=0}^{N_A} \pi(Q_A|Q_0)^{T_{Q_A}},
    \label{eq:app_xbasis_fp}
\end{align}
with $\pi(Q_A|Q_0) =  \binom{N_A}{Q_A} \binom{N_B}{Q_0-Q_A} \binom{N}{Q_0}^{-1} $. 

\subsubsection{Reverse-engineering the ensemble from the moment operators}
Plugging the approximate result Eq.~\eqref{eq:app_xbasis_fp} back into the general expression for moment operators Eq.~\eqref{eq:app_rhotarget_moment} gives
\begin{equation}
    \rho_{\rm target}^{(k)} \simeq 
    \hat{\Pi}_{\rm sym}^{(k)} 
    \left( \sum_{\vec{T} \in \mathcal{T}(k,N) } \binom{k}{\vec T}^2 \bigotimes_{Q_A=0}^{N_A} \pi(Q_A|Q_0)^{T_{Q_A}} \rho_{{\rm Haar},A,Q_A}^{(T_{Q_A})} \right) \hat{\Pi}_{\rm sym}^{(k)}
\end{equation}
(the summation over $x$ cancels the $2^{-N_B}$ prefactor). 
Following the same steps as in the case of $z$-basis measurements, we can show that this moment operator corresponds to an ensemble obtained by drawing a collection of Haar-random states $\{ \ket{\phi_{Q_A}} \}$ in each charge sector and then outputting
\begin{align}
\ket{\psi} 
& = \sum_{Q_A} \sqrt{\pi(Q_A|Q_0)} \ket{\phi_{Q_A}}
= \frac{ \rho^{1/2} \ket{\xi} }{\bra{\xi} \rho \ket{\xi}^{1/2}},
\end{align}
where
\begin{align}
\ket{\xi} & = 2^{-N_A/2} \bigoplus_{Q_A} \binom{N_A}{Q_A}^{1/2} \ket{\phi_{Q_A}}, \label{eq:app_xi2} \\
\rho & = \sum_{Q_A=0}^{N_A} \pi(Q_A|Q_0) \frac{\hat{\Pi}_{Q_A}}{\Tr(\hat{\Pi}_{Q_A})} 
= \sum_{Q_A} \binom{N_B}{Q_0 - Q_A} \binom{N}{Q_0}^{-1} \hat{\Pi}_{Q_A}. 
\end{align}
In the limit $N_A\to\infty$, the states $\ket{\xi}$ become Haar-distributed, so this is a $\rho$-distortion of the Haar ensemble, i.e., a Scrooge ensemble. 
The Scrooge ensemble is fully specified by the density matrix $\rho$, which in turn depends only on the charge of the input state. In the special case of $Q_0 = N/2$ (charge neutrality) and in the limit $N\to\infty$, the density matrix $\rho$ becomes close to the identity, and we recover the Haar ensemble. Values of $Q_0$ above or below neutrality will yield a Scrooge ensemble with a bias toward higher or lower charge, respectively. 

\subsubsection{Typical initial states}
Thus far we have computed the average of the moment operators over generator states $\Phi_{Q_0}$ belonging to a given charge sector $Q_0$, 
\begin{equation} 
\rho^{(k)}_\text{target} = \mathop{\mathbb{E}}_{\Phi_{Q_0} \sim {\rm Haar}(\mathcal{H}_{Q_0})} \left[ \rho^{(k)}_{\mathcal{E}(\Phi_{Q_0} )} \right],
\end{equation}
and conjectured that the projected ensemble constructed from a {\it typical} state $\Phi_{Q_0}$ should have moments close to $\rho_\text{target}^{(k)}$. 
However, as long as the state has a nontrivial charge density (that is, $Q_0 = \sigma_0 N$ with $0 < \sigma_0 < 1$), we can prove this conjecture by using the method based on Levy's lemma (Lemma~\ref{lemma:levy} in Appendix~\ref{app:thm1}), already applied in our proof of Theorem~\ref{thm:1} following Ref.~\cite{cotler_emergent_2023}.

The proof in Ref.~\cite{cotler_emergent_2023} carries over straightforwardly to our case by replacing the total Hilbert space dimension $d$ by the dimension of the relevant charge sector,
$d_{Q_0} = \binom{N}{Q_0} = 2^{N H_2(\sigma_0) + o(N)}$, with $H_2(\sigma_0) = -\sigma_0 \log_2(\sigma_0) - (1-\sigma_0) \log_2(1-\sigma_0)$ the binary entropy. 
This dimension is still exponential in $N$ provided $\sigma_0 \neq 0, 1$, guaranteeing the desired concentration of measure as $N\to\infty$ (with $\sigma_0$ fixed). 
For the Lipshitz constant $\eta$ we may use the same value $\eta = 2(2k-1)$ derived in Ref.~\cite{cotler_emergent_2023}. Indeed we have
\begin{equation}
    \sup_{ x\neq y \in \mathcal{H}_{Q_0} }  \frac{|f(x)-f(y)|}{\| x-y \|_2 } 
    \leq \sup_{ x\neq y \in \mathcal{H}} \frac{|f(x)-f(y)|}{\| x-y \|_2 } 
    \leq \eta,
\end{equation}
where with a slight abuse of notation we wrote $x\in \mathcal{H}$ to denote $x\in\mathbb{S}^{2 {\sf dim}(\mathcal{H})-1}$ (the real unit vector corresponding to a complex normalized wavefunction in the Hilbert space). The first inequality follows trivially from $\mathcal{H}_{Q_0} \subset \mathcal{H}$ while the second from the definition of Lipshitz constant.
In words, a charge sector subspace defines a hypersphere $\mathbb{S}^{2d_{Q_0}-1}$, which is a submanifold of the hypersphere $\mathbb{S}^{2d-1}$ associated to the whole Hilbert space; thus the moment operator restricted to symmetric generator states is still an (entry-wise) Lipshitz-continuos function with a Lipshitz constant no larger than $\eta$. 
The proof then proceeds unchanged except for the replacement $d \mapsto d_{Q_0}$. 

This proves that Haar-random states $\ket{\Phi_{Q_0}}$ give rise to projected ensembles whose moment operators are very close to the average $\rho^{(k)}_{\rm target}$ with high probability. However, note that our derivation of the average $\rho^{(k)}_{\rm target}$ itself is non-rigorous (relying on a replica limit).


\subsection{Approximation error \label{app:approximation_error}}

Here we discuss the approximation of large Hilbert space dimensions $\binom{N}{Q}, \binom{N_A}{Q_A}$ used in the derivations above. This approximation is problematic since it is taken before a replica limit: it is not clear how large these Hilbert space dimensions should be for the approximation to be reliable. 
Here we compute the first-order correction to our result in the largest of the two small parameters, $\binom{N_A}{Q_A}^{-1}$, and find that it is of order $k^2 \binom{N_A}{Q_A}^{-1}$.  This is not enough to rigorously control the approximation and ultimately we will need to rely on independent numerical confirmation of the result; but it suggests that the regime of validity of the approximation is $\binom{N_A}{Q_A} \gg k^2$. 

We start from the exact form of the $f^{(n)}_p(\vec{T})$ coefficient, Eq.~\eqref{eq:app_fpn_exact}. We observe that, for two integers $D\gg \ell$, 
\begin{equation}
    \frac{(D+\ell)!}{D!} = D^\ell \exp \left[\frac{\ell(\ell+1)}{2D} + O(\ell^3/D^2)\right].
\end{equation}
This is also known as ``birthday paradox asymptotics''. 
Our approximation to Eq.~\eqref{eq:app_fpn_exact} consisted of keeping only the $D^\ell$ term. Here we consider the effect of additionally keeping the $e^{\ell(\ell+1)/2D}$ factor, for $D = \binom{N_A}{Q_A}$ (the largest of the two Hilbert space dimensions). 

We have 
\begin{align}
f_p^{(n)}(\vec{T},z)
    & \simeq \sum_{\vec{T}'\in \mathcal{T}(n,N)} \binom{n}{\vec{T}'} 
    \prod_{Q_A=0}^{N_A} \left[ 
    p_{\rm in}(Q)^\ell
    {\binom{N_A}{Q_A}}^\ell {\binom{N}{Q}}^{-\ell}
    \exp\left( \frac{\ell(\ell+1)}{2\binom{N_A}{Q_A}} \right)
    \right]_{\substack{Q = Q_A + Q_B(z) \\ \ell= T_{Q_A} + T'_{Q_A}} } \nonumber \\
    & \simeq \sum_{\vec{T}'\in \mathcal{T}(n,N)} \binom{n}{\vec{T}'} 
    \prod_{Q_A=0}^{N_A} \left[ 
    p_{\rm in}(Q)^\ell
    {\binom{N_A}{Q_A}}^\ell {\binom{N}{Q}}^{-\ell}\right]_{\substack{Q = Q_A + Q_B(z) \\ \ell= T_{Q_A} + T'_{Q_A}} } \nonumber \\
    & \qquad \qquad \times \left(1 + \frac{1}{2} \sum_{Q_A} \frac{(T_{Q_A} + T'_{Q_A})(T_{Q_A} + T'_{Q_A}+1)}{\binom{N_A}{Q_A}} + \dots \right)
\end{align}
where $\dots$ represent neglected higher-order terms in $\binom{N_A}{Q_A}^{-1}$. 
The $1$ term in the last line yields the result of our original approximation. Thus the leading correction to our approximation is given by 
\begin{align}
\delta f_p^{(n)}(\vec{T},z)
    & \simeq \frac{1}{2} \sum_{\vec{T}'\in \mathcal{T}(n,N)} \binom{n}{\vec{T}'} \sum_{Q_A} \frac{(T_{Q_A} + T'_{Q_A})(T_{Q_A} + T'_{Q_A}+1)}{\binom{N_A}{Q_A}}   
    \prod_{Q_A=0}^{N_A} \left[ 
    p_{\rm in}(Q)
    {\binom{N_A}{Q_A}} {\binom{N}{Q}}^{-1}\right]^{T_{Q_A} + T_{Q_A}'}_{\substack{Q = Q_A + Q_B(z)} } \nonumber \\
    & = \sum_{Q_A} \frac{T_{Q_A}(T_{Q_A}+1)s_{0,Q_A} + (2T_{Q_A}+1)s_{1,Q_A} + s_{2,Q_A}}{2\binom{N_A}{Q_A}}   
    \prod_{Q_A'=0}^{N_A} \left[ p_{\rm in}(Q) {\binom{N_A}{Q_A'}} {\binom{N}{Q}}^{-1}\right]^{T_{Q_A'}}_{\substack{Q = Q_A' + Q_B(z)} }, \label{eq:app_delta_f}
\end{align}
where we introduced the sums
\begin{equation}
    s_{m,Q_A} \equiv \sum_{\vec{T}'\in \mathcal{T}(n,N)} \binom{n}{\vec{T}'} (T'_{Q_A})^m 
    \prod_{Q_A'=0}^{N_A} \left[ p_{\rm in}(Q) {\binom{N_A}{Q_A'}} {\binom{N}{Q}}^{-1}\right]^{T_{Q_A'}'}_{\substack{Q = Q_A' + Q_B(z)} }
\end{equation}    

The $s_{0}$ term (independent of $Q_A$) is straightforwardly computed by recognizing the multinomial expansion:
\begin{equation}
    s_{0} = \left[ \sum_{Q_A'} p_{\rm in}(Q_A' + Q_B(z)) \binom{N_A}{Q_A'} \binom{N}{Q_A'+Q_B(z)}^{-1} \right]^n. 
\end{equation}
For the $s_{m,Q_A}$ terms with $m>0$, we make use of the following identity: for any sequence $\{x_q\}_{q=1}^N$, we have
\begin{equation}
    x_{q^*}^m \partial^m_{x_{q^*}} \left(\sum_q x_q \right)^n
    = \sum_{\vec{T} \in \mathcal{T}(n,N)} \binom{n}{\vec{T}} x_{q^*}^m \partial^m_{x_{q^*}}  \prod_q x_q^{T_q}
    = \sum_{\vec{T} \in \mathcal{T}(n,N)} \binom{n}{\vec{T}} T_{q^*}(T_{q^*}-1)\cdots (T_{q^*}-m+1)  \prod_q x_q^{T_q}
\end{equation}
from which it follows that
\begin{align}
    \sum_{\vec{T} \in \mathcal{T}(n,N)} \binom{n}{\vec{T}} T_{q^*} \prod_q x_q^{T_q}
    & = x_{q^*} \partial_{x_{q^*}}  \left(\sum_q x_q \right)^n
    = n \frac{x_{q^*}}{\sum_q x_q} \left(\sum_q x_q \right)^{n} \\
    \sum_{\vec{T} \in \mathcal{T}(n,N)} \binom{n}{\vec{T}} (T_{q^*})^2 \prod_q x_q^{T_q}
    & = (x^2_{q^*} \partial^2_{x_{q^*}} + x_{q^*} \partial_{q^*})  \left(\sum_q x_q \right)^n
    = \left[n(n-1) \left(\frac{x_{q^*}}{\sum_q x_q}\right)^2 + n \frac{x_{q^*}}{\sum_q x_q} \right] \left(\sum_q x_q \right)^{n}.
\end{align}
Using these identities, applied to the sequence $\{ x_q \}_q \mapsto \{ p_{\rm in}(Q_A'+Q_B) \binom{N_A}{Q_A'} \binom{N}{Q_A'+Q_B}^{-1}\}_{Q_A'} $, gives
\begin{align}
    s_{1, Q_A} 
    & = n \frac{p_{\rm in}(Q_A+Q_B) \binom{N_A}{Q_A} \binom{N}{Q_A+Q_B}^{-1}}{\sum_{Q_A'} p_{\rm in}(Q_A'+Q_B) \binom{N_A}{Q_A'} \binom{N}{Q_A+Q_B}^{-1}} s_0 
    = n p(Q_A|z) s_0,  \\
    s_{2,Q_A} 
    & = n p(Q_A|z) [1+(n-1)p(Q_A|z) ] s_0. 
\end{align}

At this point we can plug the $s_{m,Q_A}$ sums into Eq.~\ref{eq:app_delta_f}, take the replica limit $n\to 1-k$, and carry out the same steps we used in previous derivations to conclude
\begin{align}
    \delta f(\vec{T},z)
    & = \binom{N_B}{Q_B}^{-1} p(Q_B(z)) \left( \prod_{Q_A} p(Q_A|z)^{T_{Q_A}}\right) \nonumber \\
    & \qquad \qquad \times \sum_{Q_A} \frac{T_{Q_A} (T_{Q_A}+1) + (1-k) p(Q_A|z) [2T_{Q_A}+2 -k p(Q_A|z) ]}{2\binom{N_A}{Q_A}}.
\end{align}
The first line equals the result of our leading-order approximation. 
We see therefore that the relative error on our result from neglecting order-$\binom{N_A}{Q_A}^{-1}$ corrections depends on factors such as
\begin{equation}
    \frac{T_{Q_A}^m}{\binom{N_A}{Q_A}}, \qquad \frac{k^m}{\binom{N_A}{Q_A}}
\end{equation}
where $m = 0,1,2$. Since $T_{Q_A}$ is an entry of the ``type'' vector $\vec{T} \in \mathcal{T}(N,k)$, it obeys $T_{Q_A}\leq k$. Thus the largest error terms are of order $k^2 \binom{N_A}{Q_A}^{-1}$. The factors of $p(Q_A|z)$ are typically of order $1/N_A$ but may be as large as 1 depending on $p$, so they do not change this scaling. 

We conclude that the validity of our approximation likely relies on the moment $k$ being sufficiently small relative to the charge sector dimensions $\binom{N_A}{Q_A}$. Another fact to note is that the system always features trivial charge sectors ($Q_A = 0, N_A$) of dimension 1, where this requirement will never be satisfied, but their contribution becomes negligible in the limit of large $N_A$. 

\section{Constraints from locality of the dynamics \label{app:locality}}

Refs.~\cite{marvian_restrictions_2022,marvian_theory_2023} have showed that not all $U(1)$-symmetric unitaries on an $N$-qubit Hilbert space can be generated from $U(1)$-symmetric interactions among $k$ qubits, if $k<N$. This applies to the random circuit models we consider in Sec.~\ref{sec:numerics}, where $k=2$ (note that geometric locality is unimportant since $U(1)$-symmetric 2-qubit gates include the SWAP gate). This calls into question the legitimacy of averaging the collection of states $\{\ket{\Phi_Q} \}_{Q=0}^N$ independently over the Haar ensemble in each charge sector $Q$, as we do in Appendix~\ref{app:ansatz} to derive the form of the target ensemble. 
Here we justify this procedure. 

Let us denote the group generated by 2-local, $U(1)$-symmetric gates on $N$ qubits by $G_{U(1)}^{N,2}$. This is a compact Lie group, so it admits a Haar measure. Our goal is to show that the Haar average over $G_{U(1)}^{N,2}$ is equivalent, for our purposes, to the Haar average over the entire $U(1)$-symmetric subgroup of $U(2^N)$, $G_{U(1)}^{N,N}$.
Ref.~\cite{marvian_theory_2023} shows that $G_{U(1)}^{N,2}$ contains $\bigoplus_Q SU(\mathcal{H}_Q)$---the group of block-diagonal unitaries that act on each charge sector $\mathcal{H}_Q$ as elements of the special unitary group (i.e., with determinant equal to 1). However, the relative phases between different sectors are subject to nontrivial constraints, so that it is not possible to generate all of $G_{U(1)}^{N,N} = \bigoplus_Q U(\mathcal{H}_Q)$. 
For our purposes, these relative phases will be unimportant, as we show next.

\begin{lemma}
For any function $f$ and any state $\ket{\Psi}$, we have
\begin{equation}
    \mathop{\mathbb{E}}_{ U\sim {\rm Haar}(G_{U(1)}^{N,2}) } \left[ f(U\ketbra{\Psi}U^\dagger) \right]
    =
    \mathop{\mathbb{E}}_{ U\sim {\rm Haar}(G_{U(1)}^{N,N}) } \left[ f(U\ketbra{\Psi}U^\dagger) \right],
\end{equation}
that is, as long as the unitary $U$ acts on a pure state, locally-generated $U(1)$-symmetric unitaries are equivalent to general $U(1)$-symmetric unitaries.
\end{lemma}

{\it Proof.}
Using the fact that $G_{U(1)}^{N,2}$ contains as a subgroup $\bigoplus_Q SU(\mathcal{H}_Q)$, and using the invariance of the Haar measure under right multiplication by elements of $G_{U(1)}^{N,2}$, we can change variable to $U \mapsto UV$, with $V = \bigoplus V_Q$, and $V_Q \in SU(\mathcal{H}_Q)$. Let us consider $V_Q = \tilde{V}_Q [(e^{i\theta_Q} \ketbra{\Phi_Q}) \oplus W_{\theta_Q}^\perp]$, where $\theta_Q$ is an arbitrary phase, $W_{\theta_Q}^\perp$ is a unitary matrix on the orthogonal complement of $\ket{\Phi_Q}$ in $\mathcal{H}_Q$ of determinant $e^{-i\theta_Q}$, and $\tilde{V}_Q$ is another unitary matrix in $SU(\mathcal{H}_Q)$. 
The action of $V$ on the state $\ket{\Psi} = \sum_Q \sqrt{p(Q)} \ket{\Phi_Q}$ is 
\begin{equation}
    V\ket{\Psi} = \sum_Q \sqrt{p(Q)} V_Q \ket{\Phi_Q}
    = \sum_Q \sqrt{p(Q)} e^{i\theta_Q} \tilde{V}_Q \ket{\Phi_Q}
\end{equation}
which in turn is equivalent to that of a general matrix $\bigoplus_Q e^{i\theta_Q} \tilde{V}_Q \in \bigoplus_Q U(\mathcal{H}_Q) = G_{U(1)}^{N,N}$. 
$\blacksquare$

\section{Charge distribution of product states \label{app:P_out}}

Here we derive the probability distribution of charge $p_{\rm out}(Q_B|\nu)$ on measurement basis states $\{ \ket{\nu} \}$ used in Sec.~\ref{sec:numerics}. These basis states are obtained from the computational basis by applying a rotation $R_x(\theta) = e^{-i\theta X/2}$ to each qubit, with the same angle $\theta$ on all qubits.

Let us denote the number of `1' measurement outcomes by $[\nu] \equiv \sum_i \nu_i$. 
Up to permutations of the qubits and overall phase factors, the basis state reads
\begin{equation}
    \left( \cos(\frac{\theta}{2})|0\rangle + \sin(\frac{\theta}{2})|1\rangle\right)^{\otimes N_B-[\nu]} 
    \otimes \left(\sin(\frac{\theta}{2})|0\rangle - \cos(\frac{\theta}{2})|1\rangle \right)^{[\nu]},
\end{equation}
where $\theta$ is the rotation angle of the measurement basis relative to the computational basis. 
Now we consider all amplitudes that contribute to total charge $Q_B$. 
These can be broken down into amplitudes that have $n_1$ charges in $(\cos(\theta/2)|0\rangle + \sin(\theta/2)|1\rangle)^{\otimes N_B-[\nu]}$ and $n_2 = Q_B-n_1$ charges from $(\sin(\theta/2)|0\rangle - \cos(\theta/2)|1\rangle)^{\otimes [\nu]}$. 
With the convention that $\binom{a}{b} = 0$ if $b < 0$ or $b>a$, we can write
 \begin{align}
    p_{\rm out}(Q_B|\nu) 
    & = \sum_{n_1,n_2} \delta_{n_1+n_2,Q_B} \binom{N_B-[\nu]}{n_1}  
    \left| \bigg(\cos(\frac{\theta}{2})\bigg)^{N_B-[\nu]-n_1}\bigg(\sin(\frac{\theta}{2})\bigg)^{n_1} \right|^2 \nonumber \\
    & \qquad\qquad \times \binom{[\nu]}{n_2} \left| \bigg(\sin(\frac{\theta}{2})\bigg)^{[\nu]-n_2} \bigg(\cos(\frac{\theta}{2})\bigg)^{n_2}\right|^2\\
    & =   \sum_{n_1} \binom{N_B-[\nu]}{n_1} \binom{[\nu]}{Q_B-n_1} 
    \left(\cos^2(\theta / 2)\right)^{N_B-[\nu]+Q_B-2n_1} \left(\sin^2(\theta/2) \right)^{[\nu] + 2n_1-Q_B}.                        
\end{align}

This formula is used to generate the target GSE moment operators for Figs.~\ref{fig:partial_revealing} and~\ref{fig:sweep}. An analogous formula (with $Q_B\to Q$, $N_B\to N$, $[\nu]\to N/2$) is also used to generate the input charge distribution $p_{\rm in}(Q)$ for Fig.~\ref{fig:sweep}.

\section{Choice of parameters for numerical simulations}\label{app:param_choice}

 \begin{figure}
    \includegraphics[width=0.45\textwidth]{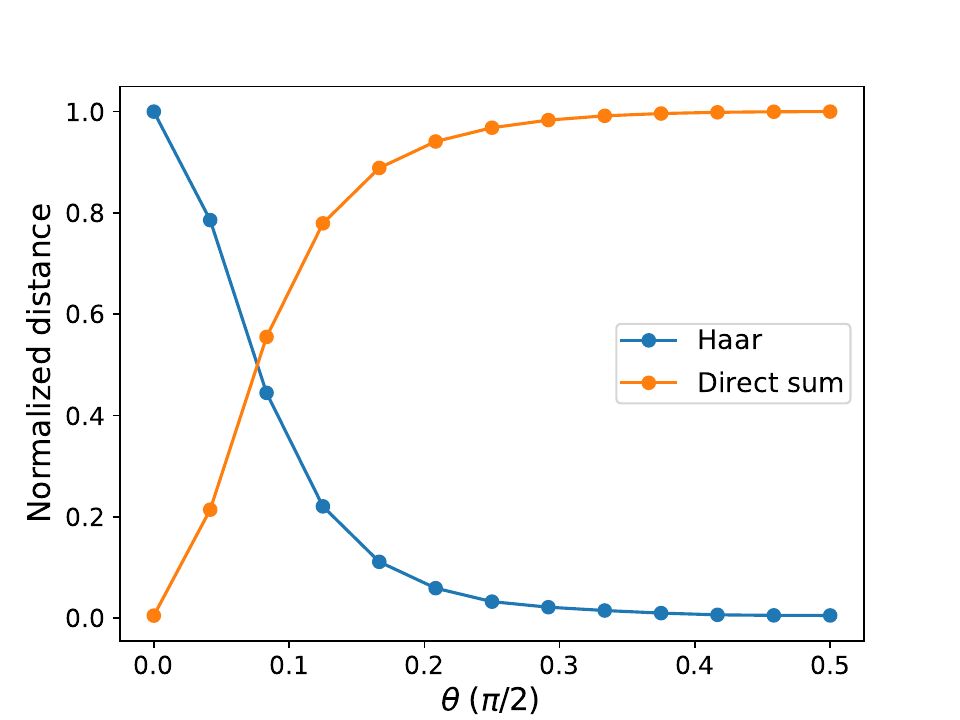}
    \caption{Late-time average of the second-moment distances between the projected ensemble and the Haar ensemble (blue) and the direct sum of Haar ensembles in the charge sectors (orange) as a function of $\theta$ (parameter of the initial state) for $N=20$, $N_A = 2$.}
    \label{fig:choice_theta}            
\end{figure}

Here we discuss parameter choices for the numerical simulations of intermediate initial states, where we test the convergence to the GSE.
First we justify the choice of $\theta = \pi/10$ as a reasonable intermediate value between the N\'eel and $X$-basis state. 
In Fig.~\ref{fig:choice_theta} we show the late-time average value of the distance $\Delta^{(2)} = \| \rho^{(2)}_{\mathcal E} - \rho^{(2)}_{\rm target} \|_{\rm tr}$ between the projected ensemble and a target ensemble chosen to be either the Haar ensemble on the whole Hilbert space, or the weighted sum of Haar ensembles on the charge sectors, as in Eq.~\eqref{eq:direct_sum_target}. 
As expected, the two target ensembles yield $\Delta^{(2)} \simeq 0$ respectively at $\theta = 0$ and $\theta = \pi/2$, but at intermediate $\theta$ they both give a large value of $\Delta^{(2)}$. 
We choose $\theta = \pi/10$ as it appears to be roughly equidistant from the two extremal ensembles. 

Finally we provide some details on the numerical construction of the GSE moment operators.
These are constructed numerically by sampling states $\ket{\psi} \sim \mathcal{E}_{GSE}$ (according to the prescription in the main text) $M$ times, and computing the approximate moment operator
\begin{equation}
    \rho^{(k)}_{\text{app.}, M} = \frac{1}{M} \sum_{m=1}^M \ketbra{\psi_m}^{\otimes k}.
    \label{eq:app_sampled_moment}
\end{equation}
It is important to ensure that the number of samples $M$ is adequate to guarantee convergence to the true GSE moment operator. 

\begin{figure}
    \includegraphics[width=0.5\textwidth]{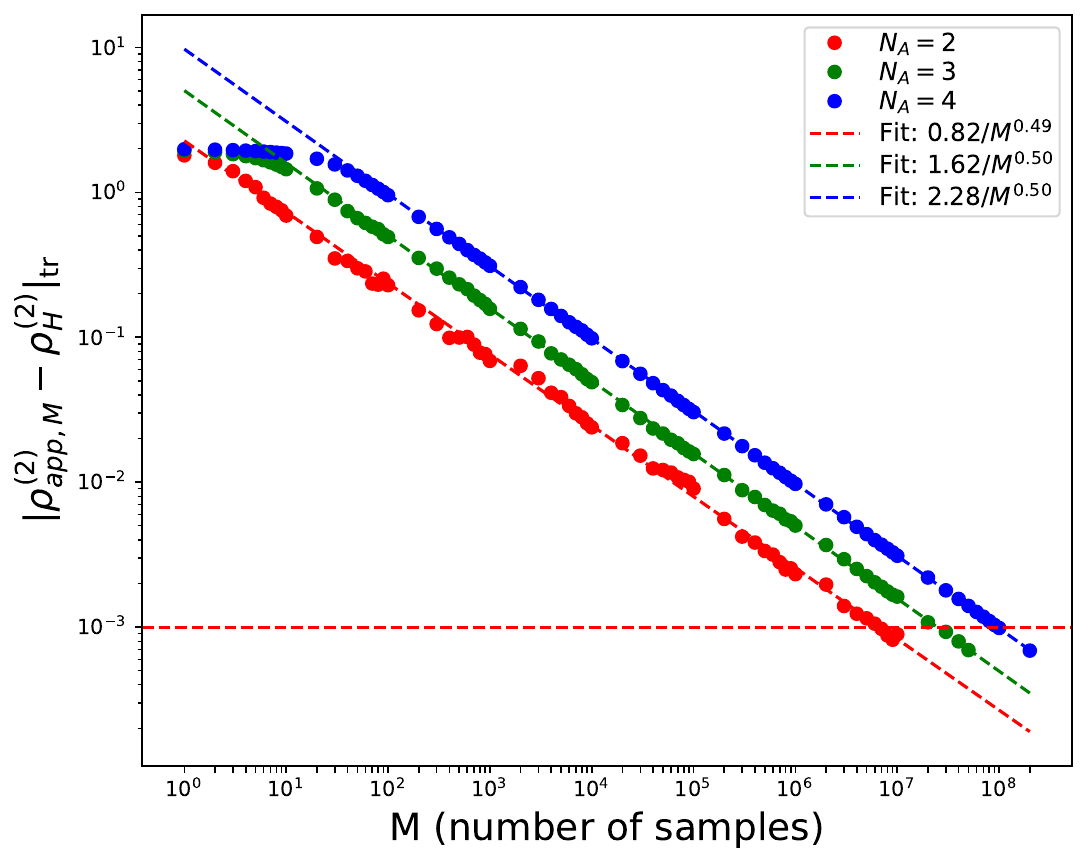}
    \caption{The distance $\|\rho^{(2)}_{\text{app.},M}-\rho^{(2)}_H\|$ between the sampled moment operator, Eq.~\eqref{eq:app_sampled_moment}, and the true Haar moment operator is plotted against the number of samples $M$, for subsystem sizes $N_A = 2,3,4$. We see the expected decay in the number of samples as $\sim M^{-1/2}$, with a prefactor roughly proportional to $2^{N_A}$. A precision of $\lesssim 10^{-3}$, which is below the finite-size floor of $\Delta^{(k)}$ for all cases studied in Fig.~\ref{fig:generalized_scrooge_combined}, is achieved for $M\gtrsim 10^8$. }
    \label{fig:choice_M}            
\end{figure}

Since the Hilbert space of the subsystem A has dimension $2^{N_A}$, it is reasonable to expect that the number of sampled states needed to ensure convergence should grow rapidly with $N_A$. 
To investigate this question quantitatively, we focused on the $|+\rangle^{\otimes N}$ initial state, where the form of the GSE reduces to the analytically tractable Haar ensemble, so as to have an exact benchmark to compare against. We calculated the distance between the sampled moment operator $\frac{1}{M} \sum_{m=1}^M \ketbra{\psi_m}^{\otimes 2}$ and the analytically constructed one, $\rho^{(2)}_H = \binom{2^{N_A}-1}{2}^{-1} \hat{\Pi}_{\rm sym}^{(2)}$. 
Results shown in Fig.~\ref{fig:choice_M} show that the statistical error due to finite sampling $M$ scales approximately as $\sim 2^{N_A}/\sqrt{M}$. This implies that reaching a target precision $\epsilon$ requires $M\sim 4^{N_A} /\epsilon^2$ samples. This is reasonable since a quantum state $k$-design on a Hilbert space of dimension $d_A$ requires $\Omega(d_A^k)$~\cite{roberts_chaos_2017}, so $4^{N_A} = d_A^2$ is the expected scaling for the case under consideration, $k = 2$. 
Since in Fig.~\ref{fig:generalized_scrooge_combined} in the main text we consider $N_A\leq 4$, to ensure a negligible error $\epsilon \lesssim 10^{-3}$ we set $M = 2\times 10^8$. 

\section{Numerical results for higher moments \label{app:higher_k}}

\begin{figure*}
    \centering
    \includegraphics[width=0.9\textwidth]{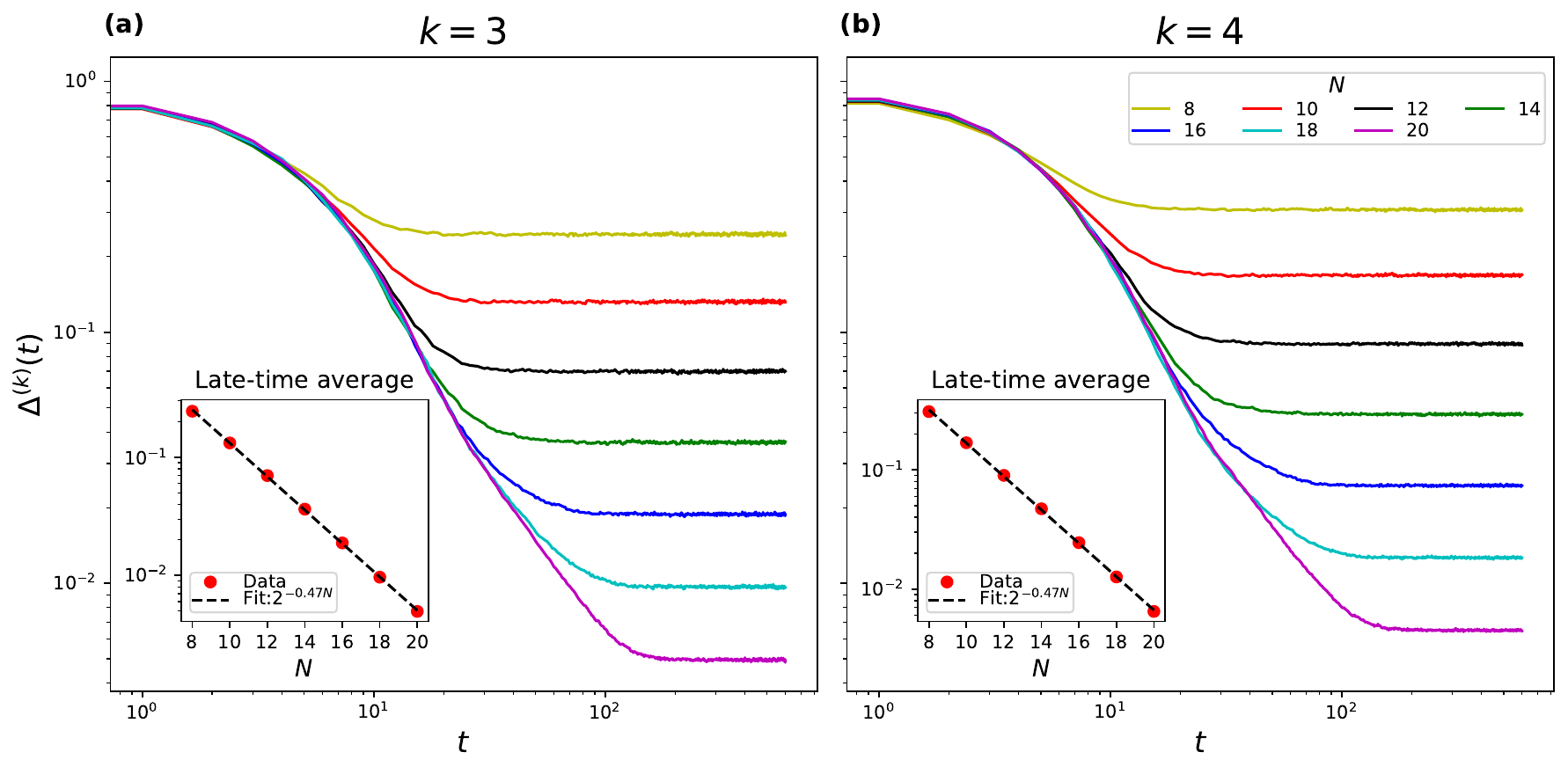}
    \caption{Trace distances $\Delta^{(3)}$ (a) and $\Delta^{(4)}$ (b) between the projected ensemble and the GSE constructed from the initial state of Eq.~\eqref{eq:init_states} with $\theta = \pi/2$ and $z$-basis measurements, for $N_A = 2$. The setting is the same as in Fig.~\ref{fig:generalized_scrooge_combined}. The late-time saturation value of $\Delta^{(3)}$ and $\Delta^{(4)}$ decays exponentially in $N$, consistent with $2^{-N/2}$ (inset).}
\label{fig:app_higher_moments}
\end{figure*}

Here we report numerical evidence for deep thermalization at the level of higher moments $k > 2$. 
Fig.~\ref{fig:app_higher_moments} shows data for moments $k = 3$ and $4$ for the setup considered in Sec.~\ref{sec:arbitrary} and Fig.~\ref{fig:generalized_scrooge_combined}: maximally charge-revealing measurements on a state that is initialized to Eq.~\eqref{eq:init_states} with $\theta = \pi/20$, i.e., a product state in a local basis intermediate between $x$ and $z$. 
The target ensemble is again set to the GSE. The third and fourth moment operators are computed by following the Monte Carlo method of Appendix~\ref{app:param_choice} with $M = 5\times 10^7$ and $2\times 10^8$ samples respectively.
Results are consistent with deep thermalization at all moments $k$. 

\end{document}